%% file: m0647JD.tex
\documentclass[iop]{emulateapj}
\usepackage{placeins}

\usepackage{amsmath}  % align equations

\usepackage{color}
\definecolor{linkcolor}{rgb}{0.85, 0, 0.5}
\usepackage[colorlinks=true,linkcolor=linkcolor,anchorcolor=blue,citecolor=blue,urlcolor=blue, draft=false]{hyperref}
\hypersetup{pdfauthor={Dan Coe}, pdftitle={CLASH: Three strongly lensed images of a candidate $\lowercase{z} \approx 11$ galaxy}}

\newcommand{\LCDM}{$\Lambda$CDM}

\newcommand{\ny}{$\tilde {\rm n}$}
\renewcommand{\~}{$\sim$}

\newcommand{\um}{$\mu$m}
\newcommand{\uJy}{$\mu$Jy}
\newcommand{\sig}{-$\sigma$}
\newcommand{\Lya}{Lyman-$\alpha$}
\renewcommand{\th}{$^{\rm th}$}

\newcommand{\tentothe}[1]{$10^{#1}$}
\newcommand{\tentotheminus}[1]{$10^{-#1}$}
\newcommand{\e}[1]{$\times 10^{#1}$}
\newcommand{\en}[1]{$\times 10^{-#1}$}

\newcommand{\super}[1]{$^{\rm #1}$}

\newcommand{\per}{$^{-1}$}
\newcommand{\inv}{\per}
\newcommand{\Mstar}{$M^*$}
\newcommand{\Lstar}{$L^*$}
\newcommand{\phistar}{$\phi^*$}

\newcommand{\LUV}{$L_{UV}$}

\newcommand{\Msun}{$M_\odot$}

\newcommand{\Hbeta}{H$\beta$}
\newcommand{\Ha}{H$\alpha$}

\newcommand{\macs}{MACSJ0647.7+7015}
\newcommand{\JD}{MACS0647-JD}

\newcommand{\Jdrop}{$J$-dropout}
\newcommand{\Jdrops}{\Jdrop s}

\newcommand{\etal}{et al.}

\newcommand{\ztwosig}{$z =10.7_{-0.4}^{+0.6}$}
\newcommand{\agetwosig}{$427_{+21}^{-30}$}

\shortauthors{Coe \etal\ 2012}
\slugcomment{Accepted for publication in the Astrophysical Journal, Monday 29$^{\rm th}$ October, 2012}
\usepackage{datetime}

\begin{document}

%\title{\large Draft \today~\currenttime}
%\title{\large Draft \monthname~\getdateday~\currenttime}
%\usdate
%\title{Draft \today~\currenttime}
%\shorttitle{Draft}

\title{CLASH: Three strongly lensed images of a candidate $\lowercase{z} \approx 11$ galaxy}
\shorttitle{CLASH: Three strongly lensed images of a candidate $z \approx 11$ galaxy}
%\title{Draft}
%\shorttitle{Draft}
%\shorttitle{M0647JD}
%\shorttitle{A strongly lensed candidate $z \sim 11$ galaxy}

% authors/affilorder.py

\author{Dan Coe\altaffilmark{1}}
\author{Adi Zitrin\altaffilmark{2}}
\author{Mauricio Carrasco\altaffilmark{2,3}}
\author{Xinwen Shu\altaffilmark{4}}
\author{Wei Zheng\altaffilmark{5}}
\author{Marc Postman\altaffilmark{1}}
\author{Larry Bradley\altaffilmark{1}}
\author{Anton Koekemoer\altaffilmark{1}}
\author{Rychard Bouwens\altaffilmark{6}}
\author{Tom Broadhurst\altaffilmark{7,8}}
\author{Anna Monna\altaffilmark{9}}
\author{Ole Host\altaffilmark{10,11}}
\author{Leonidas A.~Moustakas\altaffilmark{12}}
\author{Holland Ford\altaffilmark{5}}
\author{John Moustakas\altaffilmark{13}}
\author{Arjen van der Wel\altaffilmark{14}}
\author{Megan Donahue\altaffilmark{15}}
\author{Steven A.~Rodney\altaffilmark{5}}
\author{Narciso Ben\'itez\altaffilmark{16}}
\author{Stephanie Jouvel\altaffilmark{10,17}}
\author{Stella Seitz\altaffilmark{9,18}}
\author{Daniel D.~Kelson\altaffilmark{19}}
\author{Piero Rosati\altaffilmark{20}}

\altaffiltext{1}{Space Telescope Science Institute, Baltimore, MD, USA; \href{mailto:dcoe@stsci.edu}{DCoe@STScI.edu}} % STScI -- Dan, Marc
%\altaffiltext{2}{Institut f\"ur Theoretische Astrophysik, Zentrum f\"ur Astronomie der Universit\"at Heidelberg (ZAH), Heidelberg, Germany}  % Adi
\altaffiltext{2}{Institut f\"ur Theoretische Astrophysik, Zentrum f\"ur Astronomie, Institut f\"ur Theoretische Astrophysik, Albert-Ueberle-Str.~2, 29120 Heidelberg, Germany}  % Adi
\altaffiltext{3}{Department of Astronomy and Astrophysics, AIUC, Pontificia Universidad Cat\'olica de Chile, Santiago, Chile}  % Mauricio
\altaffiltext{4}{Department of Astronomy, University of Science and Technology of China, Hefei, China}  % Xinwen
\altaffiltext{5}{Department of Physics and Astronomy, The Johns Hopkins University, Baltimore, MD, USA}  % JHU - Wei, Larry, Holland
\altaffiltext{6}{Leiden Observatory, Leiden University, Leiden, The Netherlands}  % Rychard
\altaffiltext{7}{Department of Theoretical Physics, University of the Basque Country UPV/EHU, Bilbao, Spain}  % Tom
\altaffiltext{8}{Ikerbasque, Basque Foundation for Science, Bilbao, Spain}  % Tom
\altaffiltext{9}{Instituts f\"ur Astronomie und Astrophysik, Universit\"as-Sternwarte M\"unchen, M\"unchen, Germany}  % Stella, Anna
\altaffiltext{10}{Department of Physics and Astronomy, University College London, London, UK}  % UCL - Ole, Stephanie (since moved)
\altaffiltext{11}{Dark Cosmology Centre, Niels Bohr Institute, University of Copenhagen, Copenhagen, Denmark}  % Ole (new)
\altaffiltext{12}{Jet Propulsion Laboratory, California Institute of Technology, La Ca\ny ada Flintridge, CA, USA}  % Lexi
\altaffiltext{13}{Department of Physics and Astronomy, Siena College, Loudonville, NY, USA}  % John Moustakas
\altaffiltext{14}{Max Planck Institut f\"ur Astronomie (MPIA), Heidelberg, Germany}  % Arjen
\altaffiltext{15}{Department of Physics and Astronomy, Michigan State University, East Lansing, MI, USA}   % Megan
\altaffiltext{16}{Instituto de Astrof\'isica de Andaluc\'ia (IAA-CSIC), Granada, Spain}  % Txitxo
\altaffiltext{17}{Institut de Cincies de l'Espai (IEE-CSIC), Bellaterra (Barcelona), Spain}  % Stehpanie (new)
\altaffiltext{18}{Max-Planck-Institut f\"ur extraterrestrische Physik (MPE), Garching, Germany}  % Stella Seitz
\altaffiltext{19}{Carnegie Observatories, Carnegie Institute for Science, Pasadena, CA, USA}  % Dan Kelson
\altaffiltext{20}{European Southern Observatory (ESO), Garching, Germany}  %  Piero & old Mauricio

%\email{DCoe@STScI.edu}

%~/CLASH/team/affiliations.txt

%%%%%% ABSTRACT %%%%%%

\begin{abstract}

%We report the discovery of a candidate for the most distant galaxy known to date,
We present a candidate for the most distant galaxy known to date
with a photometric redshift 
%$z =11.0_{-0.7}^{+0.4}$ (95\% confidence)
%$z =11.0_{-0.4}^{+0.2}$
%$z =10.7_{-0.2}^{+0.4}$
%\zonesig\ observed when the universe was \ageonesig\ million years old
%(68\% confidence limits;
\ztwosig\ (95\% confidence limits;
with $z < 9.5$ galaxies of known types ruled out at 7.2\sig).  % gausstsig(3.4e-13)
%observed when the universe was \agetwosig\ million years old
%with $z < 9.7$ galaxies formally ruled out at 5.5\sig).  % gausstsig(2e-8)
%(68\% confidence)
%$\approx$400 
%$410_{+40}^{-20}$
%$410_{+20}^{-10}$
%million years old
%5.3\sig).  % gausstsig(6.3e-8)
%$z =11.0_{-0.7}^{+0.4}$ at 95\% confidence).
% the earliest galaxy yet detected,
%
This \Jdrop\ Lyman Break Galaxy, named \JD,
was discovered as part of the Cluster Lensing and Supernova survey with Hubble (CLASH).
We observe three magnified images of this galaxy
due to strong gravitational lensing by the galaxy cluster \macs\ at $z = 0.591$.
%
%The brighter two images are magnified by factors of \~8 and
The images are magnified by factors of \~8, 7, and 2,
with the brighter two observed at \~26th magnitude AB (\~0.15\uJy) in the WFC3/IR F160W filter ($\sim$1.4--1.7\um)
where they are detected at $\gtrsim$12\sig.
All three images are also confidently detected 
%at $\gtrsim$10\sig in F160W ($\sim$1.2--1.6\um)
at $\gtrsim$6\sig\ in F140W ($\sim$1.2--1.6\um),
dropping out of detection from 
%the $J$-band F125W ($\sim$1.1--1.4\um) and 
15 lower wavelength HST filters ($\sim$0.2--1.4\um),
and lacking bright detections in Spitzer/IRAC 3.6\um\ and 4.5\um\ imaging (\~3.2--5.0\um).
%and lacking bright detections in Spitzer/IRAC 3.6\um\ (\~3.2--3.9\um) and 4.5\um\ (\~4.0--5.0\um) imaging.
%
We rule out a broad range of possible lower redshift interlopers,
including some previously published as high redshift candidates.
%but later determined to be evolved and/or dusty galaxies at intermediate redshift.
%
Our high redshift conclusion is more conservative than if we had neglected a Bayesian photometric redshift prior.
%
%In spite of this, we find $P(z<9.7)$ \~  3\en{13}.
%$P(z<9.7)$ \~  2\en8.
%Our $z \sim 11$ conclusion is supported by multiple photometric redshift and strong lensing analyses.
%Our $z \sim 11$ conclusion is supported by two photometric redshift analyses and three strong lensing analyses.
%In spite of this, we find $P(z<9.7~{\rm galaxies})$ \~  2\en8.
%In spite of this, we find $P(z<9.7)$ \~  6.3\en8.
%
%Extrapolating results from lower redshifts,
%We estimate that \JD\ is toward the bright end of the luminosity function ($\sim$1--3$L^*$)
%and perhaps somewhat smaller than expected for a $z \sim 11$ galaxy, 
%and toward the small end of sizes expected for a $z \sim 11$ galaxy
%with a half-light radius $\lesssim 100$ pc,
%consistent with the sizes of star forming knots observed in lensed galaxes at $5 < z < 8$.
%(deconvolved and delensed),
%though we may only be detecting a bright star forming knot in a larger galaxy.
%
%Each of the brighter images is confidently detected in both HST filters
%observed at six different epochs over a period of 41 days.
%
%The discovery of a $z \sim 11$ galaxy magnified to 26th magnitude
%The discovery of this galaxy is in agreement with a lensed luminosity function extrapolated from lower redshifts
%
Given CLASH observations of 17 high mass clusters to date,
%including $\sim$78 square arcminutes of coverage with WFC3/IR,
our discoveries of \JD\ at $z \sim 10.8$ and MACS1149-JD1 at $z \sim 9.6$
are consistent with a lensed luminosity function extrapolated from lower redshifts.
This would suggest that low luminosity galaxies could have reionized the universe. 
%are consistent with number counts extrapolated from lower redshifts
%and convolved through our lens models for these clusters.
%
%This extrapolated luminosity function would suggest that low luminosity galaxies could have reionized the universe.
%
However given the significant uncertainties based on only two galaxies,
we cannot yet rule out the sharp drop off in number counts at $z \gtrsim 10$ suggested by field searches.

\end{abstract}

%%%%%%%%%%%%%%%%%

\keywords{early universe --
galaxies: high redshift --
galaxies: distances and redshifts --
galaxies: evolution --
gravitational lensing: strong --
galaxies: clusters: individual (MACSJ0647.7+7015)
%cosmology: dark matter --- 
%galaxies: clusters: individual (Abell 2261) ---
%gravitational lensing: strong --- 
%gravitational lensing: weak ---
%cosmology: dark energy ---
%galaxies: evolution
%galaxies: formation
}

%\vspace{-3in}

\nopagebreak[4]

\section{Introduction}
\label{sec:intro}

%FILLER
%\vspace{0.75in}

Current models of structure formation 
suggest that the first galaxies formed at $z \gtrsim 10$
when the universe was $\lesssim 470$ million years old
(\citealt{WiseAbel07,WiseAbel08,Greif08,Greif10}; 
and for recent reviews, see \citealt{BrommYoshida11} and \citealt{Dunlop12book}).
%\citep{WiseAbel07,WiseAbel08,Greif08,Greif10}
%
Observations may be closing in on these first galaxies
with one $z \sim 10$ candidate detected in the Ultra Deep Field 
%\citep{Bouwens11N}
(UDFj-39546284; \citealt{Bouwens11N})
%UDFj-39546284
and another strongly lensed by a galaxy cluster (MACS1149-JD1; \citealt{Zheng12}).

Intriguingly, the number density of $z \sim 10$ galaxies detected in unlensed fields
is several times lower than predicted based on extrapolations from lower redshifts,
assuming a luminosity function
with one or more parameters evolving linearly with redshift
\citep{Bouwens08,Bouwens11N,Oesch12a}.
This suggests that the star formation rate density built up more rapidly
from $z \sim 10$ to 8 than it did later between $z \sim 8$ and 2.
%but with significant uncertainties 
%\citep{Bouwens11N,Oesch12a}.
This is consistent with some theoretical predictions \citep{Trenti10,Lacey11}.
However, \cite{RobertsonEllis12} suggest such a sharp drop off would be in tension with $z<4$ GRB rates
as correlated with star formation rate density and extrapolated to higher redshifts.
Direct detections and confirmations of $z \gtrsim 10$ galaxies
are required to more precisely constrain the star formation rate density at that epoch.
%Our uncertainties about the $z \gtrsim 10$ universe are significant and additional observations are required.
%At stake is the reionization of the universe.
%If the lower redshift ($z \sim 7$ and 8) luminosity functions,

The observed luminosity functions at $z \sim 7$ and 8
feature steep faint end slopes of $\alpha \sim -2$ \citep{Bouwens11b,Bradley12b},
steeper than at lower redshifts, a trend consistent with model predictions 
\citep{Trenti10,Jaacks12}.
%\citep{Trenti10,Jaacks12,Shull12}.
If these luminosity functions can be extrapolated to $z \gtrsim 10$, then 
low luminosity galaxies ($M_{UV}$ fainter than $-16$ AB)
could have reionized the universe \citep{Bouwens12a,Kuhlen12},
assuming a sufficient fraction of their UV photons could escape their host galaxies to the surrounding medium
\citep[see also][]{ConroyKratter12}.
%\citep{Pawlik09,SrbinovskyWyithe10,HaardtMadau12,Kuhlen12,Bouwens12a}.
%low luminosity galaxies
%galaxies with UV luminosities fainter than $M = -16$ AB
Otherwise, a more exotic source of reionizing energy may have been required,
%explanation may be necessary, such as self-annihilating dark matter 
such as self-annihilating dark matter \citep{Iocco10, NatarajanA12}.
%\citep[e.g.,][]{NatarajanA12}.

Reionization was likely well underway by $z \gtrsim 10$
but with over half the universe still neutral 
%\citep{Pandolfi11,Mitra12}.
\citep{Robertson10,Pandolfi11,Mitra12}.
Improving our understanding of the early universe and this phase change is one of the pressing goals of modern cosmology.
%Improving our understanding of this phase change is one of the pressing goals of modern cosmology.
%\citep[see e.g.,][]{Robertson10}.

Observations with the Wide Field Camera 3 \citep[WFC3;][]{WFC3}
installed on the Hubble Space Telescope (HST)
have significantly advanced our understanding of the $z \gtrsim 7$ universe,
over 13 billion years in the past.
The Ultra Deep Field and surrounding deep fields
have yielded over 100 robust $z > 7$ candidates
%all the way down to 29th magnitude AB
as faint as 29th magnitude AB
\citep{Bunker10,Labbe10,Bouwens11b, Oesch12a}.
Analyses of wider space-based surveys
such as
CANDELS \citep{Grogin11, Koekemoer11},
BoRG \citep{Trenti12},
and HIPPIES \citep{Yan11}
have helped fill out 
the brighter end of the luminosity function
\citep{Oesch12b,Bradley12b,Yan11}.

Of the handful of $z > 7$ galaxies spectroscopically confirmed to date,
most have been discovered in even wider near-infrared surveys
%\~square degree surveys
carried out with the ground-based telescopes
Subaru \citep{Shibuya12,Ono12},
VLT \citep{Vanzella11},
and
UKIRT \citep{Mortlock11}.
Surveys with the VISTA telescope
are also beginning to yield high redshift candidates \citep{Bowler12}.

Complementary to these searches of ``blank'' fields
are searches behind strongly lensing galaxy clusters
%\citep{Kneib04,Bouwens09,Bradac09,Maizy10,Wong12}.  % Pello04
%(\citealt{Kneib04,Bouwens09,Bradac09,Maizy10,Wong12,Zackrisson12};
(\citealt{Kneib04,Bradley08,Bouwens09,Bradac09,Maizy10,Richard11,Hall12,Bradley12a,
Zitrin12MACS0329,Wong12,Zackrisson12,Bradac12};
and for a recent review, see \citealt{KneibNatarajan11}).
The drawbacks of lensed searches are
reduced search area in the magnified source planes
and some uncertainty in the estimate of that search area introduced by the lens modeling.
But the rewards are galaxies which are strongly magnified, often by factors of 10 or more.
Lensed searches are significantly more efficient
in yielding high-redshift candidates 
bright enough for spectroscopic confirmation,  % (AB mag $< 26$) 
including A1703-zD6 \citep{Bradley12a} at $z = 7.045$ \citep{Schenker12}.
%including spectroscopic follow-up.
%For a recent review, see \cite{KneibNatarajan11}.

The Cluster Lensing and Supernova survey with Hubble (CLASH; \citealt{CLASH})
is a large Hubble program imaging 25 galaxy clusters
in 16 filters including five in the near-infrared (0.9--1.7\um).
Five of these, including \macs\ ($z = 0.591$, \citealt{Ebeling07}),  % 0.5907
%MACSJ0647+70
%($z = 0.584$, \citealt{Balestra07}),
%Balestra07 do say they have spectroscopic redshifts, 
%but Ebeling07 give more details: 38 cluster galaxies with a velocity dispersion ~900 km/s.
were selected on the basis of their especially strong gravitational lensing power
as observed in previous imaging,
with the primary goal of discovering highly magnified galaxies at high redshift.
%
%To date, this survey has yielded strongly lensed images of
%To date, this survey has yielded many strongly lensed galaxy images, including
To date, some of the more notable strongly lensed galaxies found in CLASH include
a doubly-imaged galaxy with a spectroscopic redshift $z = 6.027$ observed at mag $\sim 24.6$ \citep{Richard11} 
which is possibly \~800 Myr old (although see \citealt{Pirzkal12});
a quadruply-imaged $z \sim 6.2$ galaxy observed at 24th magnitude \citep{Zitrin12MACS0329};
and 
%most recently 
the $z \sim 9.6$ candidate galaxy MACS1149-JD1 observed at 
mag \~ 25.7 \citep{Zheng12}.
%26th magnitude \citep{Zheng12}.
The $z \sim 9.6$ candidate is strongly lensed by MACSJ1149.6+2223,
another CLASH cluster selected for its high magnification strength.
%
%CLASH may even be expected to yield a few Pop III galaxies magnified to $\sim 27$th magnitude
%if some galaxies remain so pristine as late as $z \sim 8$ \citep{Zackrisson12}.

Here we report the discovery of \JD,
a candidate for the earliest galaxy yet detected
at a redshift \ztwosig\ (95\% confidence),
just \agetwosig\ million years after the Big Bang.
% 10.7 +0.4/-0.2 (68\% CL)
%at a redshift $z =11.0_{-0.7}^{+0.4}$ (95\% confidence),
%just $410_{+40}^{-20}$ million years after the Big Bang.
%at a redshift $z \approx 11$,
%just 400 million years after the Big Bang
%in the heart of the reionization epoch.
It is strongly lensed by \macs, yielding three multiple images
observed at F160W AB mag $\sim$25.9, 26.1, and 27.3,
magnified by factors of $\sim$8, 7, and 2.
%with the brightest magnified by a factor of $\sim$8.
The brightest image is similar in flux to MACS1149-JD1 (F160W mag \~ 25.7) at $z = 9.6 \pm 0.2$ (68\% confidence)
and roughly 15 times (3 magnitudes) brighter 
%than the previous robust candidate for the highest-redshift galaxy at 
than the $z = 10.3 \pm 0.8$ (68\% confidence) candidate in the UDF \citep{Bouwens11N}.

\JD\ is a \Jdrop\ as all three lensed images are securely detected in F160W and F140W
but drop out of detection in the $J$-band F125W and all 14 bluer HST filters.
%We find strong evidence that this is 
We show this photometry is most likely due to 
%absorption by neutral hydrogen between us and the source at $z \sim 11$ with 
the \Lya\ break redshifted to $\sim 1.46$\um\ at $z \sim 11$.
%with the Lyman break redshifted to $\sim 1.46$\um\ ($z \sim 11$).
This Lyman dropout technique \citep{Meier76b,Giavalisco02}
%(see \citealt{Giavalisco02} for a review)
pioneered by \cite{Steidel96} at $z \sim 3$
has been used with a high success rate
to identify high-redshift candidates later spectroscopically confirmed out to $z \sim 7$.
%\citep{Dow-Hygelund07,Vanzella09}.
%and perhaps out to $z = 8.6$ \citep{Lehnert10}.
However care must be taken not to confuse dropouts with intrinsically red (evolved and/or dusty)
galaxies at intermediate redshift 
\citep{Schaerer07,Dunlop07,Chary07,Capak11,Boone11,Hayes12}.
%with dropouts \citep[e.g.,][]{Boone11}, and we do so here.
In the case of \JD,
we show it is extremely difficult, if not impossible, for low redshift interlopers to reproduce the observed colors,
especially the $J_{125} - H_{160} \gtrsim 3$ magnitude break of \JD1.
We also test our analysis method
by reanalyzing previously published \Jdrops\ which later proved to be at intermediate redshift.
Our Bayesian photometric redshift (BPZ, \citealt{Benitez00,Coe06}) analysis
correctly shows intermediate redshift solutions are preferred for those objects,
while preferring a higher redshift for \JD.

We describe our HST and Spitzer observations in \S\ref{sec:obs}
and present photometry in \S\ref{sec:phot}.
We derive the photometric redshift in \S\ref{sec:photoz} 
and consider a wide range of possible interlopers in \S\ref{sec:loz}.
%including several previously published \Jdrops.
We present our gravitational lensing analysis in \S\ref{sec:lens}.
In \S\ref{sec:props} we derive physical properties of \JD\ based on additional photometric analysis.
In \S\ref{sec:SFRD} we compare our observed number density of $z \sim 11$ galaxies to that expected,
and we constrain the $z > 9$ star formation rate density.
Finally, we present conclusions in \S\ref{sec:conc}.

\vspace{-0.018in}

Where necessary we assume a concordance \LCDM\ cosmology
with $h = 0.7$, $\Omega_m = 0.3$, $\Omega_\Lambda = 0.7$,
where $H_0 = 100$ $h$ km s$^{-1}$ Mpc$^{-1}$.
In this cosmology $1'' = 3.93$ kpc at $z = 10.8$
and 6.62 kpc at the cluster redshift $z = 0.591$.

%Bouwens10

%detection \citep{Pawlik11,Zackrisson12}

%reionization (for a recent review, see \citealt{Robertson10}).
%\cite{Bouwens12a}.

%%%%%%%%%%%%%%%%%%%%%%%%%%%%%
% POSITIONS AND PHOTOMETRY

%\input{cat.tex}
\input{obs.tex}

%%%%%%%%%%%%%%%%%%%%%%%%%%%

% HST image + lens model
\begin{figure*}
%\centerline{\includegraphics[width = 0.8\textwidth]{figs/HSTSpitzer.png}}
%\centerline{\includegraphics[width = \textwidth]{figs/critcurveszoomhalf}}
\centerline{
\includegraphics[width = \textwidth]{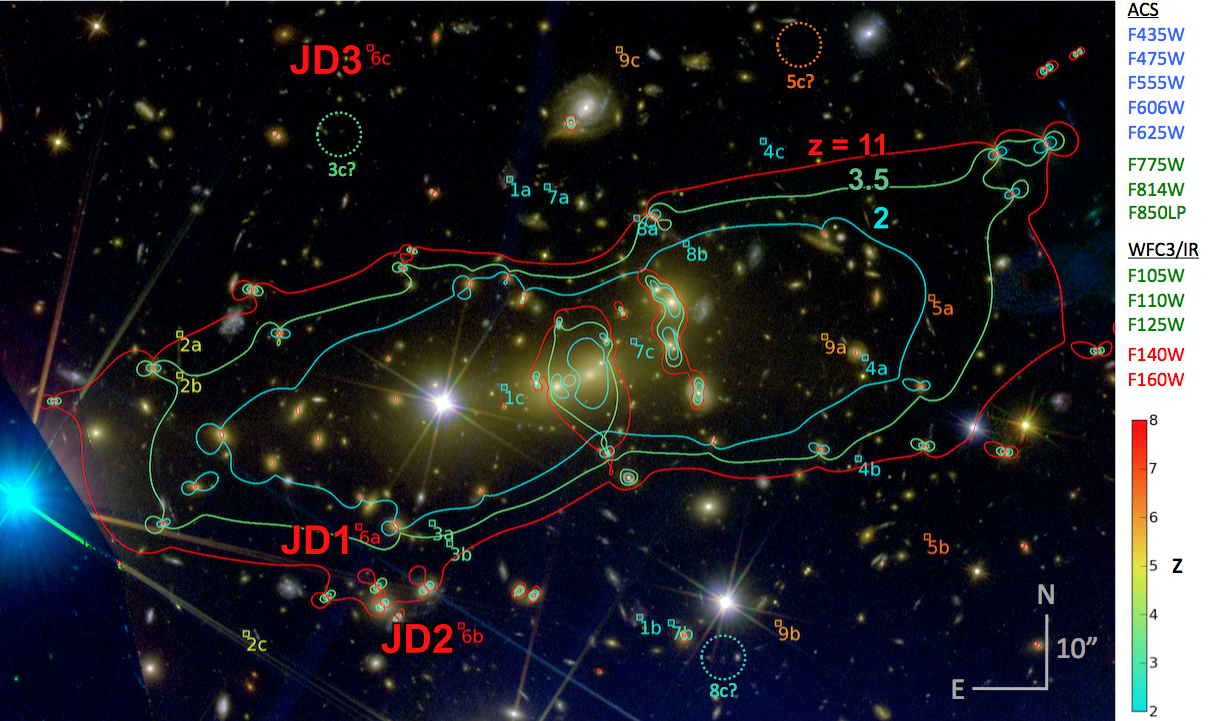}
}
%\centerline{\includegraphics[width = \columnwidth]{figs/HSTSpitzer.png}}
%\captionsetup{labelsep=pipe}
\caption{
\label{fig:HSTcluster}
Lenstool strong lensing mass model of \macs\ and multiply-imaged galaxies as identified in this work
using the \cite{Zitrin09a} method,
including two strong lensing systems identified in \cite{Zitrin11MACS}.
Each strongly lensed galaxy is labeled with a number and color-coded by redshift (scale at bottom right).
Letters are assigned to the multiple images of each galaxy.
Dashed circles indicate predicted locations of counterimages not unambiguously identified.
Overplotted are critical curves from our Lenstool model
indicating thin regions of formally infinite magnification
for background galaxies at $z = 2.0$ (cyan), 3.5 (green) and 11.0 (red).
Mirror images of galaxies straddle these critical curves.
The Hubble color image was produced using Trilogy \citep{Coe12}
and is composed of ACS and WFC3/IR filters as given at top right.
}
\end{figure*}

%%%%%%%%%%%%%%%%%%%%%%%%%%%%%
% POSITIONS AND PHOTOMETRY

%\input{cat.tex}
\input{catsum.tex}

%%%%%%%%%%%%%%%%%%%%%%%%%%%%%
% CLUSTERS

\begin{deluxetable}{rlcl}
\tablecaption{\label{tab:clusters}17 Clusters Searched in This Work}
\tablewidth{\columnwidth}
\tablehead{
\colhead{High Magnification?$^{\rm a}$}&
\colhead{Cluster$^{\rm b}$}&
\colhead{Redshift}
}
\startdata
& Abell 383  (0248.1$-$0331) & 0.187\\
& Abell 611  (0800.9+3603) & 0.288\\
& Abell 2261  (1722.5+3207) & 0.244\\
& MACSJ0329.7$-$0211 & 0.450\\
Y & MACSJ0647.8+7015 & 0.591\\
Y & MACSJ0717.5+3745 & 0.548\\
& MACSJ0744.9+3927 & 0.686\\
& MACSJ1115.9+0129 & 0.355\\
Y & MACSJ1149.6+2223 & 0.544\\
& MACSJ1206.2$-$0847 & 0.439\\
& MACSJ1720.3+3536 & 0.387\\
& MACSJ1931.8$-$2635 & 0.352\\
Y & MACSJ2129.4$-$0741 & 0.570\\
& MS2137.3$-$2353 & 0.313\\
& RXJ1347.5$-$1145 & 0.451\\
& RXJ1532.9+3021 & 0.363\\
& RXJ2129.7+0005 & 0.234\\
\vspace{-0.1in}
\enddata
\tablenotetext{1}{CLASH clusters were selected based on either X-ray or strong lensing properties.
The latter ``high magnification'' clusters are marked with Y's here.
For details, see \cite{CLASH}.}
\tablenotetext{2}{R.A.~\& Decl.~(J2000) are given in parentheses for the Abell clusters,
encoded as they are in the names of the other clusters.}
\end{deluxetable}

%\tablecomments{
%\JD\ is detected at all three positions with 
%$>10$\sig\ significance in F160W
%and $>6$\sig\ significance in F140W.
%They all drop out 
%of F125W ($J$-band) and all bluer filters
%with no 3\sig\ detections.
%JD1 is not detected at 1\sig\ in F125W or any bluer filter.
%Isophotal apertures are determined by SExtractor,
%then fluxes and uncertainties are re-measured as explained in \S\ref{sec:HSTphot}.
%Fluxes in nanoJanskys (nJy) may be converted to AB magnitudes via m$_{AB} \approx 26 - 2.5 \log_{10}(F_{\nu} / (145~{\rm nJy})$).

%%%%%%%%%%%%%%%%%%%%%%%%%%%%%

\section{Observations}
\label{sec:obs}

As part of the CLASH program,
HST observed the core of \macs\ (Fig.~\ref{fig:HSTcluster}) during 19 orbits
spread among eight different visits between October 5 and November 29, 2011
(General Observer program 12101).
Imaging was obtained with the 
Wide Field Camera 3 \citep[WFC3;][]{WFC3}
and Advanced Camera for Surveys \citep[ACS;][]{ACS}
in 15 filters spanning 0.2--1.7\um\ including 
five near-infrared WFC3/IR filters spanning 0.9--1.7\um.
These datasets were supplemented by prior ACS imaging obtained 
%with the Advanced Camera for Surveys 
in the F555W (0.56\um) and F814W (0.81\um) filters
%with depths of \appr\ 2.2 and 3.5 hours, respectively
to total depths of \~ 3.5 and 5.5 orbits, respectively
(GO 9722 P.I. Ebeling; 
GO 10493, 10793 P.I. Gal-Yam).
These observations are detailed in Table \ref{tab:obs}.
%
%Cycle 12
%http://archive.stsci.edu/proposal_search.php?mission=hst&id=9722
%Harald Ebeling
%F555W 3870 + 3870
%F814W 3960 + 3960
%
%Cycles 14,15
%http://archive.stsci.edu/proposal_search.php?mission=hst&id=10493
%http://archive.stsci.edu/proposal_search.php?mission=hst&id=10793
%Avishay Gal-Yam
%F814W 2584, 2256
%
%F555W total =  7740 ~3.37 orbits?  (2.15 hours)
%F814W total = 12760 ~5.5  orbits   (3.5 hours)

We processed the images for debias, (super)flats, and darks using standard techniques,
then co-aligned and combined them 
to a scale of 0.065"/pixel;
see \cite{Koekemoer02,Koekemoer11}
for further information on the astrometric alignment and drizzle algorithms that were used
and \cite{CLASH} for specific details on their implementation in CLASH.
We also produced inverse variance maps (IVMs) based on the 
observed sky level, identified cosmic rays, detector flat field, readnoise, dark current, and bad pixels.  
These IVMs may be used to estimate the level of uncertainty in each pixel 
before accounting for correlated noise and any Poisson source noise.

Imaging at longer wavelengths 
was obtained by the Spitzer Space Telescope
with the InfraRed Array Camera (IRAC; \citealt{IRAC}) ch1 (3.6\um) and ch2 (4.5\um)
with total exposure times of 5 hours at each wavelength
(Program 60034, P.I. Egami).
These observations were divided into two epochs separated by $\sim$5.5 months
(Nov.~10, 2009 and Apr.~23, 2010).
We combined the Basic Calibrated Data (BCD) using MOPEX \citep{MOPEX} to produce mosaicked images.

%The complete list of filters and exposure times is given in Table \ref{tab:cat}.

As of July 2012, CLASH had obtained 16-band HST observations for 17 clusters,
including \macs\ and three other ``high-magnification'' strong lensing clusters, as given in Table~\ref{tab:clusters}.
We searched for high redshift galaxies in the WFC3/IR fields of view
of all 17 of these clusters (Bradley et al.~2012c, in preparation).

Out of \~20,000 detected sources,
we identified \JD\ (Fig.~\ref{fig:HST}) as having an exceptionally high photometric redshift (\S\ref{sec:photoz}).
%We discovered \JD\ based on its high photometric redshift.
Our selection was based on SED (spectral energy distribution) fitting
as used in some previous high-redshift searches \citep[e.g.,][]{McLure06,Dunlop07,Finkelstein10}.
We did not impose specific magnitude limits, color cuts, or other detection thresholds
on our selection as in other works \citep[e.g.,][]{Bunker10,Yan11,Bouwens12b,Oesch12b}.

\section{Photometry}
\label{sec:phot}

\subsection{HST Photometry}
\label{sec:HSTphot}

\subsubsection{Photometric Analysis}
\label{sec:HSTphotanal}

We used SExtractor version 2.5.0 \citep{SExtractor}
to detect objects in a weighted sum of all five HST WFC3/IR images.
Along the edge of each object, 
SExtractor defines an isophotal aperture consisting of pixels with values above a detection threshold.
We set this threshold equal to the RMS measured locally near each object.
Isophotal fluxes (and magnitudes) are measured within these isophotal apertures.
%Around each object, SExtractor measures the local background within a rectangular annulus (default width 24 pixels).
SExtractor derives flux uncertainties
by adding in quadrature the background RMS derived from our inverse variance maps
and the Poisson uncertainty from the object flux.

Since our images are drizzled to a 0.065" pixel scale, which is 2-3 times smaller than the WFC3 PSF, 
the resulting images contain significant correlated noise. 
The weight maps produced by drizzle represent the expected variance in the absence of correlated noise.
To account for the correlated noise, one may apply a correction factor as in \cite{HDF-S}.  % Casertano00

Previous authors have also noted that 
SExtractor tends to underestimate flux uncertainties \citep{Feldmeier02,Labbe03,Gawiser06}
by as much as a factor of 2--3 \citep{Becker07}.
In this work, we obtained empirical measurements of the flux uncertainties %as follows.
using the following method which also captures the effects of correlated noise.

SExtractor has the ability to measure the local background within a rectangular annulus (default width 24 pixels) around each object.
We constructed a rectangle of the same size,
but rather than calculate the RMS of the individual pixels,
we obtained samples of the background flux within this region using the isophotal aperture shifted to new positions.
In other words, we moved the isophotal aperture to every position within this rectangle,
sampling the flux at each position.
We discarded measurements for which the aperture includes part of any object,
as we are interested in measuring the background flux.
Finally we measured the RMS of these measurements
and added in quadrature the object's Poisson uncertainty
to obtain the total flux uncertainty for that object.
We found this technique indeed yielded larger flux uncertainties than reported by SExtractor, 
typically by factors of 2--3 in the WFC3/IR filters and by lower factors in ACS and WFC3/UVIS.
%by as much as a factor of 2 or so in the WFC3/IR filters.
% ~/CLASH/data/m0647/highz/segmphotfluxplot.py

We also used this method to determine object fluxes.
The mean of the flux measurements in the nearby apertures was adopted as the local flux bias,
which we subtracted from the flux measurement in the object itself.
We found this yielded photometry very similar to that obtained using SExtractor,
agreeing well within the photometric uncertainties.
% ~/CLASH/data/m0647/highz/comparesex.py
While we used this photometry for all subsequent analyses,
we also verified that our derived photometric redshifts
did not vary significantly 
(after excluding the F110W second epoch exposures; \S\ref{sec:A9})
if we instead utilized photometry derived directly from SExtractor.

We corrected for Galactic extinction of $E(B-V) = 0.11$ in the direction of \macs\ as 
derived using the \cite{Schlegel98} IR dust emission maps.
For each filter, the magnitudes of extinction per unit $E(B-V)$ are given in \citet[their Table 5]{CLASH}.
(These values should be \~10\% lower in the NUV and optical according to \citealt{SchlaflyFinkbeiner11}.)
This extinction reddens the observed colors at the few percent level in the near-IR.
Thus the effects on the \Jdrop\ are negligible.
The extinctions range from 0.05 to 0.11 mag in the WFC3/IR images;
0.16 to 0.46 mag in the ACS images;
and 0.50 to 0.83 mag in WFC3/UVIS.
We note the extinction may be somewhat uncertain due to patchy galactic cirrus in the direction of \macs.

\subsubsection{Photometric Results}

\begin{figure*}
%\centerline{\includegraphics[width = \textwidth]{figs/Khoised.png}}
%\centerline{\includegraphics[width = \textwidth]{figs/KhoiNamased.png}}
%\centerline{\includegraphics[width = \textwidth]{figs/sed_JD_HST.png}}
\centerline{
\includegraphics[width = \textwidth]{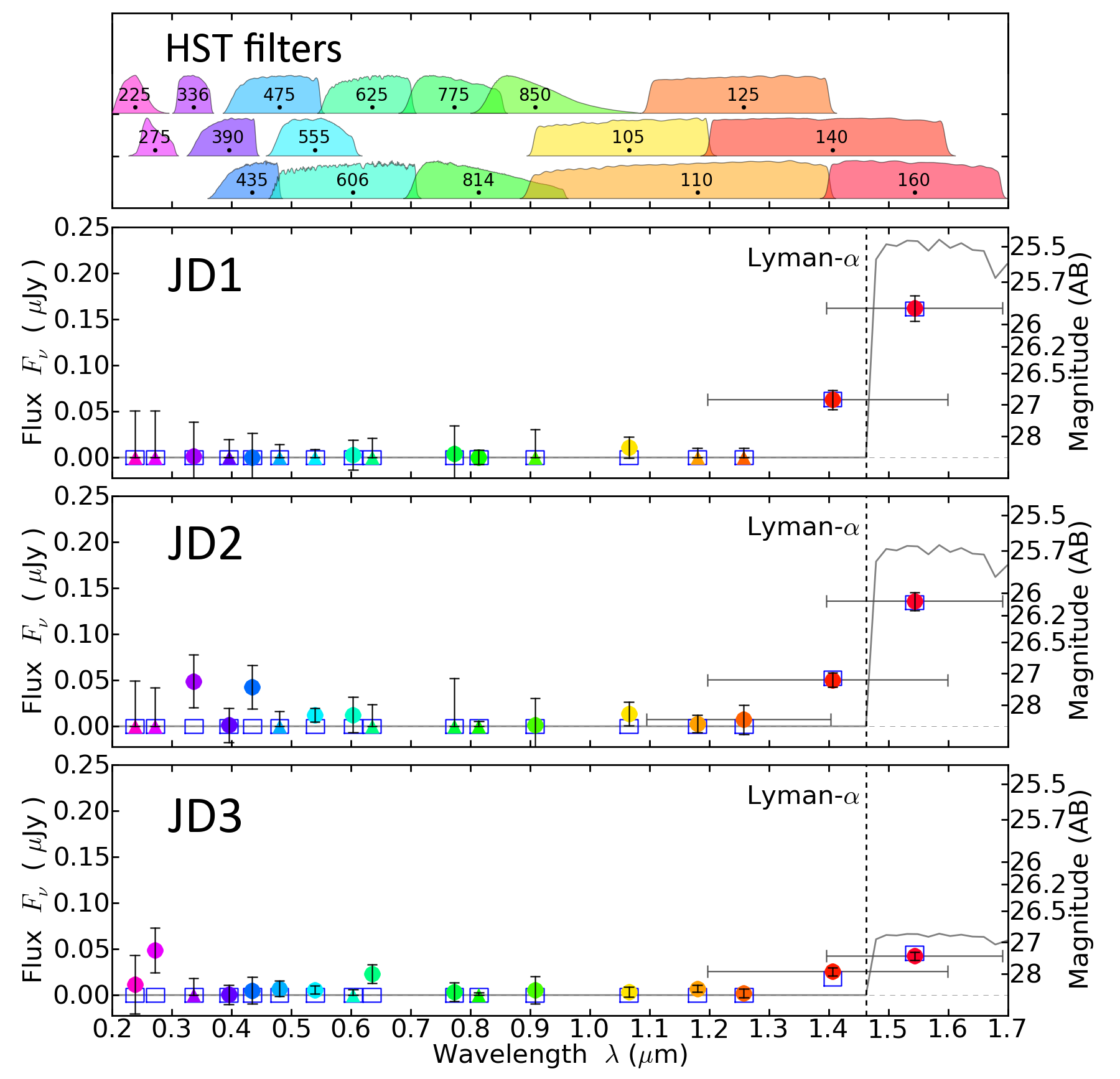}
}
%\captionsetup{labelsep=pipe}
\caption{
\label{fig:HSTsed}
Observed HST photometry (filled circles and triangles)
plotted against the expected fluxes (open blue squares)
from a young starburst galaxy spectrum (gray line) redshifted to $z \sim 11$.
HST filter transmission curves are plotted in the upper panel,
normalized to their maxima,
and with black dots indicating the effective ``pivot'' wavelengths.
Photometry of the \Jdrops\ observed through these filters (Table \ref{tab:cat})
is plotted as the larger
circles and triangles for positive and negative observed fluxes, respectively,
with 1-$\sigma$ error bars.
For some points, horizontal ``error bars'' are plotted to reiterate the filter widths.
The gray line is a model spectrum of a young starburst at $z = 11.0$, the best fit to the summed photometry.
%$z = 11.03$.
%
The integrals of this spectrum through our filters
give the model predicted fluxes plotted as blue squares.
Other galaxy types at $z \sim 11$ yield similar predicted HST fluxes,
as the shape of the spectrum cannot be constrained by the HST photometry alone.
Redshifted Lyman-$\alpha$ at $0.1216\mu$m$(1+z) \sim 1.46\mu$m is indicated by the vertical dashed line.
}
\end{figure*}

Our resulting 17-band HST photometry is given in Table~\ref{tab:cat} and Fig.~\ref{fig:HSTsed}.
All three $J$-dropouts are detected at 
$>$10\sig\ in F160W,
$>$6\sig\ in F140W,
and $<$3\sig\ in all other filters.
JD1 is not detected above 1\sig\ in any filter blueward of F140W.

\begin{figure*}
\centerline{
\includegraphics[width = 0.8\textwidth]{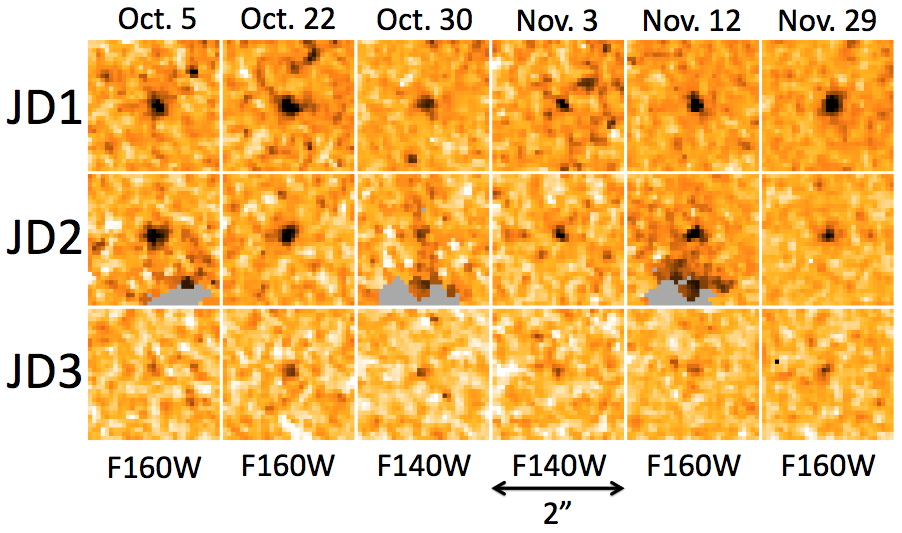}
}
%\captionsetup{labelsep=pipe}
\caption{
\label{fig:timeimages}
\JD\ as observed in each of the individual epochs of F160W and F140W obtained over a 56-day period.
These observations were obtained at two different telescope roll angles
which alternate between the stamps shown here.
A small region of the WFC3/IR images in our first roll angle
was affected by persistence due to a moderately bright star in our parallel observations immediately prior.
These pixels happen to fall within $1''$ of JD2 at that roll angle
(marked in gray here and flagged as unreliable).
Excluding this roll angle for JD2 does not significantly affect the derived photometry.
}
\end{figure*}

\begin{figure*}
\centerline{
\includegraphics[width = 0.9\textwidth]{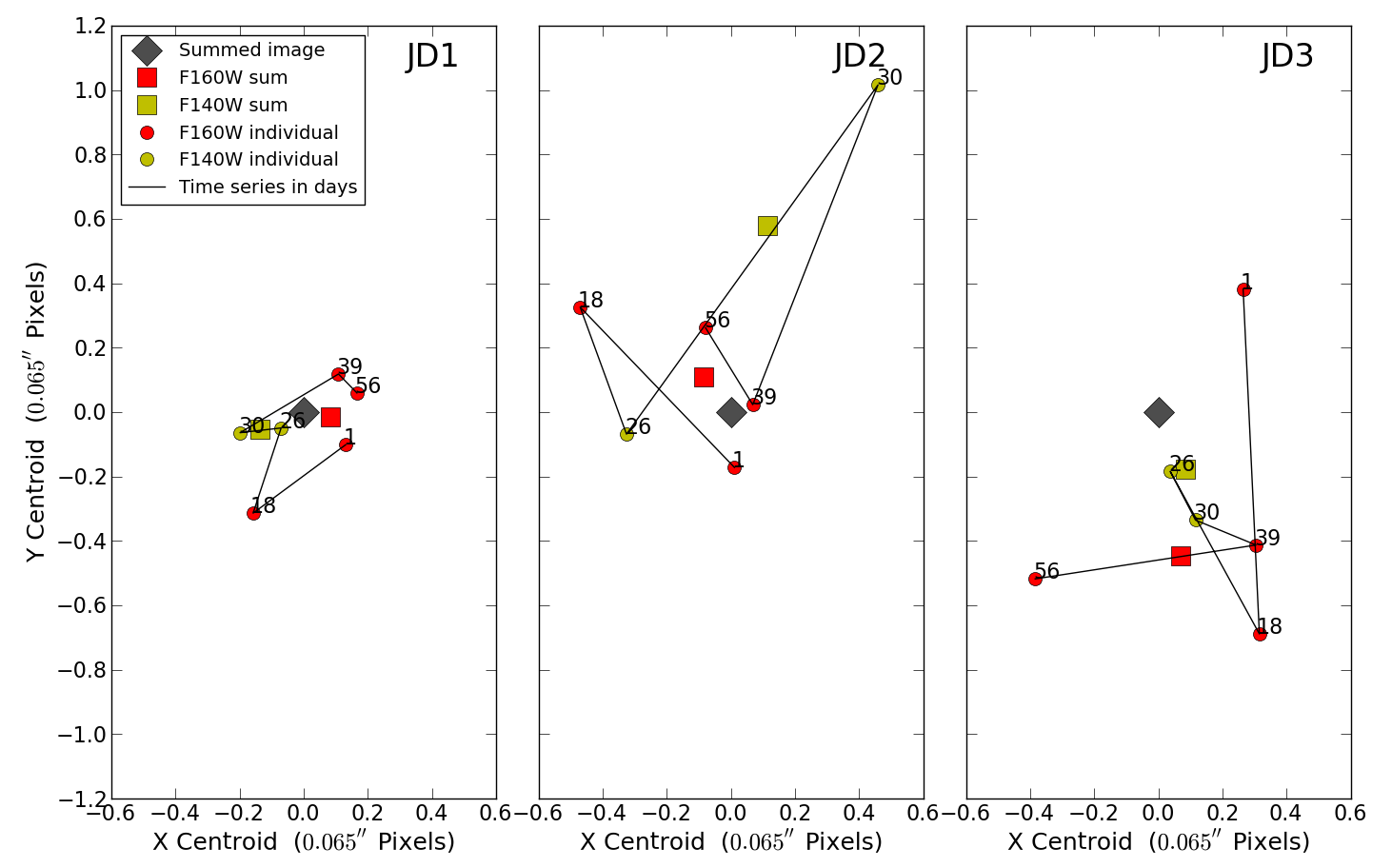}
}
%\captionsetup{labelsep=pipe}
\caption{
\label{fig:timexy}
Relative centroid measurements for the detections in F160W (red) and F140W (yellow)
in individual epochs (circles) and summed observations (squares).
Centroids measured in the summed NIR images are also plotted as gray diamonds.
The offsets are generally less than one of our drizzled pixels ($0.065''$),
roughly half the native WFC3/IR pixel size ($\sim$0.13$''$).
}
\end{figure*}

All three \Jdrops\ are confidently detected in two filters (F140W and F160W)
observed at six different epochs over a period of 56 days (Fig.~\ref{fig:timeimages}).
No significant temporal variations are observed in position or brightness,
ruling out solar system objects
and transient phenomena such as supernovae, respectively (see Figs.~\ref{fig:timeflux} and \ref{fig:timexy} and \S\ref{sec:loz}).

\subsubsection{Exclusion of F110W Second Epoch}
\label{sec:A9}

\begin{figure*}
%\centerline{\includegraphics[width = 0.95\textwidth]{figs/JD3stamps.png}}
\centerline{
\includegraphics[width = \textwidth]{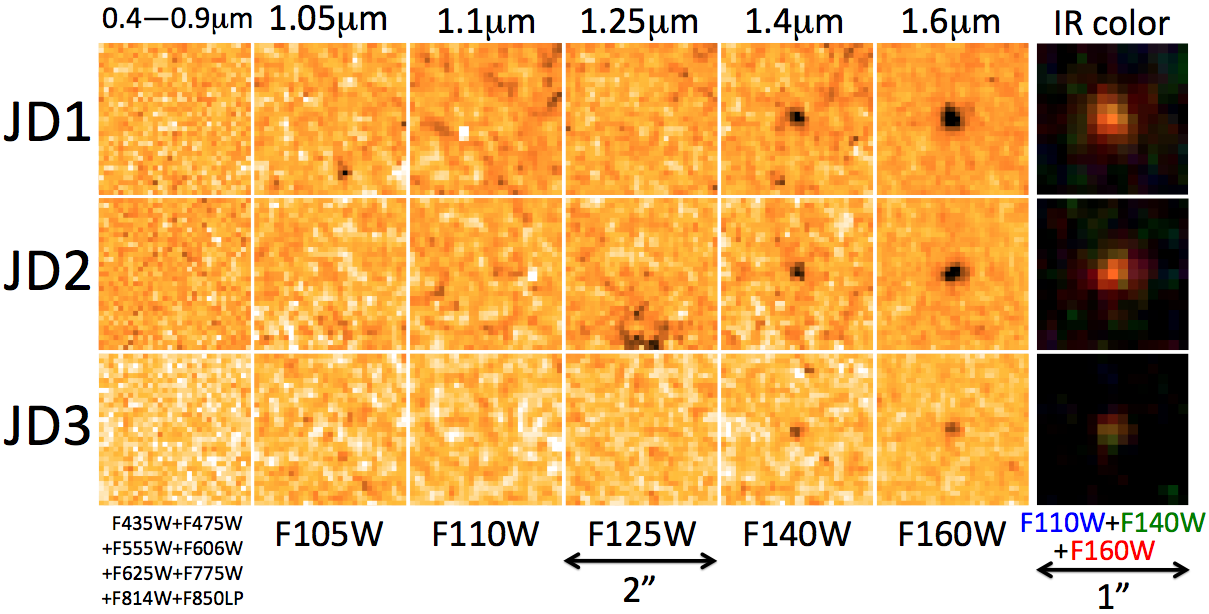}
}
%~/CLASH/data/m0647/images/stamps/stampmontageheat.py
%\captionsetup{labelsep=pipe}
\caption{
\label{fig:HST}
The three images of \JD\ as observed in various filters with HST.
The leftmost panels show the summed 11-hour (17-orbit)
exposures obtained
in 8 filters spanning 0.4--0.9\um\ with the Advanced Camera for Surveys.
The five middle columns show observations with the Wide Field Camera 3 IR channel
in F105W, F110W, F125W, F140W, and F160W,
all shown with the same linear scale in electrons per second.
The F125W images were obtained at a single roll angle,
and a small region near JD2 was affected by persistence due to a moderately bright star in our parallel observations immediately prior
(see also Fig.~\ref{fig:timeimages}).
The right panels zoom in by a factor of 2 to show F110W+F140W+F160W color images scaled linearly between 0 and 0.1 \uJy.
% ~/CLASH/data/m0647/images/stamps/colorimage.py
}
\end{figure*}

Based on an initial standard reduction of the HST images
and standard SExtractor photometry,
\JD2 was detected in F110W at 5\sig,
while JD1 and JD3 were not significantly detected (0.9\sig\ and 1.7\sig, respectively).
Our empirical rederivations of the photometric uncertainties,
including proper accounting for correlated noise (\S\ref{sec:HSTphotanal}), % in the processed images
reduced the significance of this detection to 2.5\sig.
However we ultimately we concluded this marginal detection was completely spurious
due to significantly elevated and non-Poissonian backgrounds due to Earthshine in two out of five F110W exposures,
both obtained during the second epoch (see below).
% (see \S\ref{sec:A9}).
After excluding these exposures, the detection significance drops to 0.3\sig,
consistent with background noise.
For reference, see the WFC3/IR images in Fig.~\ref{fig:HST}.

\begin{figure*}
\centerline{
\includegraphics[width = 0.9\textwidth]{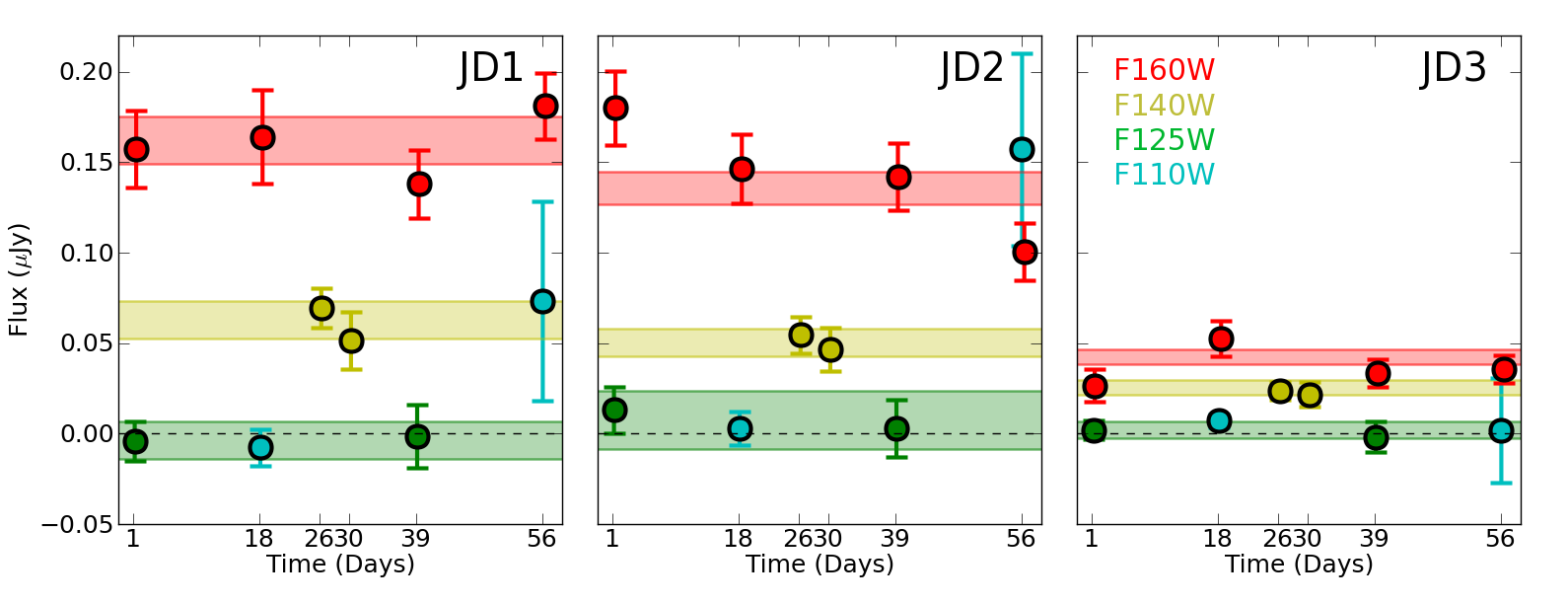}
}
%\captionsetup{labelsep=pipe}
\caption{
\label{fig:timeflux}
Flux measurements in the individual epochs observed over a period of 56 days.
Filters are colored F160W (red), F140W (yellow), F125W (green), and F110W (blue)
as both individual data points and solid bands as determined for the summed observations.
The F110W exposures obtained in the second epoch (visit A9)
were found to have significantly elevated and non-Poissonian backgrounds
due to Earthshine (\S\ref{sec:HSTphot}).
These were excluded in our analysis; we adopted the F110W fluxes measured in the first epoch (visit A2).
%or reflections from the sunlit Earth (see \S\ref{sec:HSTphot}).
%within 30 degrees of the daytime Earth horizon (limb).
}
\end{figure*}

% fullprob.py
Even based on the initial ``standard'' analysis described above, 
we determined that \JD\ is at $z < 9$
with a likelihood of \~\tentothe{-9} % 2\en5 
based on a joint photometric redshift analysis of all three images (\S\ref{sec:photoz}).
This likelihood decreased further to 3\en{13} % 6\en8 
based on our improved analysis.
These values are summarized in Table \ref{tab:A9}.
The spuriously high flux measurements may be seen in Fig.~\ref{fig:timeflux}.

The final observations of \macs\ were two 502-second exposures in F110W
obtained during visit A9, the second epoch for that filter.
We found these to have significantly elevated backgrounds
of 1.9 (sigma-clipped mean) $\pm$ 0.44 (RMS) and $6.4\pm0.27$ electrons per second, respectively,
compared to the more typical values around $1.5\pm0.08$. %e-/s.
These high backgrounds were due to Earthshine, or sunlight reflected from the Earth.
The first observation was obtained during twilight
as the telescope pointed within 67--59 degrees of the bright limb of the Earth.
%Earth horizon.
This Earth limb angle continued to steadily decrease from 47 to 24 degrees
during the second observation which was obtained during daylight.
In the observation log, the diagnostic Earth bright limb flag was raised halfway through the second exposure.
We also examined the ten individual readouts of 100 seconds each obtained over the course of both exposures
and found the mean background increased steadily from 0.9 to 7.5 electrons per second.
% in each IR exposure.
%as Hubble's viewing angle above the horizon decreased from 67 to 24 degrees and night transitioned to day.
%The diagnostic Earth bright limb flag was raised halfway through the second exposure.
The resulting elevated background RMS values of 0.44 and 0.27 electrons per second in the two exposures
are the highest and sixth highest relative to the median values for a given filter
in 1,582 CLASH observations to date of 17 clusters.
None of the three F160W observations obtained at the beginning of visit A9 exhibit elevated backgrounds
because they were obtained at night (twilight had yet to set in)
%the Earth limb angle was higher, 
and the Earth is less bright in F160W.

%%%%%%%%%%%%%%%%%%%%%%%%%%%
% HST images

%%%%%%%%%%%%%%%%%%%%%%%%%%%%%

%%%%%%%%%%%%%%%%%%%%%%%%%%%
% Time series fluxes
%%%%%%%%%%%%%%%%%%%%%%%%%%%

%%%%%%%%%%%%%%%%%%%%%%%%%%%
% HST SED

%%%%%%%%%%%%%%%%%%%%%%%%%%%

%%%%%%%%%%%%%%%%%%%%%%%%%%%
% Time series images F140W & F160W
%%%%%%%%%%%%%%%%%%%%%%%%%%%

%%%%%%%%%%%%%%%%%%%%%%%%%%%
% Time series centroids
%%%%%%%%%%%%%%%%%%%%%%%%%%%

%%%%%%%%%%%%%%%%%%%%%%%%%%%%%
% F110W effects
% ~/CLASH/data/m0647/highz/fullprob.py
% infile = 'm0647JD3.full_probs'  # -- 156, 257, 359  all obs manual
% infile = 'm0647JD3_sex_woA9.full_probs' # -- 106, 207, 309  all obs SExtractor

\begin{deluxetable}{lrrrc}
\tablecaption{\label{tab:A9}
Effects of F110W Aberrant Second Epoch
%Effects of F110W Anomalous Second Epoch
%Effects of F110W Second Epoch
%JD2 Spurious Detection in F110W Second Epoch
%Spurious Effects due to F110W Second Epoch
}
\tablewidth{\columnwidth}
\tablehead{
\colhead{F110W detection $\sigma$}&
\colhead{JD1}&
\colhead{JD2}&
\colhead{JD3}&
\colhead{P($z<9$)\super{a}}
}
\startdata
SExtractor photometry & 
 0.9  &  5.0  &  1.7  &  
 1\en9 \\
% 2\en5 \\
Empirical uncertainties & 
 0.5  &  2.5  &  1.0  &  
 4\en8 \\
% 6\en5 \\
Excluding second epoch & 
$-$0.8  &  0.3  &  1.9  &  
3\en{13} \\
%6\en8 \\
\vspace{-0.1in}
\enddata
\tablecomments{JD2 is spuriously detected in F110W images processed using standard techniques.
This is due to significantly elevated non-Poissonian backgrounds in the second epoch of observations due to Earthshine.
We exclude this epoch in our analysis.
See \S\ref{sec:HSTphot}}
\tablenotetext{1}{Based on the summed photometry of all three images,
and assuming \JD\ is a galaxy well described by our templates.  See \S\S\ref{sec:photoz}--\ref{sec:loz}.}
%See explanation in \S\ref{sec:HSTphot}}
%\tablenotetext{2}{R.A.~\& Decl.~(J2000) are given in parentheses for the Abell clusters,}
\end{deluxetable}
%%%%%%%%%%%%%%%%%%%%%%%%%%%%%

Specifically, when we compared the measured RMS values 
to what would be expected from scaling the background intensity levels, 
we found that these RMS values are several times higher 
than would be expected in the case of Poissonian statistics. 
We attribute this to the fact that the sky background was increasing in a strongly non-linear fashion during the exposure, 
whereas the up-the-ramp slope fitting algorithm implemented in ``calwf3''
implicitly assumes that the count rate is constant
when converting measured counts into counts/second
\citep[see][]{WFC3handbook4}. 
Since this assumption is violated, the the pixel-to-pixel variations in the final count-rate image 
no longer scale as expected for Poissonian statistics, 
as demonstrated by the much higher RMS values. 
Since these data no longer conform to Poissonian statistics, 
we were able to demonstrate that attempting to combine them with the other data did not yield an improvement in S/N 
but instead produced combined datasets with non-Poissonian statistics, from which we were not able to obtain reliable photometry.

We therefore exclude the two F110W visit A9 exposures from our analysis
and derive photometry instead from the weighted sum of the three visit A2 exposures.

%\vspace{0.2in}

%%%%%%%%%%%%%%%%%%%%%%%%%%%
% HST+IRAC
\begin{figure*}
%\centerline{\includegraphics[width = 0.8\textwidth]{figs/F160WIRAC.png}}
\centerline{
\includegraphics[width = 0.63\textwidth]{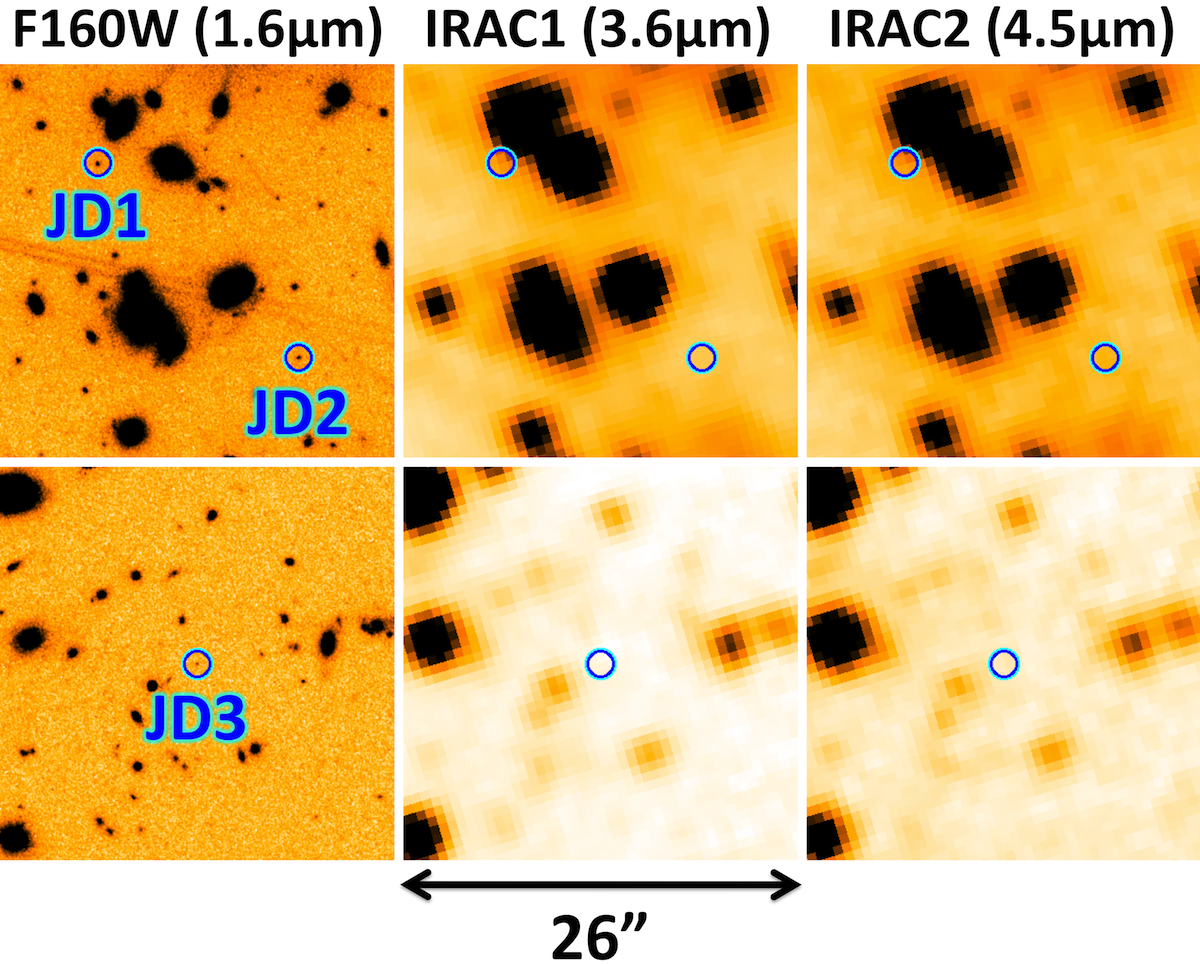}
\includegraphics[width = 0.36\textwidth]{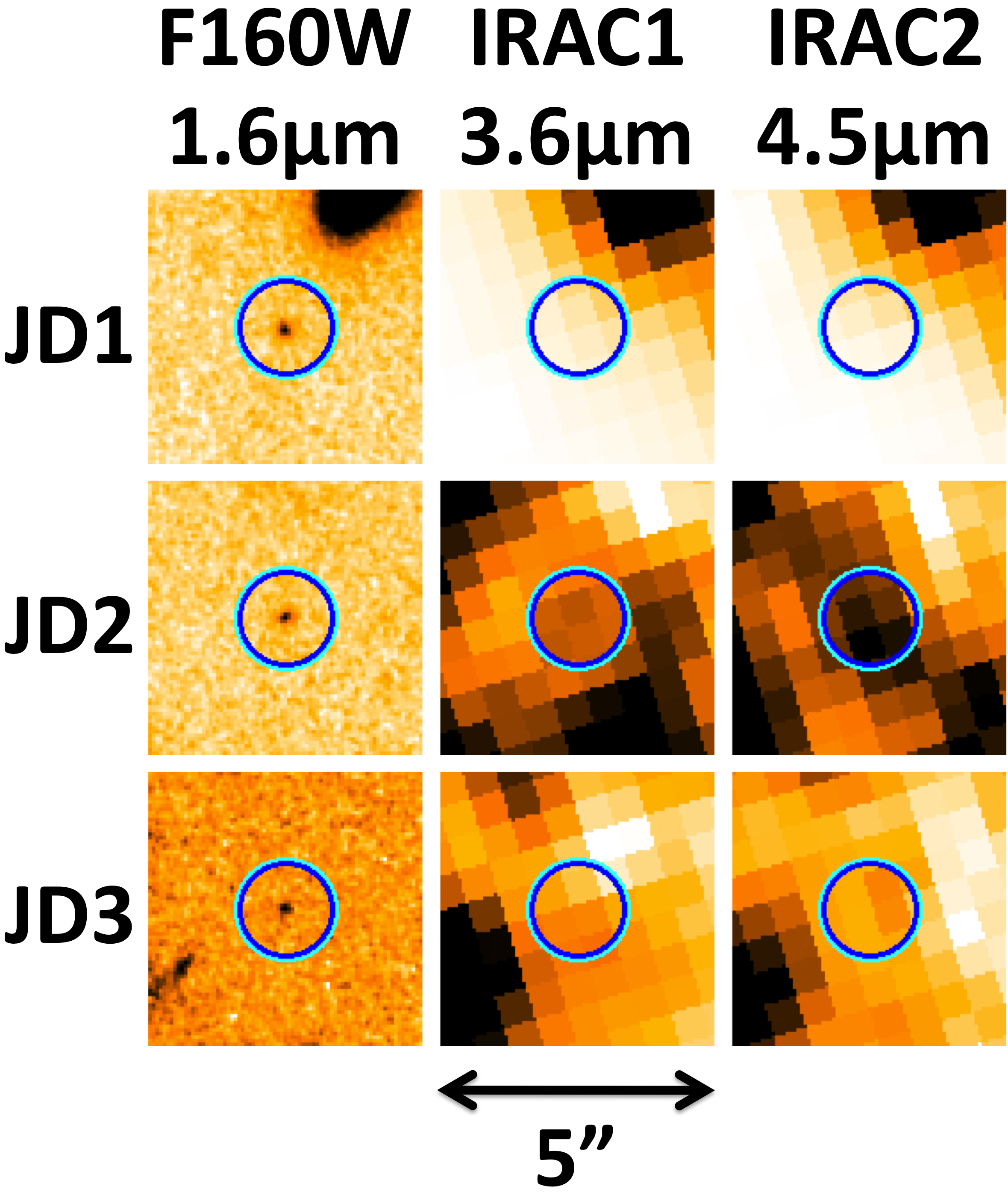}
}
%\captionsetup{labelsep=pipe}
\caption{
\label{fig:IRAC}
Spitzer IRAC ch1 (3.6\um) and ch2 (4.5\um) images of \JD\ compared to the
HST WFC3/IR F160W (1.6\um) image.
Two intensity scalings and zooms are shown.
Left: Both $26'' \times 26''$ F160W cutouts are scaled linearly in photon counts
to the same range as Fig.~\ref{fig:HST}.
And for each Spitzer filter, the same count range is used in each row.
The background photon counts are significantly higher
near JD1 and JD2 (top row) due to intracluster light and scattered starlight.
\JD\ is not detected brightly in the Spitzer images,
supporting the high-redshift solution.
The only possible detection we report is for JD2 at 3.1\sig\ in ch2 (Table \ref{tab:cat}).
JD1 is contaminated by light from other nearby galaxies
which we modeled and subtracted to estimate JD1's photometry.
%We modeled and subtracted these galaxies to estimate the photometry of JD1.
%yielding conservative upper limits.
Right: In each of these $5'' \times 5''$ closeups,
the intensity is scaled independently to the observed range within the central $3'' \times 3''$.
}
\end{figure*}
%%%%%%%%%%%%%%%%%%%%%%%%%%%

\subsection{Spitzer Photometry}
\label{sec:IRACphot}

To derive photometry in the longer wavelength Spitzer IRAC images (Fig.~\ref{fig:IRAC}),
we performed both GALFIT PSF fitting and aperture photometry on JD2 and JD3.
No significant flux is detected for either object in either channel
except for a 3\sig\ detection of JD2 in ch2: mag = $24.8 \pm 0.3$.
Aperture photometry (2.4'' diameter aperture) yields mag = $25.8 \pm 0.3$, 
subject to an approximate 0.7 mag correction, roughly consistent with the GALFIT-derived photometry.

JD1 is significantly contaminated by light from a nearby cluster galaxy.
We modeled this galaxy using GALFIT, subtracted it from the image,
and measured photometry in 2.4'' diameter apertures, yielding a null detection plus uncertainty.
We also added a simulated 25th magnitude source and used GALFIT to derive its photometry.
We conservatively combined the uncertainties from these two measurements in quadrature
to yield total uncertainties (1\sig\ upper limits) of 
277 and 245 nJy in ch1 and ch2, respectively
(3\sig\ limits of mag 24.2 and 24.1).
We also experimented with inflating these uncertainties further by one magnitude
(3\sig\ limits of mag 23.2 and 23.1).
This would increase the JD1 $z<9$ likelihood (see \S\ref{sec:photoz}) from \~3\en7 to 2\en5,
and the likelihood based on the integrated photometry of all three images from 3\en{13} to 2\en9.
% ~/m0647/highz/fullprob.py
%This would increase the JD1 $z<9$ likelihood from $P(z<9) \sim$ 3\en7 to 2\en5.
%This increases the JD1 $z<9$ likelihood by an order of magnitude or so,
%but only to $P(z<9) \sim 10^{-7}$.

%%%%%%%%%%%%%%%%%%%%%%%%%%%

% log P(z)
\begin{figure*}
\centerline{
\includegraphics[width = 0.33\textwidth]{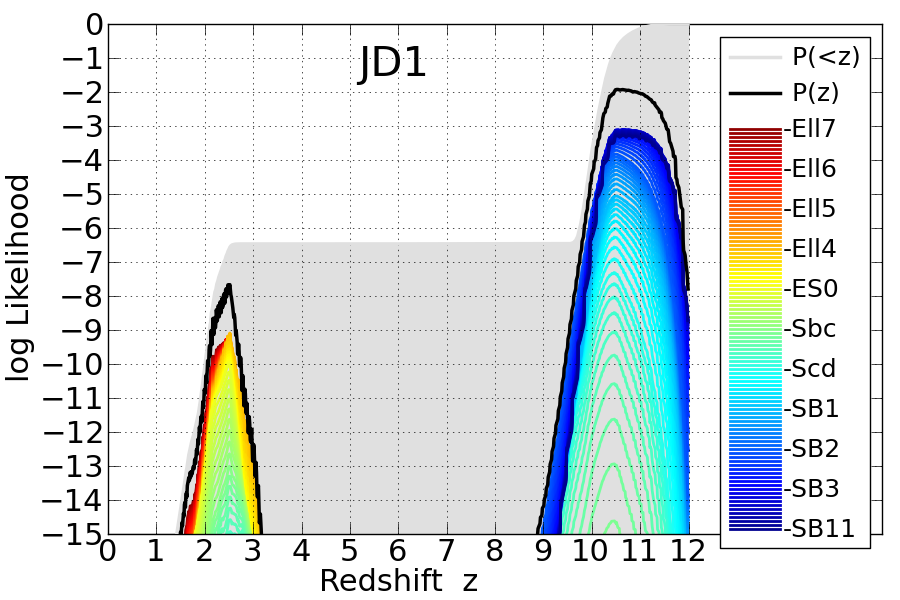}
\includegraphics[width = 0.33\textwidth]{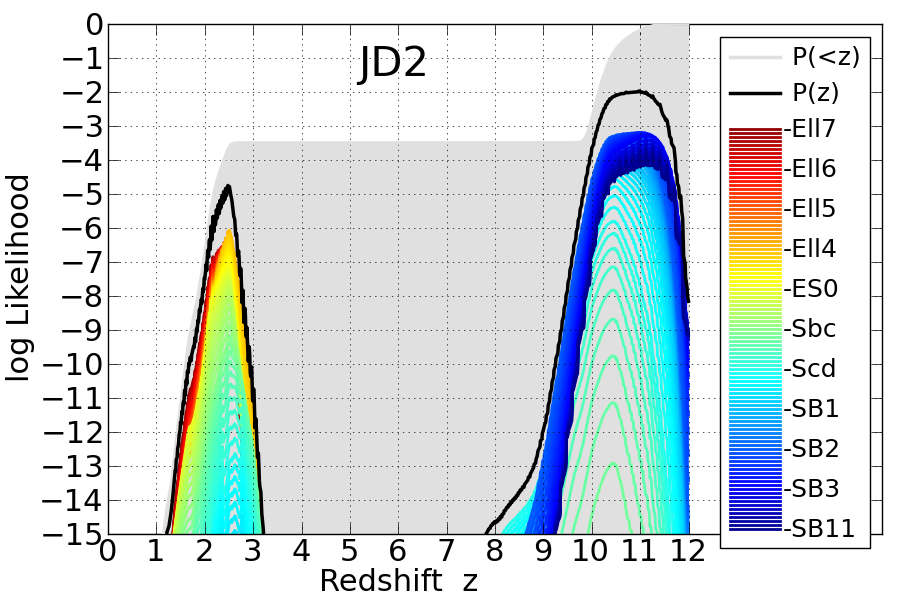}
\includegraphics[width = 0.33\textwidth]{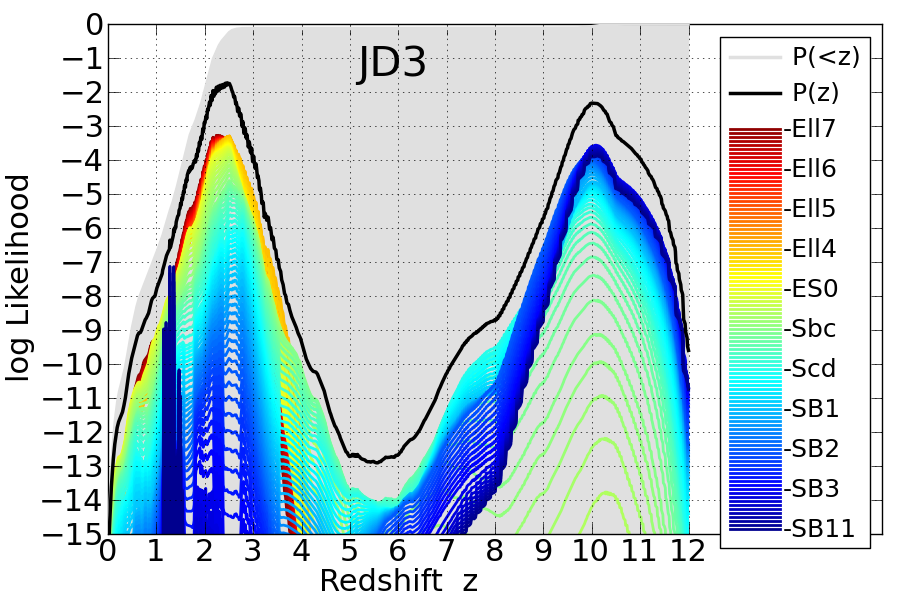}
}
\centerline{
\includegraphics[width = 0.33\textwidth]{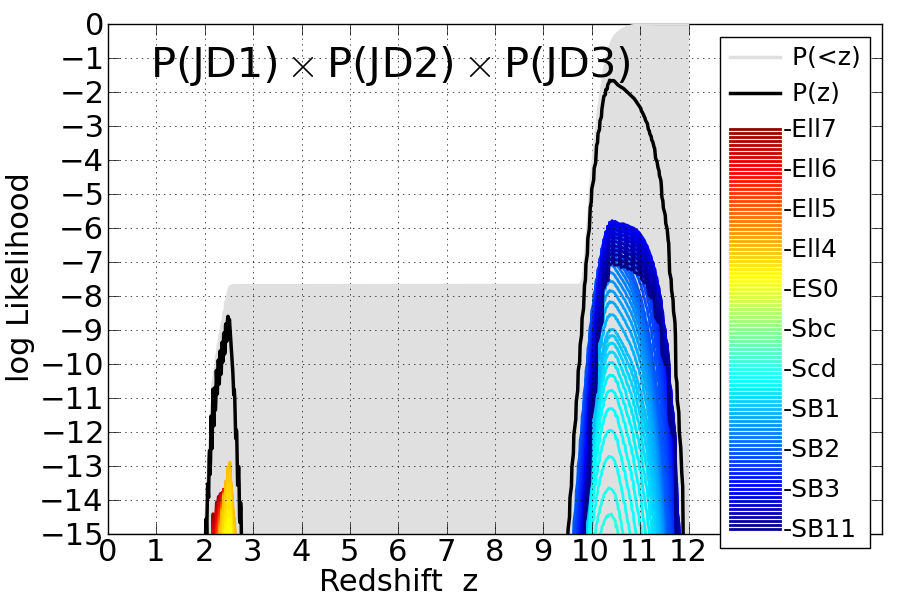}
\includegraphics[width = 0.33\textwidth]{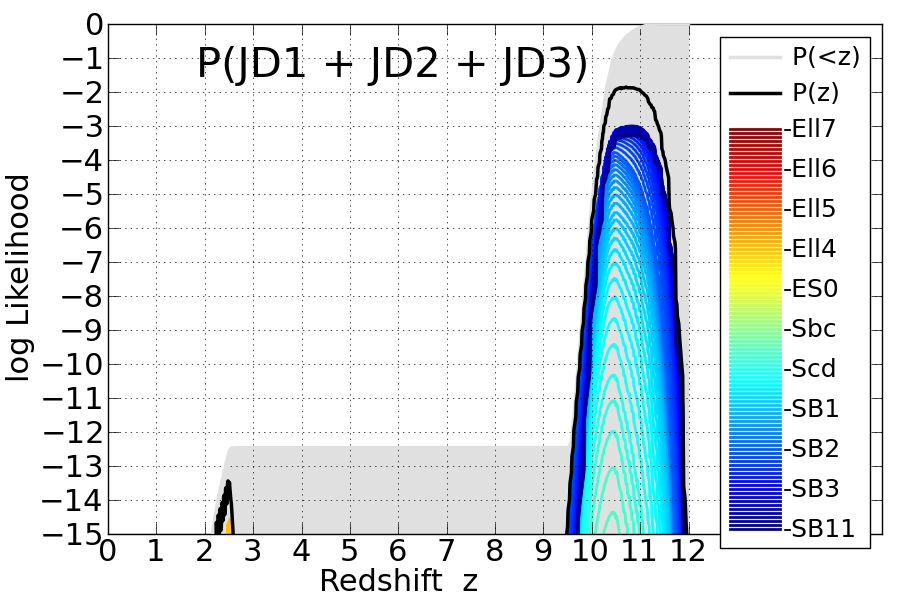}
\includegraphics[width = 0.33\textwidth]{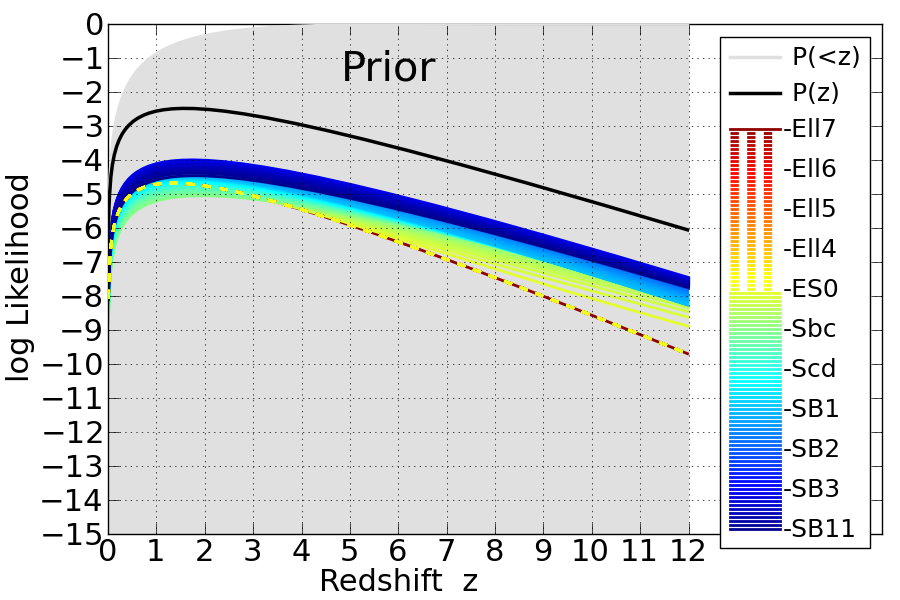}
}
%\captionsetup{labelsep=pipe}
\caption{
\label{fig:Pz}
Top row: 
photometric redshift probability distributions based on our BPZ analysis (\S\ref{sec:BPZ}) 
of HST+Spitzer photometry for each image.
Cumulative probabilities $P(<z)$ are shaded gray;
probabilities $P(z)$ per unit 0.01 in redshift are drawn as black lines;
and likelihoods for individual SED templates are drawn as colored lines.
In this work, we use 
four elliptical templates (Ell),
one E/S0,
two spirals (Sbc and Scd),
and four starbursts (SB).
We then interpolate nine templates between each pair of adjacent templates.
Bottom left:
joint likelihoods for all three images.
Bottom center:
likelihoods based on the summed photometry of all three images.
For all of these likelihoods, we assume the prior plotted at bottom right
for galaxies of intrinsic (delensed) magnitude \~28.2.
This prior was empirically derived from large surveys with photometric and spectroscopic redshifts
and extrapolated to higher redshifts.
}
\end{figure*}
%%%%%%%%%%%%%%%%%%%%%%%%%%%

%\vspace{0.5in}
%\vspace{\stretch{3}}

\section{Photometric Redshift}
%\section{Photometric Redshift derived from SED fitting}
\label{sec:photoz}

We perform two independent analyses of the HST+Spitzer photometry 
to estimate the photometric redshift of \JD.
These two methods, BPZ (\S\ref{sec:BPZ}) and LePHARE (\S\ref{sec:LePHARE})
were the top two performers out of 17 methods tested in \cite{Hildebrandt10}.
They yielded the most accurate redshifts with the fewest outliers
%in terms of minimizing photo-$z$ scatter and outlier rates 
given a photometric catalog for galaxies with known spectroscopic redshifts.

According to our gravitational lensing models (\S\ref{sec:lens}),
\JD1, 2, and 3 are likely three multiple images of the same strongly lensed background galaxy.
Thus, in this section, 
we present photometric redshift likelihoods for each individual image,
as well as jointly for the two brighter images and for all three images.

\subsection{Bayesian Photometric Redshifts (BPZ)}
\label{sec:BPZ}

We used BPZ (Bayesian Photometric Redshifts; \citealt{Benitez00, Coe06})
for our primary photometric redshift analyses.
We modeled the observed HST+Spitzer photometry of \JD\ using
model SEDs %(spectral energy distributions) 
from PEGASE \citep{Fioc97}
which have been significantly adjusted and recalibrated to match the observed photometry of galaxies
with known spectroscopic redshifts from FIREWORKS \citep{Wuyts08}.
The FIREWORKS data set includes 0.38--24\um\ photometry of galaxies
down to mag \~ 24.3 (5\sig\ $K$-band) and spectroscopic redshifts out to $z \sim 3.7$.
In analyses of large datasets with high quality spectra,
this template set yields $\lesssim$ 1\% outliers,
demonstrating that it encompasses the range of metallicities, extinctions, and star formation histories
observed for the vast majority of real galaxies.
(In \S\ref{sec:lozgal} we explore a still broader range of galaxy properties
using a synthetic template set which has not been recalibrated to match observed galaxy colors.)
%These templates include nebular emission lines as 
%Nebular emission lines are included in these templates (for details, see \citealt{Fioc97}).
%\cite{Fioc97} included nebular emission lines in the PEGASE templates.
These templates include nebular emission lines as implemented by \cite{Fioc97}.

The Bayesian analysis tempers the SED model quality of fit 
with an empirically derived prior $P(z,T|m)$
on the galaxy redshift and type given its (delensed) magnitude.
Our prior was constructed as in \cite{Benitez00}
and updated based on likelihoods $P(z,T|m)$ observed in 
COSMOS \citep{Ilbert09}, GOODS-MUSIC \citep{Grazian06,Santini09}, and the UDF \citep{Coe06}.
According to this prior (extrapolated to higher redshifts), 
all galaxy types of intrinsic (delensed) magnitude \~28.2 are over 80 times less likely to be
at $z \sim 11$ than $z \sim 2$.
%a galaxy at $z \sim 11$ is over 100 times less likely than a galaxy at $z \sim 2$.
%
Thus our analysis is more conservative regarding high redshift candidates
than an analysis which neglects to implement such a prior
(implicitly assuming a flat prior in redshift).
The prior likelihoods for \JD\ are uncertain
both due to the prior's extrapolation to $z \sim 11$
and uncertainty in \JD's intrinsic (delensed) magnitude.
Yet it serves as a useful approximation which is surely more accurate than a flat prior.

Based on this analysis, we derived photometric redshift likelihood distributions as plotted in Fig.~\ref{fig:Pz}
and summarized in Table \ref{tab:Pz}.
The images JD1, JD2, and JD3
are best fit by a starburst SED at $z \sim 10.9$, 11.0, and 10.1, respectively.
%When applying the Bayesian prior, a $z \sim 2.5$ elliptical template is slightly preferred for JD3.
After applying the Bayesian prior, 
we find JD1 and JD2 are most likely starbursts at $z \sim 10.6$ and 11.0, respectively.
A $z \sim 2.5$ elliptical template is slightly preferred for JD3,
however $z = 11$ is within the 99\% confidence limits (CL).
Observed at mag \~ 27.3, we may not expect this fainter image to yield as reliable a photometric redshift.

In Table 4 we also provide joint likelihoods based on the brighter two images and all three images equally weighted.
To properly downweight the fainter image, we also analyzed the integrated photometry of all three images
(with uncertainties added in quadrature).
Based on this analysis including our Bayesian prior,
and assuming \JD\ is a galaxy well described by our template set (see also \S\ref{sec:lozgal}),
we found \ztwosig\ (95\% CL)
with a \~3\en{13} likelihood that \JD\ is at $z < 9$.
This likelihood corresponds to a 7.2\sig\ confidence that \JD\ is at $z > 9$.
%In \S\ref{sec:loz} we explore a broad range of other possible interlopers
%and find that the assumption above is reasonable.
The joint likelihood analysis
(weighting all images equally)
yields 
a similar 95\% CL [10.2--11.1]
and a more conservative $P(z < 9)$ \~ 2\en8, or $z > 9$ at 5.5\sig.
%The 95\% (or 99\%) CL's are similar for both analyses, differing only by 0.1 or so in redshift.

The strong confidence in the high redshift solution requires the combined HST and Spitzer photometry.
Without the Spitzer photometry,
the $z > 9$ likelihood would drop to 95\% for the summed HST photometry.
Similarly, we would find $P(z > 9)$ \~ 91\% for JD1 individually.
However, the most likely solutions for JD2 and JD3 would be early types at $z \sim 4$.
We would expect such galaxies to be mag \~ 23 in the Spitzer observations,
which is extremely unlikely (as quantified above) given the measured photometry (see also \S\ref{sec:IRACphot}).

%%%%%%%%%%%%%%%%%%%%%%%%%%%%%
% P(z)
% ~/m0647/highz/fullprob.py
\begin{deluxetable}{lrrl}
\tablecaption{\label{tab:Pz}Individual and Joint Redshift Likelihoods}
\tablewidth{\columnwidth}
\tablehead{
\colhead{Image}&
%\colhead{68\% CL}&
\colhead{95\% CL}&
\colhead{99\% CL}& % includes z = 11 for JD3
\colhead{P($z<9$)}
}
\startdata
%JD1 $\times$ JD2 $\times$ JD3  &  $10.42^{+0.66}_{-0.19}$  &  [10.12--11.30]  &  2\en{8} \\
%JD1 $+$ JD2 $+$ JD3  &  $10.71^{+0.59}_{-0.37}$  &  [10.20--11.47]  &  3\en{13} \\
% 68%, 99%:
%JD1 (F160W \~ 25.9)  &  $10.62^{+0.54}_{-0.14}$  &  [10.11--11.67]  &  3\en{7} \\
%JD2 (F160W \~ 26.1)  &  $10.99^{+0.24}_{-0.52}$  &  [ 9.99--11.69]  &  3\en{4} \\
%JD3 (F160W \~ 27.3)  &  $ 2.48^{+7.54}_{-0.24}$  &  [ 1.81--11.07]  &  7\en{1} \\
%JD1 $\times$ JD2  &  $10.66^{+0.45}_{-0.14}$  &  [10.21--11.53]  &  4\en{10} \\
%JD1 $\times$ JD2 $\times$ JD3  &  $10.42^{+0.39}_{-0.07}$  &  [10.12--11.30]  &  2\en{8} \\
%99%
%JD1  &  $10.62^{+0.54}_{-0.14}$  &  [10.11--11.67]  &  3\en{7} \\
%JD2  &  $10.99^{+0.24}_{-0.52}$  &  [ 9.99--11.69]  &  3\en{4} \\
%JD3  &  $ 2.48^{+7.54}_{-0.24}$  &  [ 1.81--11.07]  &  7\en{1} \\
%JD1$\times$JD2  &  $10.66^{+0.45}_{-0.14}$  &  [10.21--11.53]  &  4\en{10} \\
%JD1$\times$JD2$\times$JD3  &  $10.42^{+0.39}_{-0.07}$  &  [10.12--11.30]  &  2\en{8} \\
%95%
%JD1 & $10.62^{+0.54}_{-0.14}$  &  [10.28--11.45]  &  3\en{7}\\
%JD2 & $10.99^{+0.24}_{-0.52}$  &  [10.22--11.49]  &  3\en{4}\\
%JD3 & $ 2.48^{+7.54}_{-0.24}$  &  [ 2.06--10.43]  &  7\en{1}\\
%JD1$\times$JD2 & $10.66^{+0.45}_{-0.14}$  &  [10.35--11.34]  &  4\en{10}\\
%JD1$\times$JD2$\times$JD3 & $10.42^{+0.39}_{-0.07}$  &  [10.23--11.08]  &  2\en{8}\\
% ~/m0647/highz/fullprob.py
JD1 (F160W \~ 25.9)\super{a}  &  $10.62^{+0.83}_{-0.34}$  &  [10.11--11.67]  &  3\en{7} \\
JD2 (F160W \~ 26.1)\super{a}  &  $10.99^{+0.50}_{-0.77}$  &  [ 9.99--11.69]  &  3\en{4} \\
JD3 (F160W \~ 27.3)\super{a}  &  $ 2.48^{+7.95}_{-0.42}$  &  [ 1.81--11.07]  &  7\en{1} \\
%JD1 $\times$ JD2  &  $10.66^{+0.68}_{-0.31}$  &  [10.21--11.53]  &  4\en{10} \\
%P(JD1) $\times$ P(JD2) $\times$ P(JD3)  &  $10.42^{+0.66}_{-0.19}$  &  [10.12--11.30]  &  2\en{8} \\
P(JD1)$\times$P(JD2)\super{b}  &  $10.66^{+0.68}_{-0.31}$  &  [10.21--11.53]  &  4\en{10} \\
P(JD1)$\times$P(JD2)$\times$P(JD3)\super{b}  &  $10.42^{+0.66}_{-0.19}$  &  [10.12--11.30]  &  2\en{8} \\
P(JD1 $+$ JD2)\super{c}  &  $10.99^{+0.43}_{-0.61}$  &  [10.22--11.59]  &  2\en{11} \\
P(JD1 $+$ JD2 $+$ JD3)\super{c}  &  $10.71^{+0.59}_{-0.37}$  &  [10.20--11.47]  &  3\en{13} \\
\vspace{-0.1in}
\enddata
%\tablecomments{JD3 is over a magnitude fainter (F160W \~ 27.3) than the other two images.
\tablecomments{
BPZ results assuming \JD\ is well modeled by our SED templates (\S\S\ref{sec:photoz}--\ref{sec:loz}).
All likelihoods include a Bayesian prior which assumes 
galaxies of this (unlensed) magnitude are over 80 times more likely to be at $z \sim 2$ than $z \sim 11$.
See also Fig.~\ref{fig:Pz}.}
\tablenotetext{1}{Approximate AB magnitudes are given in parentheses. (See also Table \ref{tab:cat}.)
Note JD3 is significantly fainter.}
\tablenotetext{2}{Joint likelihood of multiple images weighted equally.}
\tablenotetext{3}{Likelihood based on integrated photometry of multiple images.}
%$z \sim 2$ is over 100 times more likely than $z \sim 11$.}
%\tablecomments{All likelihoods include a prior which assumes $z \sim 11$ 
%is over 100 times less likely than $z \sim 2$.}
%\tablecomments{All likelihoods include a prior which assumes a $z \sim 11$ starburst galaxy
%is \~140 times less likely than a $z \sim 2.5$ early type, the next most likely candidate.}
%See explanation in \S\ref{sec:HSTphot}}
%\tablenotetext{2}{R.A.~\& Decl.~(J2000) are given in parentheses for the Abell clusters,}
\end{deluxetable}
%%%%%%%%%%%%%%%%%%%%%%%%%%%%%

\subsection{Le PHARE}
\label{sec:LePHARE}

We also used Le PHARE \citep{Arnouts99,Ilbert06,Ilbert09} 
to independently estimate the photometric redshifts.  % based on our HST+Spitzer photometry.
For this analysis we used a SED template library primarily from \cite{Ilbert09} 
as optimized for the COSMOS survey \citep{Scoville07a,Scoville07b,Koekemoer07}.
This includes three ellipticals and seven spirals as generated by \cite{Polletta07}
using the GRASIL code \citep{Silva98},
as well as 12 starburst galaxies with ages ranging from 30 Myr to 3 Gyr generated by GALAXEV based on \cite{BC03}.
We supplemented these with four additional elliptical templates for a total of seven ellipticals.

We added dust extinction in ten steps up to $E(B-V) = 0.6$.
(Stronger degrees of extinction are explored in \S\ref{sec:lozSED}.)
Four different dust laws were explored:
\cite{Calzetti00}; Calzetti plus two variations on a 2170\AA\ bump;
and \cite{Prevot84} as observed for the SMC.

%We adopted the same Bayesian prior as described above for BPZ as also implemented as an option in LePhare.
We adopted the \cite{Benitez00} prior as implemented in LePhare.
The results were consistent with those from BPZ:
$z = 10.6^{+0.6}_{-0.2}$ (JD1),
$z = 10.6^{+0.5}_{-0.3}$ (JD2), and
$z = 10.1^{+0.3}_{-0.3}$ (JD3), each at 68\% CL.
A secondary solution of $z \sim 2.5$ was reported for JD3
with a peak likelihood ten times less than that of the best fit high redshift solution.

%%%%%%%%%%%%%%%%%%%%%%%%%%%
% SED fits

\begin{figure*}
\centerline{
\includegraphics[width = \textwidth]{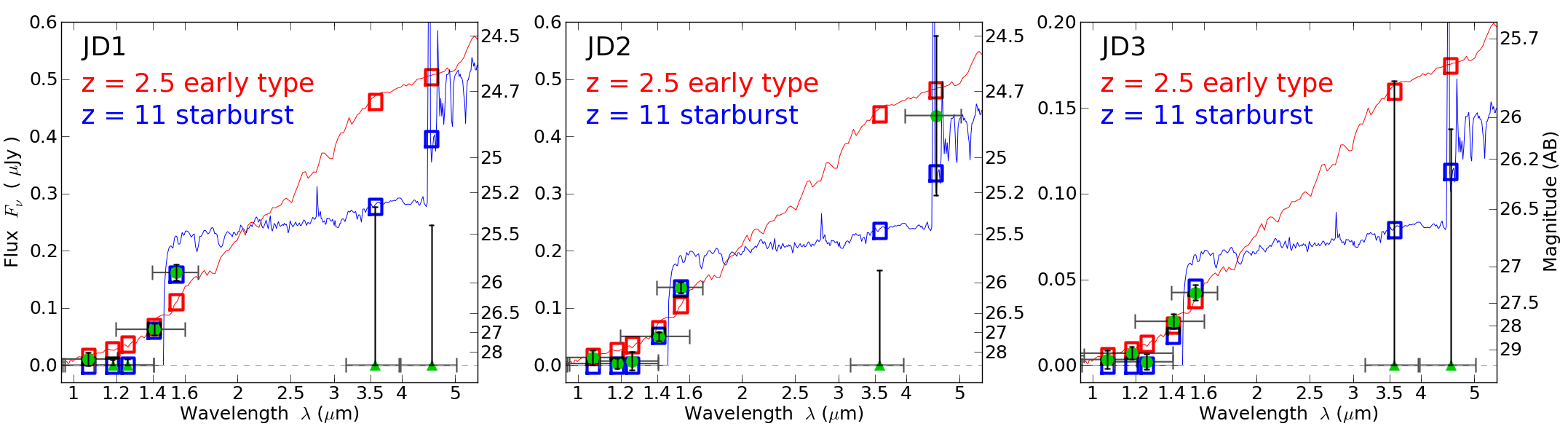}
}
\caption{
\label{fig:sedcomp}
Observed near-IR photometry from HST WFC3/IR and Spitzer IRAC (filled circles and triangles)
compared to the expected fluxes (open squares) from two SEDs:
the $z = 11.0$ starburst from Fig.~\ref{fig:HSTsed} (blue) and a $z = 2.5$ early type galaxy (red).
% 11.03
%Here the JD2 observed SED includes 
%the F110W 2.5\sig\ detection due to elevated backgrounds in the visit A9 exposures (\S\ref{sec:HSTphot}).
Note the JD3 plot is scaled differently along the y-axis.
}
\end{figure*}

%%%%%%%%%%%%%%%%%%%%%%%%%%%

% HST colors
\begin{figure*}
\centerline{
\includegraphics[width = 0.33\textwidth]{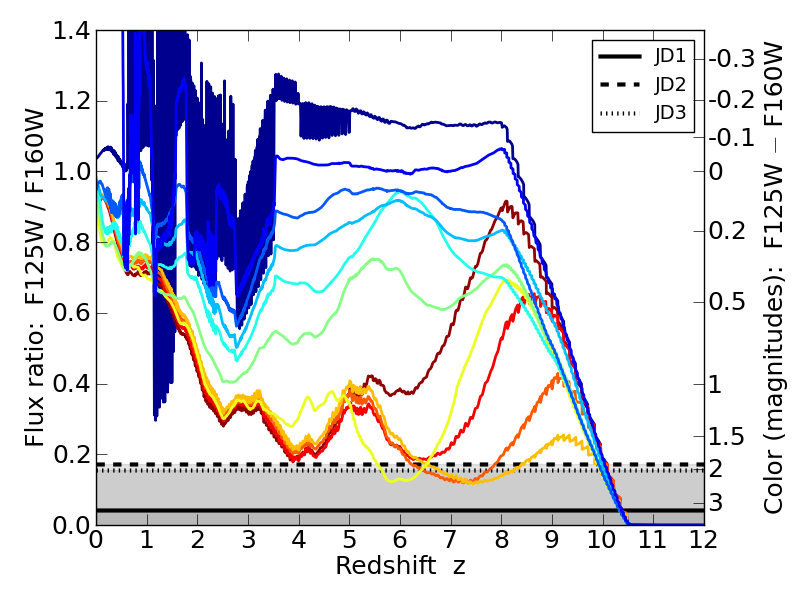}
\includegraphics[width = 0.33\textwidth]{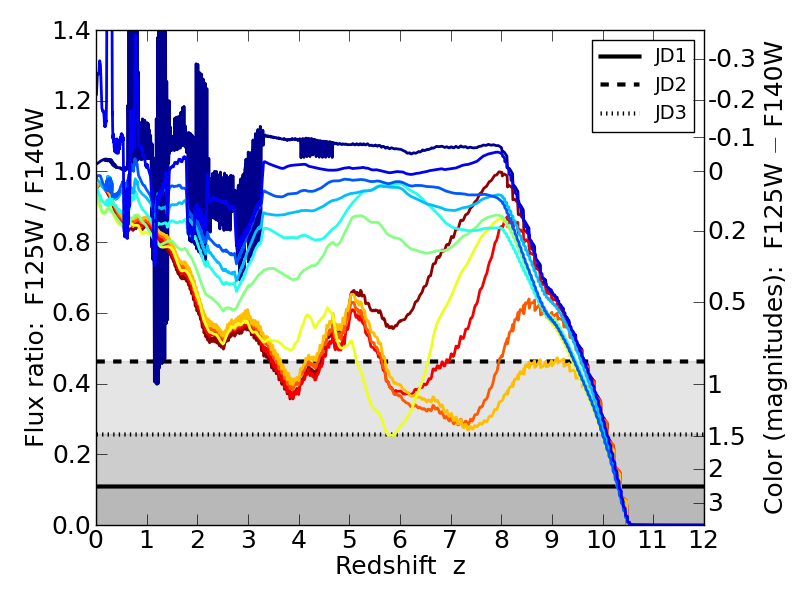}
\includegraphics[width = 0.33\textwidth]{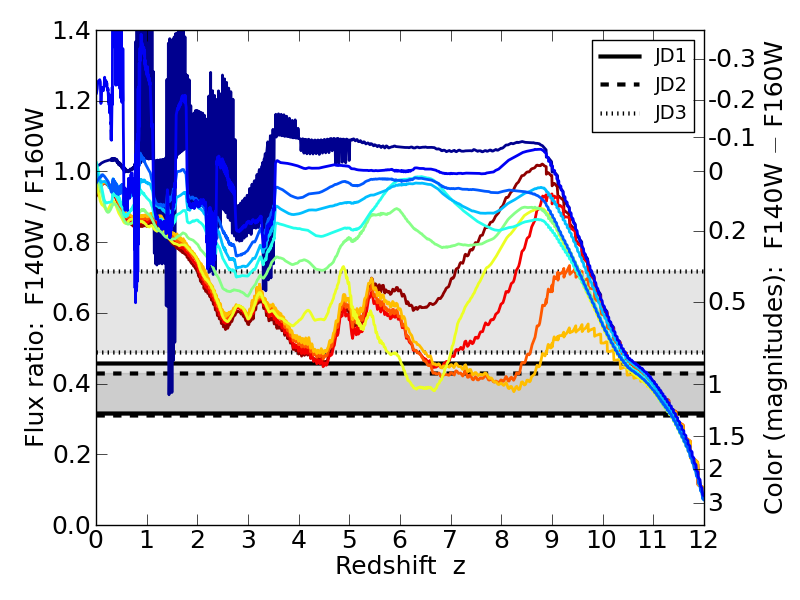}
}
%\captionsetup{labelsep=pipe}
\caption{
\label{fig:ccHST}
Observed WFC3/IR colors (shaded 68\% confidence regions)
for JD1 (solid lines), JD2 (dashed lines), and JD3 (dotted lines)
plotted against those predicted with the BPZ template set
from young starburst (blue) to early type (yellow-orange-red)
as a function of redshift.
The three panels plot flux ratios in 
F125W $/$ F160W (left),
F125W $/$ F140W (middle), and
F140W $/$ F160W (right).
The corresponding colors in magnitudes are given along the right axes.
}
\end{figure*}

%%%%%%%%%%%%%%%%%%%%%%%%%%%

% Color-color Galaxies
\begin{figure}
%\centerline{\includegraphics[width = \textwidth]{figs/timeseries.png}}
%\centerline{\includegraphics[width = \columnwidth]{figs/colorcolor.png}}
%\centerline{\includegraphics[width = \columnwidth]{figs/fluxratiosHST1.png}}
\centerline{
\includegraphics[width = \columnwidth]{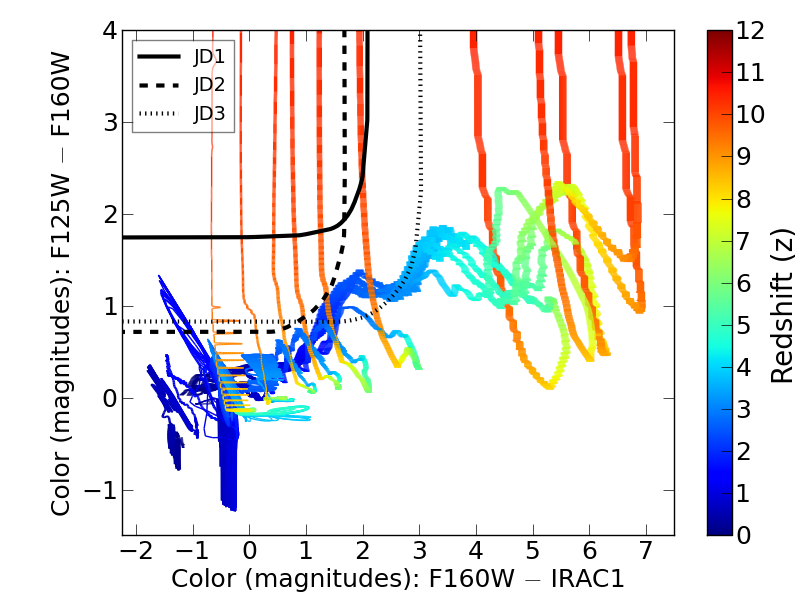}
}
%\captionsetup{labelsep=pipe}
\caption{
\label{fig:ccgal}
%{\bf Color-color.}
Observed colors in WFC3/IR F125W $-$ F160W and F160W $-$ IRAC ch1 
plotted as black lines (95\% confidence limits)
versus those predicted from the current BPZ template library 
(lines colored as a function of redshift and made thicker for earlier galaxy types).
}
\end{figure}

%%%%%%%%%%%%%%%%%%%%%%%%%%%

%\vspace{0.1in} %%%%%%%%%% XXXXX

\section{Lower Redshift Interlopers Ruled Out}
\label{sec:loz}

In this section we consider a broad range of $z < 11$ possibilities.
As found in \S\ref{sec:photoz},
the $z < 9$ likelihood is formally \~3\en{13} assuming \JD\ is a galaxy well-modeled by our SED templates.  %\~2\en8.
%\~$3.1\times10^{-7}$.
Though strongly disfavored,
a $z \sim 2.5$ early type and/or dusty galaxy is the most likely alternative,
as we discuss further in \S\ref{sec:lozgal}.
We reanalyzed previously published \Jdrops\ and found them most likely to be at intermediate redshift (\S\ref{sec:Jdrops}).
Objects within the Galaxy are less likely, as this would require three objects with extremely rare colors (Fig.~\ref{fig:catfrats})
at positions consistent with strongly-lensed multiple images according to our lens models (\S\ref{sec:lens}).
Nevertheless, we found the only stars or brown dwarfs consistent with the observed colors are rare, transient post-AGB flare-ups,
though these would be far more luminous if observed within the Galaxy (\S\ref{sec:stars}).
Solar system objects would have likely exhibited parallax motion
and are inconsistent with the observed colors (\S\ref{sec:solarsys}).
Intermediate redshift
long duration multiply-imaged supernovae (\S\ref{sec:SN})
and emission line galaxies (\S\ref{sec:ELG})
are also extremely unlikely.
We conclude that \JD\ is most likely either at $z \sim 11$
or exhibits unique photometry yet to be observed in any other known object.

%\clearpage

\subsection{Intermediate Redshift Galaxy?}
\label{sec:lozgal}

\subsubsection{SED Constraints}
\label{sec:lozSED}

%The most likely alternative to $z \sim 11$
While we found $P(z<9)$ 
\~ 3\en{13},
%\~ 2\en8,
the next best alternative to $z \sim 11$
%is an early type and/or dusty galaxy
is an early type galaxy (ETG)
at $z \sim 2.5$ (Fig.~\ref{fig:sedcomp}).
%is a galaxy which is early type (ETG) and/or dusty
At $z \sim 2.65$, the 4000\AA\ break is redshifted to 1.46\um, coinciding with \Lya\ (1216\AA) redshifted to $z \sim 11.0$.
However, 4000\AA\ breaks are not expected to be as strong as observed for 
\JD\ (Figs.~\ref{fig:ccHST} and \ref{fig:ccgal}).
JD1 features a $J_{125} - H_{160} \gtrsim 3$ magnitude break between F125W and F160W
as well as a \~1 magnitude break between F140W and F160W.
%(Fig.~\ref{fig:ccHST}).
Thus low redshift ETGs yield a significantly worse SED fit than $z \sim 11$
for all three images as quantified in Fig.~\ref{fig:Pz} and Table~\ref{tab:Pz}.

To explore an even broader range of galaxy SED models than used in \S\ref{sec:photoz},
we utilized the flexible stellar population synthesis (FSPS) models from \cite{Conroy09} and \cite{ConroyGunn10}.
They provide simple stellar population (SSP) models
which span ages of $5.5 \leq$ log(age/yr) $\leq 10.175$
and metallicities of $0.0002 \leq Z \leq 0.03$ (where $Z_\odot = 0.019$).
Nebular emission lines are not included.
We convolved their SSP models with star formation histories (SFH)
ranging from the single early burst (SSP) to exponentially declining (``$\tau$ models''),
continuous (constant rate), and exponentially rising (``inverted $\tau$ models'').
The latter rising SFH likely describes high-redshift galaxies best 
according to both observations \citep{Maraston10,Papovich11,Reddy12} and simulations \citep{Finlator11}.
Finally we added a variable degree (up to $A_V = 30$ magnitudes)
of \cite{Calzetti00} dust extinction with $R_V = A_V / E_{B-V} = 4.05$.

% ~/fsps/tauAV1z.py
To uncover the most likely solutions in different regions of this multi-dimensional parameter space,
we began with relatively coarse grid searches 
with redshift intervals of 0.1 and \~9 steps in each of the four other free parameters.
We then zoomed in on the higher likelihood regions,
found again to be roughly $z \sim 2.5$ and 11.
Finally, we ran \cite{Powell64} minimizations to find the best fitting SEDs at each of these redshifts.

% ~/m0647/highz/EplusA/
We supplemented these SEDs
with a suite of smooth $\tau$ models with stochastic bursts superposed \citep[e.g.,][]{Kauffmann03a, Salim07}, 
as well as truncated (``quenched'') SFHs designed to reproduce the colors of post-starburst (K+A) galaxies.

% ~/CLASH/data/m0647/highz/fsps/opt_JD1.py
Our results with this combined template set confirm that a $z \sim 11$ model fits \JD\ best,
while evolved and/or dusty galaxies at $z \sim 2.5$ provide the best alternatives
but are still significantly worse statistically.
The best fitting intermediate-redshift template to the summed photometry
($z \sim 2.7$; 
\~400 Myr old; 
$A_V \sim 0.8$ mag)
with $\chi^2 = 57.6$
is only \~\tentotheminus9 times as likely as the best fitting $z \sim 11$ template
($z \sim 10.9$; \~6 Myr old; $A_V = 0$)
yielding $\chi^2 = 16.9$
with $\gtrsim 14$ degrees of freedom 
given the 19 photometric measurements and $\lesssim 5$ free parameters
%three of which provide the most freedom
(see discussion in \citealt{Andrae10}).

The uncertainties on the $z \sim 11$ SED parameters are quantified in \S\ref{sec:sedprops}.
A proper calculation of the redshift likelihoods based on these templates
would require an estimate of the prior likelihoods in this multidimensional parameter space,
which is beyond the scope of this work.
And while these templates probe a broad parameter space,
we derive our primary photometric redshift estimates in \S\ref{sec:photoz} 
from templates which have been well calibrated to match the observed photometry of galaxies with spectroscopic redshifts.

We note there is no evidence that $z > 2$ ETGs have significantly different SEDs 
than our ETG models calibrated at lower redshifts.
The highest redshift ETG observed to date is HUDF-1446 with a spectroscopic redshift $z = 2.67$ \citep{Damjanov11}.
\cite{Coe06} published a photometric redshift $z = 2.74 \pm 0.44$ for this object using BPZ,
in good agreement with the true redshift.
Their ETG templates yielded a good fit to the 
ACS ($B_{435}V_{606}i'_{775}z'_{850}$) and NICMOS ($J_{110}H_{160}$) photometry,
including the $J_{110} - H_{160} = 1.89 \pm 0.13$ break
with $H_{160} = 23.074 \pm 0.098$
and significant detections in all ACS filters.

%\subsubsection{Physical Size Compared to Other $z > 2$ ETGs}  % red nuggets
\subsubsection{Lower Stellar Mass Than Observed $z > 2$ ETGs}  % red nuggets

If \JD\ were at $z \sim 2.5$
(despite the low likelihood of this from SED fitting)
it would likely be the least massive early type host galaxy observed to date at $z > 1$.
Spectroscopically confirmed $z > 1.4$ ETGs to date
have stellar masses $>$ 2\e{10}\Msun\ \citep{Damjanov11}.
HUDF-1446 at $z = 2.67$, for example, is \~8\e{10}\Msun.

Our subset of lens models that allow for \JD\ to be at $z \sim 3$ (\S\ref{sec:lensresults})
suggest that the magnification of the brightest two images would be $\mu > 30$.
Thus it would be intrinsically \~300 times fainter than HUDF-1446,
with a correspondingly lower stellar mass on the order of  
\~2\e8\Msun\ (and still $<$ \tentothe9\Msun\ if we assume a more conservative magnification factor of $\mu \sim 10$;
%\~2\e8\Msun\ (or $<$ \tentothe9\Msun\ more conservatively, assuming $\mu \sim 10$; 
see also $z \sim 11$ mass estimates in \S\ref{sec:mass}).

Quiescent galaxies of such low masses at $z > 2$ would be a surprising discovery.
Observations to date demonstrate \citep[e.g.,][]{Peng10}
that star formation is only significantly quenched by feedback in more massive galaxies,
or alternatively as a galaxy is harassed as a satellite of a larger halo.
\JD\ is not observed to be a satellite of a galaxy group.
%more massive group of galaxies.

% ~/papers/Damjanov11_table2.txt
%                                            RA              Dec         spec-z    selection   Mstel[1e11Msun]            r_1/2 kpc     Sersic index
% HUDF-1446              53.163333 -27.809000 2.6700 photometry    0.782 0.266  F850LP  0.760 0.120 0.800 0.200           
% GN/DEIMOS-1323        189.024020  62.167500 0.9362 morphology    0.750        F606W   0.377       4.000            
% SA22-0455             334.455917   0.246961 1.3130 spectroscopy  0.352 0.096  F160W   0.420 0.011 4.560 0.285          

%%%%%%%%%%%%%%%%%%%%%%%%%%%
% A2667-J1 SED & P(z)

\begin{figure*}
\centerline{
\includegraphics[width = \columnwidth]{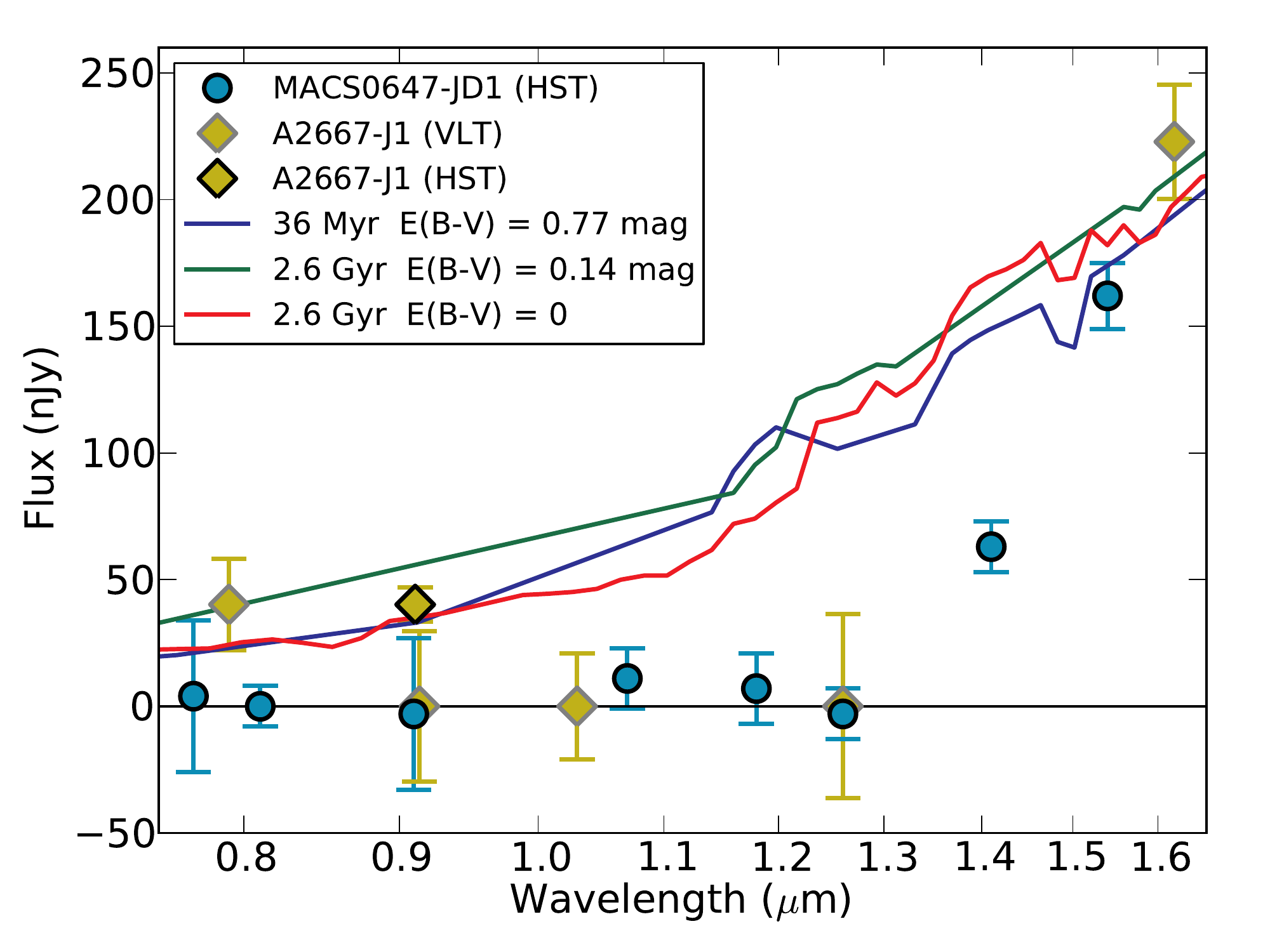}
\includegraphics[width = 1.07\columnwidth]{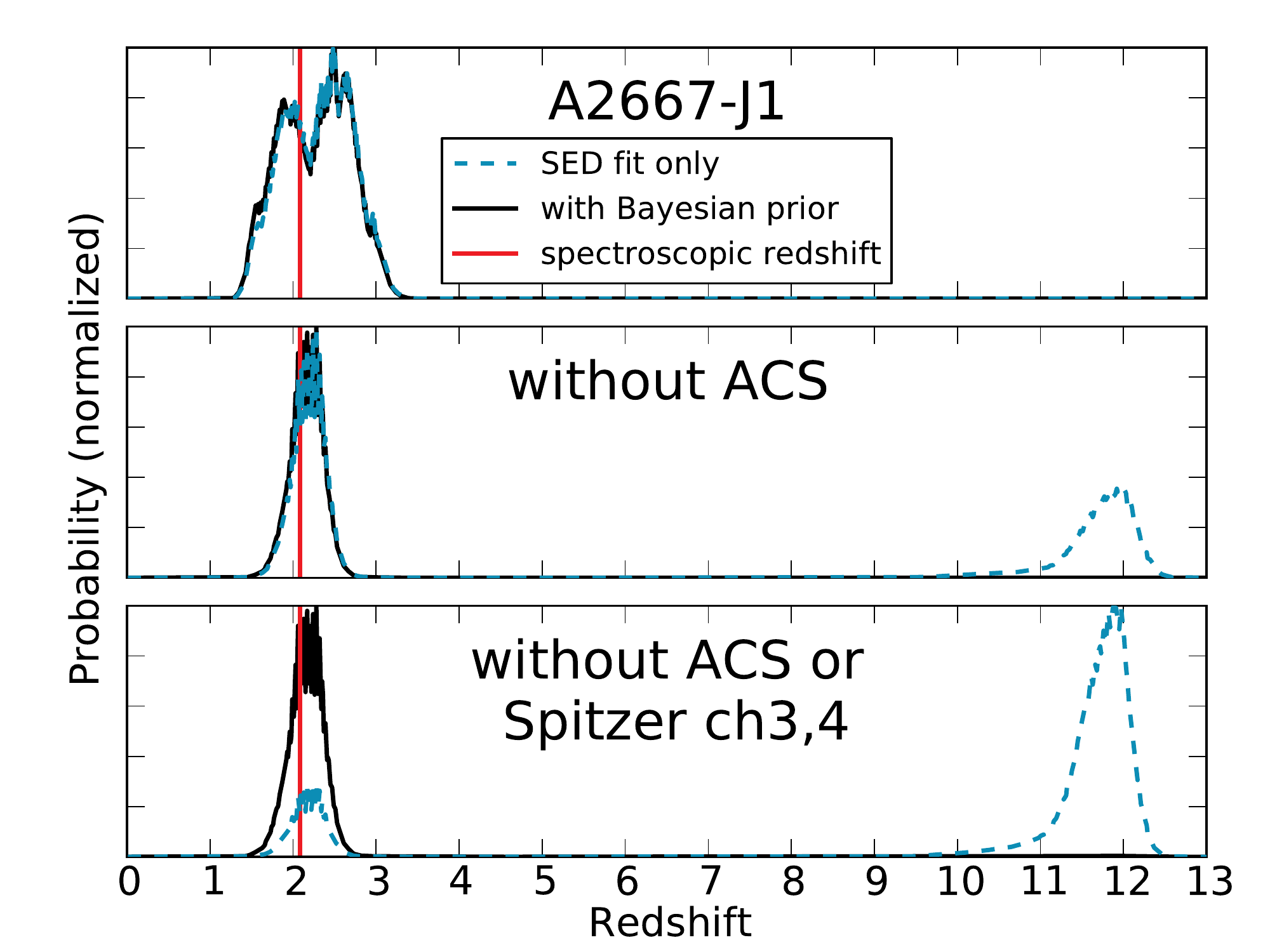}
}
\caption{
\label{fig:Hayes}
\label{fig:PzLH}
Left:
comparison of near infrared photometry of lensed \Jdrops\ \JD1 (this work) and A2667-J1 \citep{Laporte11,Hayes12}.
Also overplotted are three SED fits from Figure 2 of \cite{Hayes12} 
to the photometry of A2667-J1 at its spectroscopic redshift $z = 2.082$.
The \cite{Hayes12} photometry plotted here is all from VLT (FORS2 and HAWK-I)
except for the 6.0\sig\ detection in HST/ACS F850LP,
the upper diamond with a darker border at 0.91\um.
Right:
photometric redshift probability distribution for A2667-J1 
based on our reanalysis of the photometry provided in \cite{Hayes12}
with and without a Bayesian prior.
%(as in Fig.~\ref{fig:Pz} for \JD).
%with (dark lines) and without (light lines) a Bayesian prior on galaxy redshift and type given the (delensed) magnitude $P(z,T|m)$.
%as determined by our analysis
%The top panel uses all available photometry from \cite{Hayes12}.
The top panel uses all the available photometry.
The middle panel omits the 6.0\sig\ detection in ACS/F850LP.
The bottom panel omits both ACS and Spitzer IRAC ch3 and ch4.
The spectroscopic redshift $z = 2.082$ is indicated by the red vertical lines.
}
\end{figure*}

%%%%%%%%%%%%%%%%%%%%%%%%%%%

%\subsection{Comparisons to Previous \Jdrop\ High Redshift Candidates}
\subsection{Comparisons to Previously Published \Jdrops}
\label{sec:Jdrops}

The previous highest redshift candidate, UDFj-39546284 \citep{Bouwens11N},
was detected at 5.8\sig\ in a single HST band (WFC3/IR F160W)
dropping out of F125W and bluer filters
also with non-detections in Spitzer
yielding a photometric redshift of $z = 10.3 \pm 0.8$.
%The availability of the F140W filter on WFC3 \citep{BrownBaggett06} and CLASH's use of it
The ultimate inclusion of the F140W filter on WFC3 \citep{BrownBaggett06} 
%and in CLASH observations
%and CLASH's use of it
and in the CLASH observing program
enable us to securely identify \JD\ as the highest redshift galaxy candidate to date.
%The additional detection 
At $z \sim 11.0$, \Lya\ is redshifted to \~1.46\um, 
causing the galaxy light to drop out of \~2/3 of the F140W bandpass
as well as \~1/5 of F160W.
The ratio between these two filling factors 
($0.8 / 0.33 \sim 2.4$, corresponding to \~1.0 mag)
places tight, model-independent constraints on the wavelength of the (redshifted) Lyman break
and thus the redshift of \JD\ (Fig.~\ref{fig:ccHST}).
The five NIR HST filters used by CLASH
also enabled \cite{Zheng12} to discover a \Jdrop\ lensed by MACSJ1149.6+2233
and robustly measure its photometric redshift to be $z = 9.6 \pm 0.2$ (68\% CL).

\cite{Laporte11} identified a \Jdrop\ lensed by Abell 2667
based on VLT (FORS2 and HAWK-I), ACS/F850LP, and Spitzer IRAC (ch1 through ch4) photometry.
\cite{Hayes12} then measured a spectroscopic redshift of $z = 2.082$ for that galaxy, A2667-J1.
\cite{Laporte11} had already emphasized that $z > 9$ possibilities were excluded
based on the significant (6.0\sig) ACS detection.
We concur with this conclusion
after reanalyzing their photometry as provided in \cite{Hayes12}.
Only by excluding the ACS data point and assuming no Bayesian redshift prior
do $z > 9$ solutions have significant probability (Fig.~\ref{fig:PzLH}).
If Spitzer IRAC ch3 and ch4 were not available (as is the case with \JD)
{\em in addition} to the ACS detection being unavailable,
then the $z > 9$ likelihood would rise further
yet still be insignificant once the prior is included.
The $z > 9$ likelihood is enhanced further, but only modestly, 
if the IRAC ch1 and ch2 uncertainties are inflated to yield only 3\sig\ detections
(as is the case for our JD2 IRAC ch2).
In Fig.~\ref{fig:Hayes} we compare the observed NIR photometry of A2667-J1 and \JD1.
Our multiband HST photometry of the latter yields significantly tighter upper limits on the non-detections
and adds a key data point at 1.4\um, resulting in a far greater $z \sim 11$ likelihood
%which is significant for all three of our images 
even when accounting for the Bayesian prior which disfavors them (Fig.~\ref{fig:Pz}).

We also applied our analysis methods to the photometry of other \Jdrops\ in the literature.
\cite{Schaerer07} showed A1835-\#17 was fit well by a dusty ($A_V \sim 3.6$ mag) starburst at $z \sim 0.8$.
\cite{Dickinson00} presented both $z \gtrsim 2$ and $z \gtrsim 10$ solutions for HDF-N J123656.3+621322.
And HUDF-JD2 \citep{Mobasher05} has since been shown to likely be a $z \sim 1.7$ LIRG \citep{Chary07}.
For all three of these \Jdrops, our analysis yields low redshift ($2 \lesssim z \lesssim 4$ or very dusty $z \lesssim 1$)
solutions which are strongly preferred given our Bayesian prior.

%%%%%%%%%%%%%%%%%%%%%%%%%%%

% Catalog Flux ratios
\begin{figure}
\centerline{
\includegraphics[width = \columnwidth]{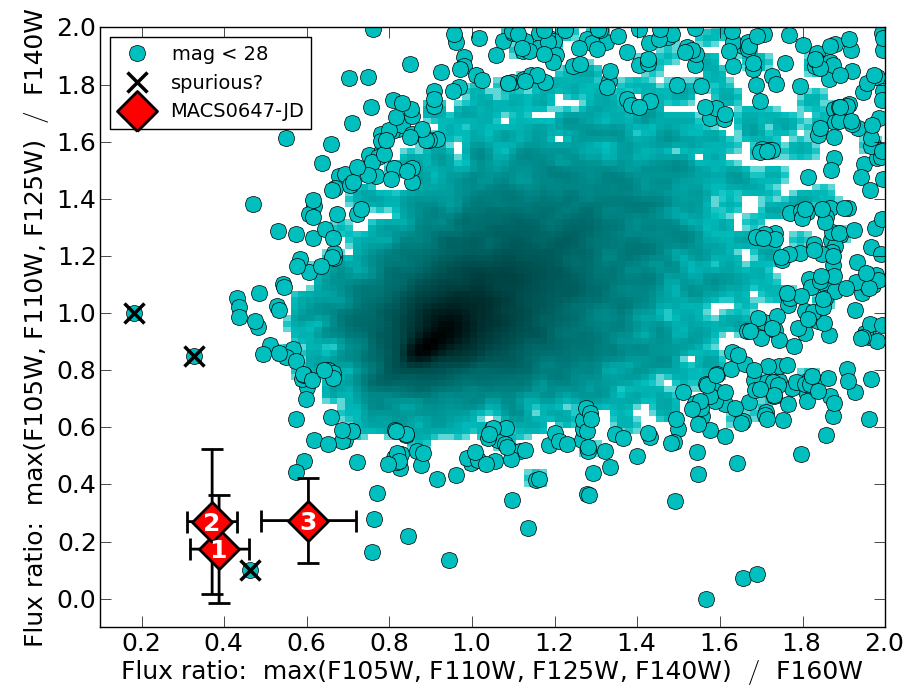}
}
%\captionsetup{labelsep=pipe}
\caption{
\label{fig:catfrats}
Observed NIR colors of the \Jdrops\ (red diamonds with 1\sig\ uncertainties) 
plotted against those observed for all other 20,746 CLASH sources
brighter than 28th magnitude in both F160W and F140W
and also observed in F125W (filled circles and density map).
The horizontal axis gives the ratio of F160W flux to the maximum flux in all bluer WFC3/IR filters.
The vertical axis gives a similar flux ratio but for F140W.
Three objects with colors similar to the \Jdrops\ appear to be spurious IR artifacts based on visual inspection,
and we mark these with X's.
}
\end{figure}

%%%%%%%%%%%%%%%%%%%%%%%%%%%

% Color-color Stars
\begin{figure}
\centerline{
\includegraphics[width = \columnwidth]{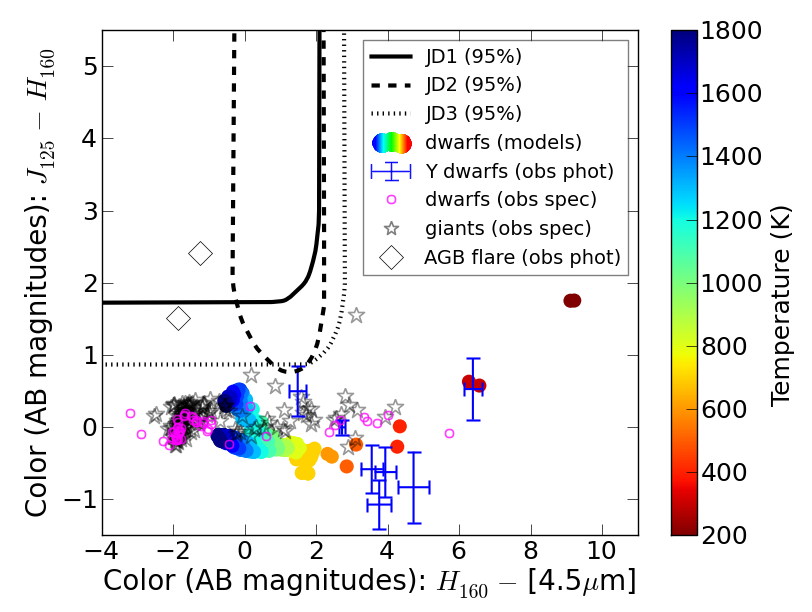}
}
%\captionsetup{labelsep=pipe}
\caption{
\label{fig:ccstars}
%{\bf Color-color.}
%Observed HST+Spitzer colors in F125W, F160W, and IRAC ch2 plotted as black lines (95\% confidence contours)
Observed colors in $J_{125}$, $H_{160}$, and at 4.5\um\ plotted as black lines (95\% confidence contours)
versus those observed and predicted for stars and brown dwarfs.
Colors derived from stellar spectra observed with IRTF \citep{Cushing05,Rayner09} are plotted as 
open magenta circles for dwarfs and
open black star symbols for giants and supergiants.
Blue error bars are observed photometry (ground-based and WISE) for Y dwarfs \citep{Kirkpatrick12}.
Open black diamonds are post-AGB flare ups with dust ejecta observed with 2MASS and WISE;
the upper diamond is ``Sakurai's object'' \citep{DuerbeckBenetti96}
and the lower diamond is WISE J1810-3305 \citep{Gandhi12}.
Simulated dwarf spectra from \cite{HubenyBurrows07} are plotted as filled circles colored as a function of temperature.
}
\end{figure}

%%%%%%%%%%%%%%%%%%%%%%%%%%%

\subsection{Stars or Brown Dwarfs?}
\label{sec:stars}

\JD1, JD2, and JD3 are most likely multiple images of a strongly lensed background galaxy, well behind the $z = 0.591$ cluster.
Their observed colors are extremely rare in our multiband HST catalogs of 17 clusters observed to date (Fig.~\ref{fig:catfrats}).
And they lie at or near the predicted positions of multiply-lensed images (\S\ref{sec:lens}).
It would be highly unlikely to find three foreground (unlensed) objects with such rare colors coincidentally at these positions.
Still we consider here possible interlopers within the Galaxy,
namely stars, brown dwarfs, and (in \S\ref{sec:solarsys}) solar system objects 
including KBOs (Kuiper Belt Objects) and Oort Cloud objects.

JD1 and JD2 are perhaps marginally resolved with deconvolved FWHM $\lesssim 0.2\arcsec$
($0.3\arcsec$ observed with a $0.2\arcsec$ PSF).
We performed two independent analyses attempting to determine whether the observed FWHM
was large enough to definitively distinguish it from the stellar locus.
These analyses reached different conclusions.
Therefore we turn to other lines of evidence to rule out stars and smaller objects.

Stars are relatively plentiful in this field as the Galactic latitude is relatively low ($+25.1\arcdeg$).
We used the online tool 
TRILEGAL\footnote{\href{http://stev.oapd.inaf.it/trilegal}{http://stev.oapd.inaf.it/trilegal}} \citep{Girardi05}
to calculate that we may expect \~5 late type M dwarfs of \~26th magnitude or fainter within our FOV.
However the predicted colors are $J_{125} - H_{160} \sim 0.4$, a break significantly weaker than that observed.

In Fig.~\ref{fig:ccstars},
observed and expected colors of stars and brown dwarfs (including types M, L, T, and Y)
are plotted versus those observed for the \Jdrops.
No dwarf color is able to reproduce the observed \Jdrop\ colors.
According to models, the colors of extremely cold (\~200K) Y dwarfs
%late type (M,L,T,Y) dwarfs 
come close to matching the red observed HST NIR colors,
but these are expected to be significantly brighter in IRAC by up to 10 magnitudes.
The coldest dwarfs yet discovered are Y dwarfs including WISEP J1828+2650 at \~300 K
\citep{Cushing11,Kirkpatrick12}
with colors as plotted in Fig.~\ref{fig:ccstars}
from ground-based $JH$ and WISE W2 4.6\um\ observations.
%with ground-based ($JH$) and WISE (W2 4.6\um) colors

%
%with F140W (HST) $-$ W2 (WISE 4.6\um) colors ranging from 
%\~3 to 7 AB magnitudes,
%or \~5 to 9 Vega
%\citep{Cushing11,Kirkpatrick12}.

Of the stellar spectra observed with IRTF \citep{Cushing05,Rayner09},
the M8III red giant WX Piscium (IRAS 01037+1219; \citealt{Ulrich66,Decin07}) comes closest
to matching the observed colors of \JD.
However such a large, bright star ($M \sim -4$)
would need to be well outside the Galaxy ($\sim 10$ Mpc distant) to be observed at 26th magnitude in F160W
(as argued in \citealt{Dickinson00} and \citealt{Bouwens11N} for two previous $z \sim 10$ candidates).
%
%The very red SED is "one of the most extreme infrared (IR) AGB objects known (Ulrich66)" (Decin07):
%IRAS 01037+1219
%WX Piscium
%
If \JD\ were within the Galaxy (out to \~10 kpc), it would have an absolute magnitude of $M \sim +11$ or fainter,
consistent with a red dwarf in terms of magnitude but not color as shown above.

A few red giants in the post-AGB phase have been observed to flare up
apparently as the result of a helium burning ``thermal pulse''
which triggers the ejection of a dust shell.
``Sakurai's object'' \citep{DuerbeckBenetti96}
and WISE J1810-3305 \citep{Gandhi12}
do have similar colors to our \Jdrops.
But again these are very bright events,
observed at $0.34 \pm 0.01$ Jy and $2.74 \pm 0.06$ Jy, respectively, in the $H$-band with 2MASS.
Believed to be a few kpc distant,
they would need to be removed to several Mpc
to be observed at \~0.1 \uJy\ as our \Jdrops.
These are also rare events, lasting on the order of 100 years (but varying more rapidly),
such that only these two have been reported to date.
It would be highly unlikely to detect three such events occurring at the same time in the same HST field.

\subsection{Solar System Objects?}
\label{sec:solarsys}

If the \Jdrops\ were solar system objects, we would expect to have detected their proper motions
in our 6 epochs of F160W/F140W imaging spanning 56 days (Figs.~\ref{fig:timeimages} and \ref{fig:timexy}).
Only an Oort cloud object at \~50,000 AU would be orbiting the Sun sufficiently slowly
for us not to have detected its motion.
But Oort cloud objects are expected to be significantly fainter 
%($\sim$35th magnitude assuming a diameter of \~20 km; \citealt{Sheppard10})
($\sim$58th magnitude assuming a diameter of \~20 km; \citealt{Sheppard10})
and have different colors than those observed here.
%and the colors are not as expected.
\cite{Benecchi11} measured HST/NICMOS F110W$-$F160W colors of 80 Trans-Neptunian Objects
and found they have HST F110W$-$F160W colors clustered around \~0.6
with none redder than 0.8.
For Oort cloud objects to be observed at \~26th magnitude 
due to reflected sunlight alone, even with 100\% albedo,
their radii would have to be \~\tentothe{10} m which is larger than the Sun.
%an Oort cloud object would have to be the size of a small moon.
%Even if such objects exist, they are almost certainly rare, or they would have been discovered by now.
%It would be highly improbable to discover the first three within a single HST FOV.
Oort cloud brown dwarfs emitting thermally would be far brighter than observed
and, as discussed in \S\ref{sec:stars}, have different colors.

\subsection{Lensed Supernova?}
\label{sec:SN}

The \Jdrops\ do not exhibit any significant temporal variations in brightness either over our 56 days of observations
(Fig.~\ref{fig:timeflux}),
ruling out most transient phenomena such as supernovae.
However, Type IIP supernovae can plateau to a roughly constant magnitude
for $\sim$100 days \citep[e.g.,][]{Arcavi12}
which we would observe to last $\sim$$100(1+z)$ days due to cosmic time dilation \citep{Blondin08}.
A Type Ia supernova at $z \sim 4$ would be observed to have magnitudes and colors
similar to those observed in HST for \JD.
A Type IIP plateau supernova would likely be bluer,
but could perhaps match the observed HST colors.
We would expect to detect it as a bright object at longer wavelengths,
but the Spitzer images were obtained 1.5 years earlier, perhaps before the star went supernova.

This intriguing scenario is ruled out by the gravitational lens time delays due to \macs,
which we estimate to be on the order of 1--10 years between JD1 and JD2
and $\sim$50 years between these and JD3.
Our subset of lens models (\S\ref{sec:lens}) which allow for $z \sim 4$
also suggest that the intrinsic fluxes of all three images are roughly consistent with one another
(at least to within a magnitude).
Thus the supernova plateau (several magnitudes brighter than the host galaxy)
would have to have lasted $50 / (1+z) \sim 10$ years
for us to observe it simultaneously in all three images.
Even if this were somehow possible, 
we would then expect to have detected the earlier (least time-delayed) images with Spitzer.

\subsection{Emission Line Galaxy?}
\label{sec:ELG}

In principle, an AGN / starburst galaxy with an undetected continuum and
two or more extremely strong nebular emission lines redshifted into F140W and F160W
could reproduce the observed HST colors.
The only plausible configuration is that 
\Hbeta\ (4861\AA) and [OIII] (4959\AA) are redshifted to within F140W and F160W, 
while [OIII] (5007\AA) is redshifted beyond F140W but within F160W.
This is possible for the narrow redshift range $2.20 < z < 2.22$.
At this redshift, F140W and F160W have rest-frame widths of \~1229\AA\ and \~889\AA, respectively.
In the case of JD1, we measure the flux blueward of F140W to be $-3.4 \pm 3.7$ nJy.
We conservatively adopt $< 7.4$ nJy as the 2\sig\ upper limit on the continuum flux.
Boosting the F140W flux to the observed \~63 nJy 
would require emission lines with a combined equivalent width
EW $> 1229 \times (63 / 7.4 - 1) \sim 9525$\AA.
Similarly, increasing the F160W flux to the observed \~162 nJy,
would require a combined
EW $> 889 \times (162 / 7.4 - 1) \sim 19114$\AA\ (\~\tentotheminus{15} erg/s/cm$^2$).
Thus in our configuration assuming a continuum flux of 7.4 nJy:

\vspace{-0.12in}

% Khoi/EW.py
\begin{align}
& \rm EW (H\beta + [OIII]_{4959}) \approx & \rm 9234\AA, \label{eq:140}\\
& \rm EW (H\beta + [OIII]_{4959} + [OIII]_{5007}) \approx & \rm 18573\AA, \label{eq:160}\\
& \rm EW ([OIII]_{5007}) \hspace{0.05in} \approx (\ref{eq:160}) - (\ref{eq:140}) \approx & \rm 9339\AA, \label{eq:5007}\\
& \rm EW ([OIII]_{4959}) \hspace{0.05in} \approx (\ref{eq:5007}) ~ / ~ 3 \approx & \rm 3113\AA, \label{eq:OIII}\\
& \rm EW (H\beta)           \hspace{0.05in} \hspace{0.38in} = (\ref{eq:140}) - (\ref{eq:OIII}) \approx & \rm 6121\AA, \label{eq:Hb}
\end{align}
where the line ratio in Equation \ref{eq:OIII} is dictated by the relative transition probabilities.
% e.g., Osterbrock book: Astrophysics of Gaseous Nebulae and Active Galactic Nuclei
% search for OIII on Google online version
% p. 309 Table 12.6 -- predicted ratios are almost exactly 3
% also http://www.astr.ua.edu/keel/galaxies/emission.html

An [OIII] (5007\AA) line with EW $>$ 9000\AA\ would be several times greater
than the strongest emission lines observed to date,
approaching EW \~ 2000\AA, for [OIII] (5007\AA) and \Ha\ (6563\AA)
\citep{Atek11, vanderWel11, Shim11, Fumagalli12}.
If we consider instead the 5\sig\ continuum limit of 18.4 nJy,
we would still require EW \~ 3959\AA\ for [OIII] (5007\AA) and EW \~ 1659\AA\ for \Hbeta.
The strongest \Hbeta\ lines are robustly predicted to have 
EW $\lesssim 800$\AA\ even for extremely young stellar populations,
according to models like 
Starburst99\footnote{\href{http://www.stsci.edu/science/starburst99/}{http://www.stsci.edu/science/starburst99/}}
\citep{Leitherer99,Leitherer10}.
Finally, given such bright lines, one would also expect 
a significant contribution of [OII] (3727\AA) in F110W and F125W, which is not observed.

%\vspace{0.1in} %%%%%%%%%% XXXXX

\section{Gravitational Lens Modeling}
\label{sec:lens}

We identified 24 strongly-lensed images of 9 background galaxies (\S\ref{sec:SLimages}),
used them to derive lens models using three different methods (\S\ref{sec:SLmethods}),
and derived results including magnifications in (\S\ref{sec:lensresults}).
Importantly, our lens models show that \JD1, 2, and 3 are observed in relative positions 
as expected if they are strongly-lensed multiple images of the same galaxy at $z \sim 11$.

\subsection{Strongly Lensed Multiple Images}
\label{sec:SLimages}

\cite{Zitrin11MACS} presented a preliminary gravitational lens mass model of the \macs\ cluster core
based on pre-CLASH HST/ACS F555W+F814W imaging
and their identifications of two background galaxies strongly lensed to produce multiple images.
Based on CLASH imaging in 15 additional HST filters
and additional lens modeling using the \cite{Zitrin09a} method,
we have now identified 7 more galaxies which have been multiply imaged,
and we have measured robust photometric redshifts for all 9 galaxies.
This enables us to model the mass distribution (primarily dark matter) and thus lensing properties in greater detail.

In addition to the three images of \JD\ at $z \sim 11$, 
we observe 21 multiple images of 8 background galaxies
with photometric redshifts ranging from $2 \lesssim z \lesssim 6.5$ 
(Table \ref{tab:arcs} and Fig.~\ref{fig:HSTcluster}).
The candidate $z \sim 6.5$ system is notable in its own right,
consisting of two images observed at magnitudes $\sim 26.3$ and 27.3 in the NIR.

For each of systems 3, 5, and 8, our lens models predict a third faint counterimage,
but we cannot unambiguously identify it among several possible candidates.
%To be conservative, we exclude these uncertain identifications from our lens model.
To be conservative, we do not include these uncertain identifications as constraints on our lens models.
Inclusion of these candidates does not significantly affect the lens models.
In Fig.~\ref{fig:HSTcluster} we indicate the predicted locations of these counterimages.

\subsection{Strong Lens Modeling Methods}
\label{sec:SLmethods}

Based on the observed positions of all 24 strongly lensed images, 
we model the mass of \macs\ using three different methods:
the \cite{Zitrin09a} method, Lenstool \citep{Kneib93,Jullo07}, and LensPerfect \citep{Coe08,Coe10}.
The first two methods are ``parametric'' in that they assume light traces mass,
which has proven to be a very good prior.
For example, some of the earliest efforts to model cluster lenses
found that assigning masses to individual luminous cluster galaxies
significantly improved the reproduction of strongly lensed images \citep{Kassiola92,Kneib96}.
LensPerfect makes no assumptions about light tracing mass, 
exploring a broader range of mass models
and perfectly reproducing the observed positions of all multiple images positions as input.

The \cite{Zitrin09a} mass model parameterization
consists of three components:
the cluster galaxies, a smooth cluster halo, and an external shear.
Cluster galaxies were identified according to the ``red sequence''
in F814W-F555W color-magnitude space,
then verified with photometric redshifts.
Each cluster galaxy was modeled as a power law density profile,
its mass scaling with flux observed in F814W.
In this work we also allowed the masses of the two brightest central galaxies to vary independently.
The cluster halo component was derived from this galaxy component
by smoothing the latter with either a polynomial spline or a Gaussian.
The two components were allowed to scale independently before being added.
In all, there were eight free parameters:~the 
mass scalings of the galaxy and halo components,
the masses of the two brightest central galaxies,
the power law of the galaxy density profiles,
the degree of the polynomial spline or Gaussian smoothing width,
and the amplitude and direction of the external shear.

The Lenstool model consisted of an ellipsoidal NFW halo \citep{NFW96}
plus cluster galaxies modeled as
truncated PIEMDs (pseudo-isothermal elliptical mass distributions; \citealt{KassiolaKovner93}).
We assumed core radii $r_{core} = 300$ pc and luminosity scaling relations as in 
\cite{Jullo07}:~velocity dispersion $\sigma_0 \propto L^{1/4}$
and cutoff radius $r_{cut} \propto L^{1/2}$,
resulting in all galaxies having equal mass-to-light ratios. %$M/L$).
The normalizations of these two scaling relations were free parameters
along with the cluster halo position ($x$, $y$), ellipticity ($e$, $\theta$), scale radius, and concentration.
There were eight free parameters in all.

%\vspace{0.2in}
%\clearpage

%%%%%%%%%%%%%
% ARCS
\input{arctable.tex}
%%%%%%%%%%%%%

\subsection{Lens Model Results}
\label{sec:lensresults}

Given the observed position of any one of the \JD\ images,
%By delensing and relensing any one of the \Jdrops,
the Lenstool model accurately predicts and reproduces the positions of the other two images
to an RMS of 1.3\arcsec, as minimized for $z = 11.59 ^{+0.12}_{-1.53}$.
%to an RMS of 1.1\arcsec, as minimized for $z = 11.84 \pm 0.55$.
This scatter is consistent with the \~1.4\arcsec\ expected due to
lensing by line of sight structures and variation in the mass-to-light ratio of cluster galaxies \citep{Jullo10,Host12}.
%$z \sim 12.0^{+0.2}_{-2.9}$.

The lens model and inferred redshift for \JD\ do not change significantly
if the \JD\ images are excluded as constraints.
In this case, the 8-parameter lens model remains well constrained by the 21 other multiple images
which provide 26 constraints (see discussion below).

Using the \cite{Zitrin09a} method,
two sets of acceptable models are found in different regions of the model parameter space.
One set prefers $z \sim 11$ for \JD, while the other prefers $z \sim 3.5$.
The latter mass models have flatter profiles.

Our LensPerfect analysis confirms this is a degeneracy between the \JD\ redshift
and the cluster mass distribution.
A wide range of redshifts including $z = 3.5$, 11.0, and 11.6 is permitted given the strong lensing data.
When fixing the redshift to any of these values,
LensPerfect produces reasonable lens models
(physical and with light approximately tracing mass)
which perfectly reproduce all 24 observed positions of the 9 strongly lensed galaxies.
When the \JD\ redshift is set lower, the cluster mass distribution is more spread out yielding a flatter profile.
We confirm that the parametric models have similar differences,
in part due to their parameterizations of the cluster mass distribution.

Including the redshift of \JD, 
both parametric models have $\lesssim 9$ free parameters.
% when we include the redshift of \JD.
(This number should be considered a maximum given covariances among the parameters.
See discussion in \citealt{Andrae10}.)
There are 30 constraints
%= 2 coordinates ($x$,$y$) $\times$ (20 images $-$ 8 galaxies).
$= 2 \times (24 - 9)$,
where the constraints are the two coordinates ($x$,$y$) from each of the 24 multiple images
minus the 9 unknown source positions.
Thus each model has $\gtrsim 21$ degrees of freedom (30 constraints $-$ 9 parameters).
%Thus each model has $24 - 8 = 16$ degrees of freedom.
The Lenstool model reproduces all lensed image positions to an RMS of 1.17\arcsec.
Assuming a scatter of 1.4\arcsec\ as explained above,
this yields a $\chi^2 \approx 24 \times (1.17'' / 1.4'')^2 = 16.8$
with 21 degrees of freedom, for a reduced $\chi^2_\nu \lesssim 16.8 / 21 \approx 0.8$.

The Zitrin spline model with the flat mass profile preferring $z \sim 3.5$ for \JD\ obtains 
an RMS of 1.1\arcsec\ for a $\chi^2 \approx 15$,
also with 21 degrees of freedom, yielding $\chi^2_\nu \lesssim 0.7$.
When the models are forced to adopt $z \sim 11$,
the best fit is found with a Gaussian-smoothed model,
yielding an RMS of 2.9\arcsec,
for a $\chi^2 \approx 103$
and a reduced $\chi^2_\nu \lesssim 4.9$.

Assuming \JD\ is at $z = 11.0$, the Lenstool model (Fig.~\ref{fig:HSTcluster}) 
estimates magnifications of $\sim$ 8.4, 6.6, and 2.8 for JD1, JD2, and JD3, respectively,
with uncertainties of $\sim 20\%$.
These magnifications are consistent with a F160W $= 20 \pm 4$ nJy ($28.2 \pm 0.2$ mag)
%(28.15 mag) 
source magnified by factors of $\sim$ 8.1, 6.8, and 2.1 
to match the observed F160W fluxes within their $\sim 10\%$ uncertainties (Table~\ref{tab:cat}).

LensPerfect models perfectly reproduce all 24 observed image positions as input.
These data constrain the mass distribution and profile well globally
but only to a resolution of \~20'', or \~130 kpc,
roughly the average separation between the strongly lensed images.
This resolution is insufficient to obtain robust estimates of the image magnifications
which are strong functions of local mass gradients.
Here these magnifications are better estimated by adopting priors of light tracing mass as in the parametric methods.

%%%%%%%%%%%%%%%%%%%%%%%%%%%

% SED constraints parameters
\begin{figure*}
\centerline{
\includegraphics[width = 0.9\textwidth]{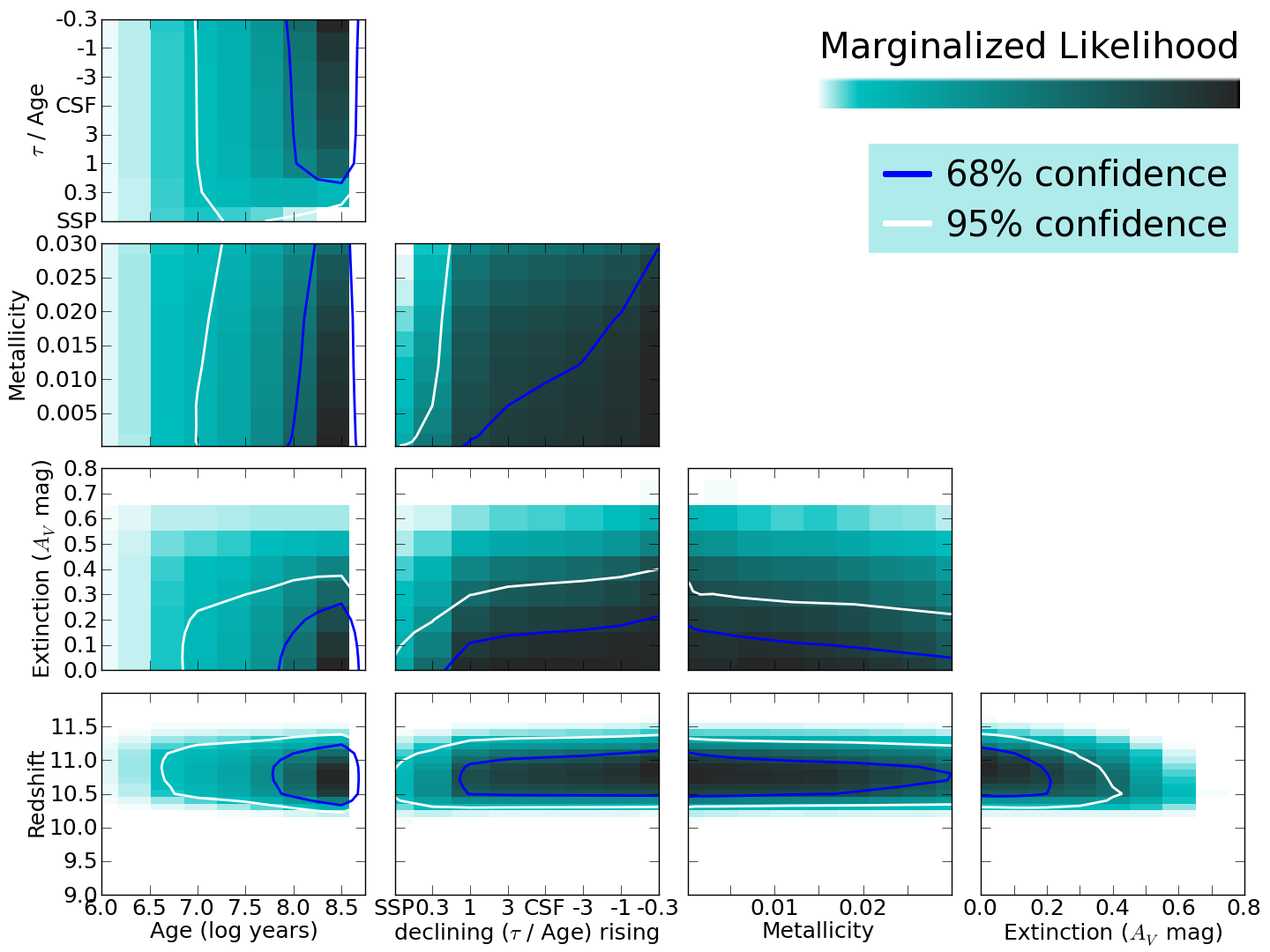}
}
%\captionsetup{labelsep=pipe}
\caption{
\label{fig:params}
Constraints on redshift, age, star formation history (exponential scale factor $\tau$),
metallicity (where $Z_\odot \approx 0.019$), and dust extinction (in $V$-band magnitudes).
%
%\JD\ likely has recent star formation ($< 100$ Myr) and low dust extinction ($A_V < 0.3$ mag).
%\JD\ is likely a fairly young galaxy with low metallicity and low dust extinction.
%The light from \JD\ is likely dominated by recent star formation with low metallicity and low dust extinction.
This is based on fitting the HST+Spitzer photometry 
integrated over the three images
to the flexible stellar population synthesis models of \cite{Conroy09} and \cite{ConroyGunn10}
convolved with an exponential star formation history $\propto \exp{(-t / \tau)}$
and with \cite{Calzetti01} dust extinction added ($R_V = 4.05$).
The star formation history was 
either decaying ($\tau / {\rm age} = 0.3, 1, 3$),
rising ($\tau / {\rm age} = -0.3, -1, -3$),
%staying constant ($\tau = \infty$),
constant star formation rate (CSF; $\tau = \infty$),
or occurring in a single burst at ``birth'' (simple stellar population, SSP; $\tau = 0$),
with equal likelihoods for all eight possibilities.
We assumed a flat linear prior for age (though it's plotted as log here)
up to the age of the universe (as a function of redshift).
Within each panel, the marginalized likelihood is plotted as a color map (scaled linearly)
and confidence contours of 68\% and 95\% are overplotted as blue and white lines, respectively.
%1- and 2\sig\ 
%Deeper Spitzer imaging would significantly improve these constraints.
}
\end{figure*}

%%%%%%%%%%%%%%%%%%%%%%%%%%%

\section{Physical Properties of \JD}
\label{sec:props}

We estimate the rest-frame UV luminosity and star formation rate (\S\ref{sec:UV}),
infer a rough stellar mass (\S\ref{sec:mass}),
and place upper limits on the physical size (\S\ref{sec:sizes}).
In \S\ref{sec:sedprops} we explore SED parameter degeneracies and place modest constraints on other properties of \JD.

%Observed at 26th magnitude, JD1 and JD2 are $\approx 15$ times brighter than the 29th magnitude $z \approx 10$ candidate
%UDFj-39546284 \citep{Bouwens11N}.

\subsection{Rest-frame UV Luminosity; Star Formation Rate}
\label{sec:UV}

As described above, we estimate the intrinsic (unlensed) magnitude of \JD\ to be $28.2 \pm 0.2$ in F160W.
At $z \sim 11$, the \Lya\ break falls within the F160W bandpass, attenuating the observed flux;
% the F160W flux is attenuated by the \Lya\ break,
the rest-frame UV (0.16\um) continuum flux is \~0.25 mag brighter.
To convert this flux to rest-frame UV absolute magnitude $M_{UV,AB}$, we add three terms.
Most significantly, the magnitude is brighter by the distance modulus \~50.3.
The flux per unit frequency is also dimmer by a factor of $1+z$ (\~2.7 mag)
simply because the rest frame samples a higher frequency.
We also derive a small color term of \~0.1 mag
as we switch from the blueshifted F160W filter (\~0.13\um)
to a tophat filter centered on 0.16\um\ for comparison with previous measurements.
Combining these terms, we find $M_{UV} \sim -19.5$.
Converting this to UV luminosity at a distance of 10 pc, we find
$L_{UV} \sim 2.8 \times 10^{28}$ erg s\per\ Hz\per.

This rest-frame UV luminosity can be generated
by a star formation rate (SFR) of \~4$M_\odot$ yr$^{-1}$
assuming a \cite{Salpeter55} IMF with mass limits 0.1--100 \Msun\ \citep{Kennicutt98a}.
The ionizing efficiency could be increased by a factor of 
\~1.8 for a \cite{Chabrier03} IMF
or \~3 for a top-heavy IMF \citep{BC03, Schaerer03, Stiavelli04},
with the latter generally realized in simulations of high-redshift galaxies \citep[e.g.,][]{Abel02,Bromm02}.
Stellar rotations may also increase this efficiency by a factor of \~2--5 \citep{Levesque12}.
%and perhaps suggested (among other possibilities) by their observed blue rest-frame UV slopes
%\citep[e.g.,][]{Stanway05,Bouwens12b}.
%
%and/or an additional factor of \~3 for zero metallicity.
%
Given these and other uncertainties, 
a star formation rate of \~1 $M_\odot$ yr$^{-1}$ or lower could generate the $L_{UV}$ derived for \JD.
%Given these higher efficiencies, a lower star formation rate of \~1--2$M_\odot$ yr$^{-1}$ could generate the $L_{UV}$ derived for \JD.
%Alternatively, assuming a \cite{Chabrier03} IMF, 
%a lower star formation rate of \~2$M_\odot$ yr$^{-1}$ could generate the same $L_{UV}$.
%or $\sim$1/3 lower assuming a \cite{Kroupa01} IMF.
%top-heavy (shallower) IMF

This luminosity \LUV\ is \~$L^*$ or perhaps a few times brighter than $L^*$,
depending on which extrapolation we assume to estimate this characteristic luminosity at $z \sim 11$
\citep{Bouwens08,Robertson10, Bouwens11N, Bradley12b}.
Based on the estimated luminosity function (\S\ref{sec:SFRD}) and our lens magnification model,
we find that a $z \sim 11$ galaxy lensed to 26th magnitude
does in fact have an $\sim$80\% likelihood of being intrinsically brighter than $L^*$.
% ~/CLASH/papers/Khoi/counts/countsplots.py -> sourcemag3.png
%is in fact likely to be observed at $\sim 3L^*$
%and has roughly an $80\%$ probability of being intrinsically brighter than $L^*$.
%is most likely sampling the bright end of the luminosity function.
%It is not surprising for this galaxy to be toward the bright end of the luminosity function.
% ~/CLASH/data/m0647/highz/SFR.py

\subsection{Stellar Mass}
\label{sec:mass}

Meaningful observational constraints on the stellar mass of \JD\ would require 
rest-frame optical photometry redward of 0.4\um\ (beyond the Balmer and 4000\AA\ breaks), or 4.8\um\ observed.
However, we may infer a stellar mass estimate as follows.

%Galaxies at $2 \lesssim z \lesssim 7$ all, remarkably, have average 
Specific star formation rates (sSFR) 
of 2--3 Gyr$^{-1}$ 
(that is, 2--3 $M_\odot$ formed per year per $10^9 M_\odot$ total stellar mass)
are observed on average for galaxies
over a remarkably broad range of redshifts 
($2 \lesssim z \lesssim 7$;
see e.g., \citealt{Stark09,Gonzalez10,McLure11,Bouwens12b}).
If this ``plateau'' continues out to $z \sim 11$
and \JD\ has a typical sSFR of \~2 or 3 Gyr\per,
then this combined with our derived SFR 
would imply a stellar mass on the order of \~$10^{9} M_\odot$.
The average stellar mass of \Lstar\ galaxies
was \~$10^9 M_\odot$ at $z \sim 7$--8
and rose to a few times $10^{10} M_\odot$ by $z \sim 2$
\citep{Gonzalez10,Labbe10,Finkelstein10}.
Based on this trend, we may expect the average stellar mass of \Lstar\ galaxies at $z \sim 11$
to be less than $10^9 M_\odot$.
If this holds true for \JD, it would suggest a higher sSFR, more in line with expectations from simulations
which are in some tension (but perhaps only mild tension) with the observed sSFR plateau
\citep[e.g.,][]{KhochfarSilk11, Dave11a, Weinmann11,Behroozi12}.

We conclude the stellar mass of \JD\ is most likely on the order of $10^8$--$10^9 M_\odot$.
The lower end of this mass range is more compatible with 
expectations from cosmological simulations and galaxy formation models.
Based on simulations \citep[e.g.,][]{Klypin11},
we may expect to find a dark matter halo of 
virial mass \~\tentothe{10}\Msun\ or so within our search volume of a few times 1000 Mpc$^3$.
This would comfortably host a galaxy of stellar mass \tentothe8\Msun\ or so.
A stellar mass of \tentothe9\Msun\ would be larger than expected.

\subsection{Other SED-based Constraints}
\label{sec:sedprops}

%Frame it more as exploring additional SEDs, plus degeneracies in (z, dust), etc.

In \S\ref{sec:lozgal} we explored a broad range of galaxy properties
to rule out lower redshift interlopers with a high degree of confidence.
In this section we quantify the degeneracies in those parameters
\citep[see also][]{Pirzkal12,Pforr12}
along with the modest constraints we obtain on them,
assuming that \JD\ is indeed at high redshift.
%given that the high redshift solution is most likely.

%To further constrain the properties of \JD, we modeled the HST+Spitzer SEDs of all three images
We modeled the observed photometry  % of all three images
using the flexible stellar population synthesis (FSPS) models from \cite{Conroy09} and \cite{ConroyGunn10}.
As described in \S\ref{sec:lozgal},
we convolved their simple stellar population (SSP) models with a range of star formation histories,
including a single early burst, exponentially declining, constant, and exponentially rising.
Though little dust is expected at these redshifts \citep[e.g.,][]{Bouwens12b},
we tested this assumption explicitly by adding a variable degree of
\cite{Calzetti01} dust extinction.
We did not add nebular emission lines.
Only [OII] (3727\AA) redshifted to $\sim$4.5\um\ might be a significant concern,
though we only modestly detect 4.5\um\ flux in JD2.
We assume a flat linear prior in age up to the age of the universe at each redshift.

Our constraints on redshift, age, SFH ($\tau$), metallicity, and extinction
are shown in Fig.~\ref{fig:params} 
based on the integrated photometry of all three images.
%for each lensed image individually as well as jointly.
We confirm that significant dust extinction is unlikely,
with a rest-frame $V$-band extinction of 
$A_V < 0.25$ mag (95\% confidence).
%$A_V < 0.26$ mag (68\% confidence).
Constraints on the other parameters are modest,
with slight preferences for low metallicity and rising or continuous SFH in a maximally old galaxy
(limited to $\lesssim$400 Myr by the age of the universe).
% > 200 Myr (68%) vs. 50% from flat prior
% > 100 Myr (95%) vs. 75% from flat prior
% ~/CLASH/data/m0647/highz/fsps/constraints.py
% ~/CLASH/data/m0647/highz/fsps/multitausu.py

%%%%%%%%%%%%%%%%%%%%%%%%%%%

% Sizes
\begin{figure}
\centerline{
\includegraphics[width = \columnwidth]{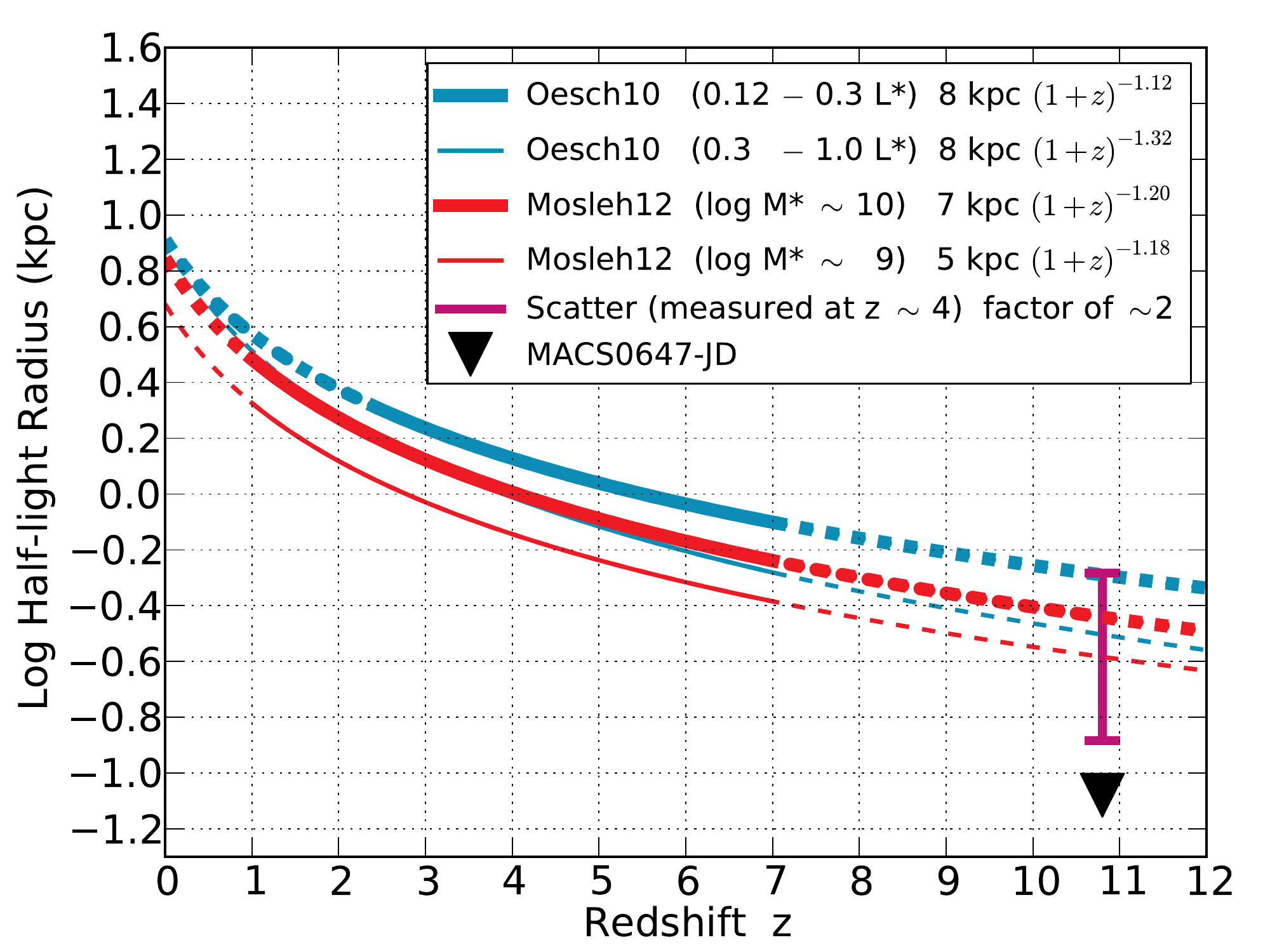}
}
%\captionsetup{labelsep=pipe}
\caption{
\label{fig:sizes}
Upper limit on \JD\ half-light radius (deconvolved and delensed) compared to observed mean galaxy sizes 
from \cite{Oesch10b} and \cite{Mosleh12}
extrapolated to $z \sim 11$.
The intrinsic scatter in galaxy sizes is a factor of \~ 2 as measured for well-studied samples at $z \sim 4$ \citep{Ferguson04}.
}
\end{figure}

%%%%%%%%%%%%%%%%%%%%%%%%%%%

%\vspace{-0.07in}  % XXX
%\vspace{0.2in}  % XXX
%\clearpage

\subsection{Physical Size}
\label{sec:sizes}

After correcting for the observed PSF, 
JD1 and JD2 have observed half-light radii $\lesssim 0.1''$, or delensed $\lesssim 0.03\arcsec$ ($\lesssim 0.1$ kpc).
Based on extrapolations from lower redshifts \citep{Oesch10b,Mosleh12},
we expect an average half-light radius of roughly $\langle r_{1/2} \rangle \sim 0.26$ kpc
for a galaxy with a stellar mass \~\tentothe9\Msun.
\JD\ is likely somewhat less massive (\S\ref{sec:mass}).
Scatter in galaxy sizes is large: \~0.3 dex, or a factor of \~2,
as found for well-studied samples at $3 \lesssim z \lesssim 5$ \citep{Ferguson04}.
So our derived $r_{1/2} \lesssim 0.1$ kpc is on the small side, though not beyond expectations (see Fig.~\ref{fig:sizes}).
Furthermore, we may only be detecting a bright star forming knot in a larger galaxy.
These knots typically have sizes of $\sim 0.1$ kpc
as observed in high-redshift ($5 < z < 8$) lensed galaxies \citep{Franx97,Bradley08,Bradley12a}.
%The predicted sizes of $z \sim 11$ galaxies from simulations are XXX.

%Assuming each has a mass between 1--10 billion times solar or so\cite{Pawlik11},

%%%%%%%%%%%%%%%%%%%%%%%%%%%

% High-z counts
\begin{figure}
\centerline{
\includegraphics[width = \columnwidth]{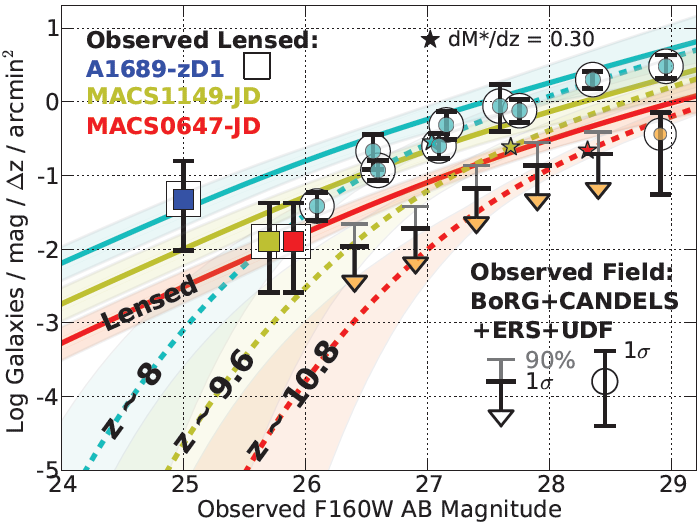}
}
%\captionsetup{labelsep=pipe}
\caption{
\label{fig:counts}
Observed number counts for both lensed (squares) and field (circles / arrows) galaxies at $z \approx 8$, 9.6, and 10.8
as a function of observed F160W AB magnitude.
CLASH discoveries of candidates 
at $z \approx 10.8$ (\JD; this work)
and $z \approx 9.6$ (MACS1149-JD; \citealt{Zheng12})
in $\sim$78 square arcminutes
(17 cluster lensing fields)
are plotted as the red and yellow squares, respectively,
in units of counts per unit magnitude, redshift, and square arcminute.
The blue square indicates the single robust $z \approx 7.5$ candidate (A1689-zD1; \citealt{Bradley08})
identified in 11 non-CLASH cluster fields covering 21 square arcminutes \citep{Bouwens09}.
Cyan and orange circles represent counts in unlensed fields
at $z \sim 8$ and 10.3, respectively
from the BoRG survey \citep{Bradley12b}
as well as CANDELS, ERS, and HUDF09 \citep{Oesch12a}
including UDFj-39546284 \citep{Bouwens11N}.
%are counts in unlensed fields: the HUDF09, ERS, and BoRG survey \citep{Bouwens11N,Bouwens11b,Bradley12b}.
%The black arrows correspond to $z \sim 10$ and
The arrows give both 1\sig\ and 90\% confidence upper limits from \cite{Oesch12a}.
The bottom dashed curves are expected counts with uncertainties from
the \cite{Bradley12b} $z \sim 8$ luminosity function
and extrapolated to $z \sim 9.6$ and 10.8
assuming an evolution of $dM^*/dz = 0.30$,
similar to that found in e.g., \cite{Bouwens08,Bouwens12a}
(also see Fig.~\ref{fig:SFRD}).
This yields F160W $m^* \sim 27.0$, 27.7, and 28.3 at these redshifts,
plotted as a small star along each curve.
We then simulate the lensing of these expected counts 
in our CLASH WFC3/IR survey area
using mass models of the 17 clusters
and excluding the area (\~17\%) covered by foreground objects.
These are given as the upper solid curves
which are consistent with the CLASH discoveries (red and yellow squares).
%
%Note the significant advantage lensing provides 
%toward the discovery of bright high-redshift galaxies.
%at bright magnitudes.
%Even if we were to survey the entire 40,000 square degree sky of unlensed fields,
%we may only have a 0.1\% chance 
%our odds would only be about 1 in 1,000
%of finding another galaxy as bright as MACS0647-JD at $z \approx 11$.
}
\end{figure}

%%%%%%%%%%%%%%%%%%%%%%%%%%%%%

\begin{figure*}
%\centerline{\includegraphics[width = 1.05\columnwidth]{figs/SFRD.png}}
\centerline{
\includegraphics[width = 0.7\textwidth]{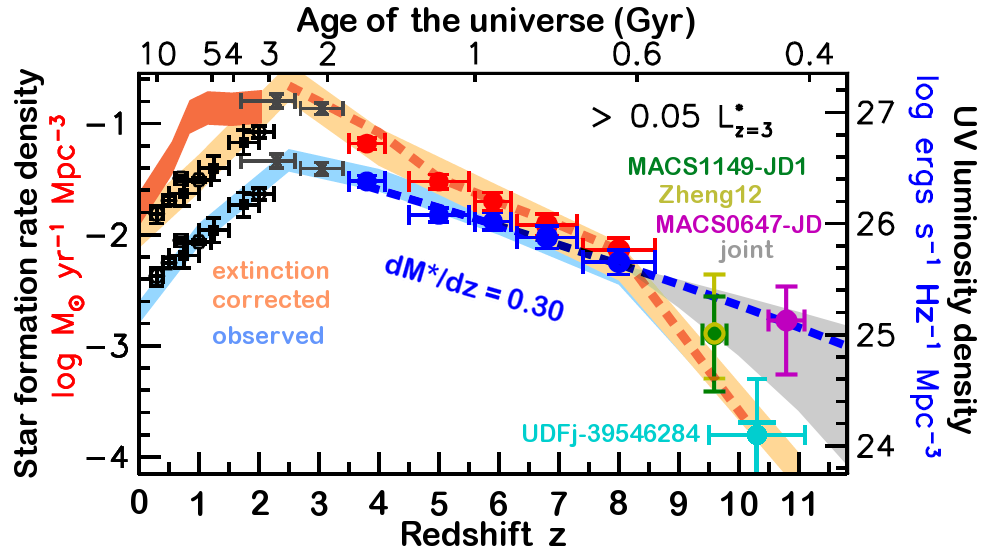}
}
%\captionsetup{labelsep=pipe}
\caption{
\label{fig:SFRD}
CLASH constraints on the 
%$z \gtrsim 10$ 
cosmic star formation rate density 
at $z \sim 9.6$ and 10.8
compared to previous estimates from $0 \lesssim z \lesssim 10$
as compiled and corrected for dust extinction by \cite{Bouwens12b}.
We adopt the $z \sim 8$ LF from \cite{Bradley12b}
and assume \Mstar\ evolves with linearly redshift.
\JD\ constrains 
%$dM^* / dz ~ (z \sim 11) = 0.24^{+0.37}_{-0.32}$,
%$dM^* / dz = 0.24^{+0.37}_{-0.32}$,
$dM^* / dz = 0.30^{+0.25}_{-0.18}$,
while MACS1149-JD1 \citep{Zheng12} constrains
%$dM^* / dz ~ (z \sim 9.6) = 0.62^{+0.69}_{-0.59}$.
%$dM^* / dz = 0.62^{+0.69}_{-0.59}$,
$dM^* / dz = 0.62^{+0.43}_{-0.33}$,
yielding a joint constraint of
%$dM^* / dz = 0.35\pm 0.30$.
$dM^* / dz = 0.40^{+0.23}_{-0.17}$.
We integrate these LFs down to $0.05 L^* (z=3)$, 
or $M_{UV,1400} \approx -17.7$ AB mag,
to obtain rest-frame UV luminosity densities 
(magenta and green points with the gray shaded region giving the joint constraint).
The yellow point is an alternative estimate of the $z \sim 9.6$ SFRD from \cite{Zheng12}  % indepdendent
rescaled by $12/17$ to account for the larger search volume behind 17 clusters.
An extrapolation from $z \sim 8$ assuming $dM^* / dz = 0.30$ is shown as the dashed blue line.
Points along the bottom blue curve are observed
%assume no dust extinction, as we assume here for $z > 9$,
while the upper orange curve is corrected for extinction by \cite{Bouwens12b}.
We assume a star formation rate of one solar mass per year (left axis)
produces a UV (0.14\um) luminosity of $8 \times 10^{27}$ ergs s\per\ Hz\per\ (right axis),
as from a \cite{Salpeter55} IMF truncated between 0.1 and 125 $M_\odot$.
A \cite{Chabrier03} IMF would yield lower SFRD by a factor of $\sim$1.8.
%$\sim$55\% of the values shown here.
%
Previous data are from
\cite{Schiminovich05, Oesch10c, Reddy09, Bouwens07, Bouwens11N, Oesch12a}
as described in \citet[see their Fig.~19]{Bouwens12b}.
}
\end{figure*}

%%%%%%%%%%%%%%%%%%%%%%%%%%%%%

%\section{Expected $z \sim 11$ Number Counts and Star Formation Rate Density}

\section{Number Count and Star Formation Rate Densities at $z \gtrsim 10$}
\label{sec:SFRD}

%As we demonstrate below,
Our discovery of a 26\th\ magnitude $z \approx 11$ candidate
in 17 cluster lensing fields ($\sim$78 square arcminutes)
agrees with rough expectations
given observed luminosity functions (LFs) at lower redshifts
extrapolated to higher redshift and propagated through our lens models.

The LF at $z \sim 8$ has recently been robustly constrained 
at both faint \citep{Bouwens11b} and bright \citep{Bradley12b} magnitudes (see also \citealt{Oesch12b}).
Based on the combined HST data from the HUDF09, ERS, CANDELS, and BoRG pure parallel fields,
\cite{Bradley12b} find the $z \sim 8$ LF follows a \cite{Schechter76} function with
a normalization $\phi^* = 4.3^{+3.5}_{-2.1} \times 10^{-4}$ Mpc$^{-3}$,
characteristic rest-frame UV absolute magnitude $M^*_{UV} = -20.26^{+0.29}_{-0.34}$,
and faint end slope $\alpha = -1.98^{+0.23}_{-0.22}$.
These data and LF with uncertainties are plotted in blue in Fig.~\ref{fig:counts}.

We then assumed an evolving LF 
in which \Mstar\ varies linearly with redshift
while \phistar\ and $\alpha$ are fixed.
Previous work \citep[e.g.,][]{Bouwens08,Bouwens12b} has shown that $dM^* / dz \sim 0.3$
yields good agreement to data at $4 \lesssim z \lesssim 8$ (see also Fig.~\ref{fig:SFRD}).
So we first assumed that this holds out to $z \sim 11$.

We convolved this evolving LF through our lens models for the 17 CLASH clusters studied in this work,
accounting for both the brightening of sources and the reduction 
in search area due to the magnifications.
Some of these models have been published 
(\citealt{Zitrin11A383, Zitrin12MACS1206, Zitrin12MACS0329, Coe12, Zheng12}; and this work)
% Zitrin11MACS
and the rest will be detailed in upcoming work.
We applied masks to the lensed regions, restricting our search area to the WFC3/IR observations
and excluding regions covered by foreground objects 
($\sim$17\% of the total area)
according to our SExtractor segmentation maps.
Our total search area for 17 clusters is $\sim$78 square arcminutes (as observed and lensed).

The resulting expected lensed number counts for $z \sim 8$, 9.6, and 10.8
are plotted in Fig.~\ref{fig:counts}.
These are consistent with CLASH observed \~26th magnitude number counts at $z \sim 10.8$ (this work)
as well as $z \sim 9.6$ (MACS1149-JD1; \citealt{Zheng12}).

However, the observed $z \sim 10$ number counts in the field are a factor of $\sim$4 lower than expected
based on an evolving LF such as this,
suggesting a sharp drop off in star formation density at these redshifts \citep{Bouwens11N, Oesch12a}.
%
%In contrast, \cite{RobertsonEllis12} suggest such a dropoff would be in tension with GRB rates at $z < 4$
%correlated with SFRD and extrapolated to higher redshifts.
To test for such a dropoff, 
we allowed for more (or less) rapid evolution in \Mstar\ at $z > 8$,
still as extrapolated from the \cite{Bradley12b} LF at $z \sim 8$.
%prior to $z \sim 8$.
We found that \JD\ constrains 
%$dM^* / dz ~ (z \sim 11) = 0.24^{+0.37}_{-0.32}$,
%$dM^* / dz = 0.24^{+0.37}_{-0.32}$,
$dM^* / dz = 0.30^{+0.25}_{-0.18}$,
and MACS1149-JD1 constrains
%$dM^* / dz ~ (z \sim 9.6) = 0.62^{+0.69}_{-0.59}$.
%$dM^* / dz = 0.62^{+0.69}_{-0.59}$,
$dM^* / dz = 0.62^{+0.43}_{-0.33}$,
yielding a joint constraint of
%$dM^* / dz = 0.35\pm 0.30$.
$dM^* / dz = 0.40^{+0.23}_{-0.17}$.

We then integrated these LFs (with uncertainties) down to 
%$M_{UV,1400} \approx -17.7$ ($0.05 L^*_{z=3}$)
%$M_{UV} = -17.7$ ($0.05 L^*_{z=3}$)
$0.05 L^*_{z=3}$ ($M_{UV} = -17.7$)
to obtain star formation rate densities
which can be compared directly to previously published estimates (Fig.~\ref{fig:SFRD}).
We found
SFRD $= (1.9^{+5.4}_{-1.6}) \times 10^{-3} M_\odot ~ {\rm yr}^{-1} ~ {\rm Mpc}^{-3}$ at $z \sim 11$
and
$(1.2^{+3.8}_{-1.1}) \times 10^{-3} M_\odot ~ {\rm yr}^{-1} ~ {\rm Mpc}^{-3}$ at $z \sim 9.6$.
%SFRD $(z \sim 11) = (1.9^{+5.4}_{-1.6}) \times 10^{-3} M_\odot ~ {\rm yr}^{-1} ~ {\rm Mpc}^{-3}$
%SFRD $(z \sim 9.6) = (1.2^{+3.8}_{-1.1}) \times 10^{-3} M_\odot ~ {\rm yr}^{-1} ~ {\rm Mpc}^{-3}$.
For consistency with the other measurements derived and compiled by \cite{Bouwens12b},
we assumed a star formation rate of one solar mass per year
produces a UV (0.14\um) luminosity of $8 \times 10^{27}$ ergs s\per\ Hz\per,
as from a \cite{Salpeter55} IMF truncated between 0.1 and 125 $M_\odot$.

Our SFRD estimate at $z \sim 9.6$ 
is consistent with an independent estimate of 
$(1.8^{+4.3}_{-1.1}) \times 10^{-3} M_\odot ~ {\rm yr}^{-1} ~ {\rm Mpc}^{-3}$
%$(1.0^{+2.4}_{-0.6}) \times 10^{-3} M_\odot ~ {\rm yr}^{-1} ~ {\rm Mpc}^{-3}$
based on MACS1149-JD1
presented in \cite{Zheng12}.
Only 12 clusters were searched in that work.
To account for the larger volume now searched without additional \~26th magnitude $z \sim 10$ candidates,
we rescaled this estimate by a factor of $12/17$, yielding
$(1.3^{+3.0}_{-0.8}) \times 10^{-3} M_\odot ~ {\rm yr}^{-1} ~ {\rm Mpc}^{-3}$.
This value is plotted in Fig.~\ref{fig:SFRD}.

Our joint constraint from both galaxies on SFRD at $z > 8$ is also plotted as the gray shaded region in Fig.~\ref{fig:SFRD}.
Given the large uncertainties,
we cannot confidently discriminate between the trend observed at lower redshifts ($dM^* / dz \sim 0.30$)
and the sharp drop off suggested by the paucity of $z \sim 10$ galaxies detected in the field.

The dominant uncertainty in our SFRD measurement is
the Poisson uncertainty of \~0.7 dex (a factor of \~5) given our single detection.
Subdominant uncertainties include cosmic variance in 17 independent fields \citep[e.g.,][]{TrentiStiavelli08};
uncertainties in the lens models \citep[e.g.,][]{Bradac09};
and incompleteness in our ability to detect \~26th magnitude galaxies (25.4 $<$ F160W $<$ 26.4)
after masking out areas corresponding to foreground galaxies.
Accounting for overlapping regions in the source plane (multiple images)
would slightly increase our derived SFRD.

\section{Conclusions}
\label{sec:conc}

We have discovered a candidate for the earliest galaxy yet known at
%$z =11.0_{-0.7}^{+0.4}$ 
\ztwosig\ (95\% confidence limits)
when the universe was 
%$410_{+40}^{-20}$ 
\agetwosig\ million years old.
The galaxy is strongly lensed by \macs\ producing three magnified images,
including two observed at \~26th magnitude AB (\~162 and 136 nJy)
%($\sim$0.15\uJy)
in HST WFC3/IR F160W imaging ($\sim$1.4--1.7\um).
The intrinsic (delensed) magnitude is \~20 nJy (mag \~ 28.2)
based on our lens models for a galaxy at $z \sim 11$.
%\~8 times fainter (\~20 nJy; mag \~ 28.2).
The unattenuated continuum is $\sim$0.25 mag brighter (lensed \~0.2 \uJy\ in the brightest image).
%($\sim$0.19\uJy).

%Spitzer/IRAC upper limits and our lens modeling further support the high redshift solution.
Spitzer/IRAC upper limits further support the high redshift solution.
We tested a broad range of lower redshift interlopers,
including some previously published as high-redshift candidates,
and showed that none is able to reproduce the observed HST+Spitzer photometry.
% fullprob.py
Galaxies of known types at $z < 9.5$ are formally ruled out at 7.2\sig,
with the next most likely alternative being an early type and/or dusty galaxy at $z \sim 2.5$.
Our Bayesian priors assume a $z \sim 2$ galaxy is over 80 times more likely than a $z \sim 11$ galaxy,
making our $z \sim 11$ claim more conservative than if such a prior were neglected.
For \JD\ to be at $z < 9.5$,
it appears it would have to belong to a new class of objects not yet observed.

The discoveries of both \JD\ at $z \sim 10.8$ (this work)
and MACS1149-JD1 at $z \sim 9.6$ \citep{Zheng12} 
in CLASH observations of 17 clusters to date
are consistent with extrapolations of luminosity functions observed at lower redshifts
\citep{Bouwens12a,Bradley12b},
assuming a linear evolution of \Mstar\ with redshift,
and as convolved through our lens models.
If these extrapolations are valid to $z \gtrsim 10$,
then low luminosity galaxies could have reionized the universe \citep{Bouwens12a,Kuhlen12}.
However these extrapolations are in conflict with the paucity of $z \sim 10$ galaxies discovered in unlensed fields,
suggesting a rapid buildup in star formation density between $z \sim 10$ and 8 \citep{Bouwens11N,Oesch12a}.
Our data do not allow us to discriminate between these two scenarios,
given the large uncertainties dominated by the Poisson statistics of these two detections.

\JD\ is likely close to the characteristic luminosity for a $z \sim 11$ galaxy ($\sim$1--3 $L^*$),
producing a few \Msun\ yr\inv,
with an inferred stellar mass of roughly
\~\tentothe8--\tentothe9\Msun,
%\~ \tentothe{8-9}\Msun.
and a half-light radius of $\lesssim 100$ pc (deconvolved and delensed).
This is smaller by a factor of a few than the average size expected as
extrapolated from lower redshifts \citep{Oesch10b,Mosleh12}
with an intrinsic scatter in sizes of perhaps a factor of \~2 \citep{Ferguson04}.
The size of $\lesssim 100$ pc is similar to the sizes of bright knots 
observed in lensed galaxies at $5 < z < 8$ \citep{Franx97,Bradley08,Bradley12a}.

Thanks to the magnified views afforded us by gravitational lensing,
this galaxy may be studied further with existing
%including the Atacama Large Millimeter Array (ALMA),
%and the telescopes of tomorrow, 
and future large telescopes,
including the James Webb Space Telescope (JWST; \citealt{Gardner06JWST,Stiavelli09book})
and extremely large ground-based telescopes constructed in the northern hemisphere.
Unfortunately due to its high declination of +70, it is not accessible to southern telescopes such as 
the Atacama Large Millimeter Array (ALMA).

This $z \sim 11$ candidate approaches the redshift limit of galaxies
detectable by Hubble's WFC3/IR camera.
Galaxies at $z > 12$ would drop out completely of the F140W filter
and to an increasing degree in F160W until $z \sim 13$,
when all the light redward of Lyman-$\alpha$ would be 
redshifted beyond the observable wavelength range.

% Pawlik11

%%%%%%%%%%%%%%

%Then the body of the main text appears after the intro paragraph.
%Figure environments can be left in place in the document.
%\verb|\includegraphics| commands are ignored since Nature wants
%the figures sent as separate files and the captions are
%automatically moved to the end of the document (they are printed
%out with the \verb|\end{document}| command. However, tables must
%be manually moved to the end of the document, after the addendum.

%\caption{Each figure legend should begin with a brief title for
%the whole figure and continue with a short description of each
%panel and the symbols used. For contributions with methods
%sections, legends should not contain any details of methods, or
%exceed 100 words (fewer than 500 words in total for the whole
%paper). In contributions without methods sections, legends should
%be fewer than 300 words (800 words or fewer in total for the whole
%paper).}

%\clearpage

%--

%building blocks

%bridge gap in mass between obs and sims

%--

%Two-prong approach: field + clusters.
%JWST

%==

%The first galaxies\cite{BrommYoshida11}.

%Cluster Lensing\cite{KneibNatarajan11}.

%JWST\cite{Gardner06JWST,Stiavelli09book}.

%Reionization\cite{Robertson10}.

%comoving (scaled up to the current size of the universe).

\acknowledgements{
We thank Nor Pirzkal for useful discussions regarding the WFC3/IR backgrounds
and contributing the single emission line galaxy scenario.
We thank the referee for useful comments which helped us improve the manuscript.

The CLASH Multi-Cycle Treasury Program is based on observations made with the NASA/ESA Hubble Space Telescope. 
The Space Telescope Science Institute is operated by the Association of Universities for Research in Astronomy, 
Inc.~under NASA contract NAS 5-26555. 
ACS was developed under NASA contract NAS 5-32864.

This work is also based in part on observations made with the Spitzer Space Telescope, 
which is operated by the Jet Propulsion Laboratory, California Institute of Technology under a contract with NASA.
%Support for this work was provided by NASA through an award issued by JPL/Caltech.

AZ is supported by contract research Internationale Spitzenforschung II-1 of the Baden-W\"urttemberg Stiftung.
% some of the lens modeling:
We wish to acknowledge the support of the 
Michigan State University High Performance Computing Center and the Institute for Cyber-Enabled Research.
SS was supported by the DFG cluster of excellence Origin and Structure of the Universe. 
The work of LAM was carried out at Jet Propulsion Laboratory, California Institute of Technology, under a contract with NASA.
The Dark Cosmology Centre is funded by the DNRF.
}

%% Put the bibliography here, most people will use BiBTeX in
%% which case the environment below should be replaced with
%% the \bibliography{} command.

%\newpage

{\it Facilities:}
\facility{HST (WFC3, ACS)};
\facility{Spitzer (IRAC)}
%\facility{Subaru (Suprime-Cam)};
%\facility{KPNO (Mayall)};
%\facility{Chandra (ACIS)};
%\facility{MMT (Hectospec)}

\bibliographystyle{astroads}
\bibliography{papersb}

%\begin{thebibliography}{1}
%\bibitem{dummy} Articles are restricted to 50 references, Letters to 30.
%\bibitem{dummyb} No compound references -- only one source per reference.
%\end{thebibliography}

%% Here is the endmatter stuff: Supplementary Info, etc.
%% Use \item's to separate, default label is "Acknowledgements"

%\begin{addendum}
% \item[Supplemental Information] is linked to the online version of the paper at www.nature.com/nature
% \item Put acknowledgements here.
% \item[Author Contributions] 
%D.A.C. performed most of the data analysis, wrote all of the text, and created all the figures presented here.
%R.J.B., W.Z., L.B., and D.A.C.~all independently identified this object as a high-redshift candidate
%using their respective well-tested pipelines.
%%and D.A.C.~first identified it as a $z \approx 11$ candidate.
%Other...
% \item[Author Information] 
%Reprints and permissions information is available at
%www.nature.com/reprints. The authors declare no competing financial interests.
%Readers are welcome to comment on the online version of this article at
%www.nature.com/nature.
%Correspondence and requests for materials
%should be addressed to D.A.C.~(email: dcoe@stsci.edu).
%\end{addendum}

%%
%% TABLES
%%
%% If there are any tables, put them here.
%%

\end{document}

%% file: obs.tex
\begin{deluxetable}{lcr}
\tablewidth{0.63\columnwidth}
\tablecaption{\label{tab:obs}Observed Filters and Integration Times}
\tablehead{
\colhead{Filter}&
%\colhead{$\lambda$}&
\colhead{wavelength\super{a}}&
\colhead{exposure}
}
\startdata
F225W&
0.24$\mu$m&
3805 sec
\\
F275W&
0.27$\mu$m&
3879 sec
\\
F336W&
0.34$\mu$m&
2498 sec
\\
F390W&
0.39$\mu$m&
2545 sec
\\
F435W&
0.43$\mu$m&
2124 sec
\\
F475W&
0.47$\mu$m&
2248 sec
\\
F555W&
0.54$\mu$m&
7740 sec
\\
F606W&
0.59$\mu$m&
2064 sec
\\
F625W&
0.63$\mu$m&
2131 sec
\\
F775W&
0.77$\mu$m&
2162 sec
\\
F814W&
0.81$\mu$m&
12760 sec
\\
F850LP&
0.90$\mu$m&
4325 sec
\\
F105W&
1.06$\mu$m&
2914 sec
\\
F110W$^{\rm b}$&
1.15$\mu$m&
1606 sec
\\
F125W&
1.25$\mu$m&
2614 sec
\\
F140W&
1.39$\mu$m&
2411 sec
\\
F160W&
1.54$\mu$m&
5229 sec
\\
IRAC ch1&
3.55$\mu$m&
18000 sec
\\
IRAC ch2&
4.50$\mu$m&
18000 sec
\\
\vspace{-0.1in}
\enddata
%\tablecomments{}
%\tablenotetext{a}{Effective ``pivot'' wavelength (defined as in \citealt{TokunagaVacca05}).}
\tablenotetext{1}{Effective ``pivot'' wavelength \citep{TokunagaVacca05}.}
%\tablenotetext{b}{Visit A2 only, excluding visit A9 which exhibits significantly elevated and non-Poissonian backgrounds due to Earthshine (\S\ref{sec:HSTphot}).}
\tablenotetext{2}{Visit A2 only, excluding visit A9 (\S\ref{sec:HSTphot}).}
\end{deluxetable}

%% file: catsum.tex
% ~/m0647/highz/cattable.py
% modified manually
% ~/m0647/highz/catsummary.py
\begin{deluxetable*}{lcccc}
\tablecaption{\label{tab:cat}Coordinates, Observed Filters, and Photometry of the $J$-dropouts}
\tablehead{
\colhead{}&
%\colhead{$\lambda$}&
%\colhead{wavelength$^{\rm a}$}&
%\colhead{exposure}&
\colhead{JD1}&
\colhead{JD2}&
\colhead{JD3}&
%\colhead{sum}
%\colhead{JD1+JD2+JD3}
\colhead{JD1 + JD2 + JD3\super{a}}
}
\startdata
R.A. (J2000)&
\tt~~06:47:55.731&
\tt~~06:47:53.112&
\tt~~06:47:55.452\\
Decl. (J2000)&
\tt +70:14:35.76~&
\tt +70:14:22.94~&
\tt +70:15:38.09~\\
F225W&
\tt~-129~$\pm$~51~nJy (-2.5$\sigma$)&
\tt~~-40~$\pm$~50~nJy (-0.8$\sigma$)&
\tt~~~12~$\pm$~32~nJy ( 0.4$\sigma$)&
\tt~-157~$\pm$~78~nJy (-2.0$\sigma$)
\\
F275W&
\tt~~-95~$\pm$~51~nJy (-1.9$\sigma$)&
\tt~~-31~$\pm$~42~nJy (-0.8$\sigma$)&
\tt~~~49~$\pm$~24~nJy ( 2.0$\sigma$)&
\tt~~-77~$\pm$~70~nJy (-1.1$\sigma$)
\\
F336W&
\tt~~~~2~$\pm$~37~nJy ( 0.0$\sigma$)&
\tt~~~49~$\pm$~29~nJy ( 1.7$\sigma$)&
\tt~~-25~$\pm$~18~nJy (-1.4$\sigma$)&
\tt~~~25~$\pm$~50~nJy ( 0.5$\sigma$)
\\
F390W&
\tt~~~-8~$\pm$~20~nJy (-0.4$\sigma$)&
\tt~~~~1~$\pm$~19~nJy ( 0.1$\sigma$)&
\tt~~~~1~$\pm$~10~nJy ( 0.1$\sigma$)&
\tt~~~-6~$\pm$~29~nJy (-0.2$\sigma$)
\\
F435W&
\tt~~~~0~$\pm$~26~nJy ( 0.0$\sigma$)&
\tt~~~43~$\pm$~24~nJy ( 1.8$\sigma$)&
\tt~~~~5~$\pm$~14~nJy ( 0.4$\sigma$)&
\tt~~~48~$\pm$~38~nJy ( 1.3$\sigma$)
\\
F475W&
\tt~~~-2~$\pm$~14~nJy (-0.1$\sigma$)&
\tt~~-27~$\pm$~16~nJy (-1.7$\sigma$)&
\tt~~~~7~$\pm$~~8~nJy ( 0.9$\sigma$)&
\tt~~-22~$\pm$~23~nJy (-1.0$\sigma$)
\\
F555W&
\tt~~~-3~$\pm$~~9~nJy (-0.3$\sigma$)&
\tt~~~12~$\pm$~~7~nJy ( 1.7$\sigma$)&
\tt~~~~6~$\pm$~~4~nJy ( 1.4$\sigma$)&
\tt~~~15~$\pm$~12~nJy ( 1.3$\sigma$)
\\
F606W&
\tt~~~~3~$\pm$~16~nJy ( 0.2$\sigma$)&
\tt~~~13~$\pm$~20~nJy ( 0.6$\sigma$)&
\tt~~~-1~$\pm$~~6~nJy (-0.1$\sigma$)&
\tt~~~15~$\pm$~26~nJy ( 0.6$\sigma$)
\\
F625W&
\tt~~-35~$\pm$~21~nJy (-1.7$\sigma$)&
\tt~~-52~$\pm$~24~nJy (-2.2$\sigma$)&
\tt~~~23~$\pm$~10~nJy ( 2.3$\sigma$)&
\tt~~-64~$\pm$~33~nJy (-1.9$\sigma$)
\\
F775W&
\tt~~~~4~$\pm$~30~nJy ( 0.2$\sigma$)&
\tt~~-16~$\pm$~52~nJy (-0.3$\sigma$)&
\tt~~~~4~$\pm$~10~nJy ( 0.3$\sigma$)&
\tt~~~-8~$\pm$~61~nJy (-0.1$\sigma$)
\\
F814W&
\tt~~~~0~$\pm$~~8~nJy ( 0.1$\sigma$)&
\tt~~~-2~$\pm$~~5~nJy (-0.3$\sigma$)&
\tt~~~-2~$\pm$~~3~nJy (-0.8$\sigma$)&
\tt~~~-3~$\pm$~10~nJy (-0.3$\sigma$)
\\
F850LP&
\tt~~~-3~$\pm$~30~nJy (-0.1$\sigma$)&
\tt~~~~1~$\pm$~29~nJy ( 0.0$\sigma$)&
\tt~~~~6~$\pm$~15~nJy ( 0.4$\sigma$)&
\tt~~~~4~$\pm$~45~nJy ( 0.1$\sigma$)
\\
F105W&
\tt~~~11~$\pm$~12~nJy ( 0.9$\sigma$)&
\tt~~~14~$\pm$~13~nJy ( 1.1$\sigma$)&
\tt~~~~3~$\pm$~~5~nJy ( 0.6$\sigma$)&
\tt~~~28~$\pm$~18~nJy ( 1.6$\sigma$)
\\
F110W\super{b}&
\tt~~~-8~$\pm$~10~nJy (-0.8$\sigma$)&
\tt~~~~3~$\pm$~~9~nJy ( 0.3$\sigma$)&
\tt~~~~7~$\pm$~~4~nJy ( 1.9$\sigma$)&
\tt~~~~2~$\pm$~14~nJy ( 0.1$\sigma$)
\\
F125W&
\tt~~~-3~$\pm$~10~nJy (-0.3$\sigma$)&
\tt~~~~7~$\pm$~16~nJy ( 0.5$\sigma$)&
\tt~~~~2~$\pm$~~5~nJy ( 0.4$\sigma$)&
\tt~~~~6~$\pm$~20~nJy ( 0.3$\sigma$)
\\
F140W&
\tt~~~63~$\pm$~10~nJy ( 6.0$\sigma$)&
\tt~~~50~$\pm$~~8~nJy ( 6.7$\sigma$)&
\tt~~~26~$\pm$~~4~nJy ( 6.1$\sigma$)&
\tt~~139~$\pm$~14~nJy ( 9.9$\sigma$)
\\
&
~~~~$= 26.90 \pm 0.17$ mag AB&
~~~~$= 27.15 \pm 0.17$ mag AB&
~~~~$= 27.86 \pm 0.17$ mag AB&
~~~~$= 26.04 \pm 0.11$ mag AB
\\
F160W&
\tt~~162~$\pm$~13~nJy (12.4$\sigma$)&
\tt~~136~$\pm$~~9~nJy (15.1$\sigma$)&
\tt~~~42~$\pm$~~4~nJy (10.1$\sigma$)&
\tt~~341~$\pm$~16~nJy (21.3$\sigma$)
\\
&
~~~~$= 25.88 \pm 0.09$ mag AB&
~~~~$= 26.07 \pm 0.07$ mag AB&
~~~~$= 27.34 \pm 0.10$ mag AB&
~~~~$= 25.07 \pm 0.05$ mag AB
\\
IRAC ch1&
\tt~~~~$<$~277~nJy\super{c}&
\tt~~~~$<$~166~nJy&
\tt~~~~$<$~166~nJy&
\tt~~~~$<$~363~nJy
%\tt~~~~0~$\pm$~363~nJy ( 0.0$\sigma$)
\\
IRAC ch2&
\tt~~~~$<$~245~nJy\super{c}&
\tt~~436~$\pm$~139~nJy ( 3.1$\sigma$)&
\tt~~~~$<$~138~nJy&
\tt~~436~$\pm$~314~nJy ( 1.4$\sigma$)
\\
\vspace{-0.1in}
\enddata
\tablecomments{
Fluxes in nanoJanskys (nJy) may be converted to AB magnitudes via m$_{AB} \approx 26 - 2.5 \log_{10}(F_{\nu} / (145~{\rm nJy})$).
Magnitude uncertainties, where given, are non-Gaussian but are approximated as $2.5 \log_{10}(e)$ times the fractional flux uncertainties.
}
%\tablenotetext{1}{``Pivot'' wavelength (weighted average of the filter response function).}
%\tablenotetext{a}{Effective ``pivot'' wavelength (defined as in \citealt{TokunagaVacca05}).}
\tablenotetext{1}{Sum of all three images with uncertainties added in quadrature.}
\tablenotetext{2}{Visit A2 only, excluding visit A9 which exhibits significantly elevated and non-Poissonian backgrounds due to Earthshine (\S\ref{sec:HSTphot}).}
\tablenotetext{3}{Includes uncertainties from modeling and subtracting a nearby brighter galaxy.  
More conservative estimates of these uncertainties were also considered in the analysis (\S\ref{sec:IRACphot}).}
%\tablenotetext{3}{Conservative 1-\sig\ upper limits.  IRAC photometric measurements of JD1 are contaminated by a nearby galaxy.}
% Corresponding to mag = 23 yielding only 3-sigma
\end{deluxetable*}

%% file: arctable.tex
\begin{deluxetable}{crccl}
\tablecaption{\label{tab:arcs}24 Multiple Images of 9 Strongly Lensed Galaxies}
\tablewidth{0.9\columnwidth}
\tablehead{
\colhead{}&
\colhead{}&
\colhead{R.A.}&
\colhead{Decl.}&
\colhead{Photometric}\\
\colhead{}&
\colhead{ID}&
\colhead{(J2000.0)}&
\colhead{(J2000.0)}&
\colhead{Redshift\super{a}}
}
\startdata
&
1a&
06 47 51.87&
+70 15 20.9&
$2.2 \pm 0.1$\\
&
b&
06 47 48.54&
+70 14 23.9&
$2.2 \pm 0.1$\\
&
c&
06 47 52.01&
+70 14 53.8&
$2.2 \pm 0.1$
\vspace{0.02in}\\
\hline
&
2a&
06 48 00.33&
+70 15 00.7&
$4.7 \pm 0.1$\\
&
b&
06 48 00.33&
+70 14 55.4&
$4.7 \pm 0.1$\\
&
c&
06 47 58.62&
+70 14 21.8&
$4.7 \pm 0.1$
\vspace{0.02in}\\
\hline
&
3a&
06 47 53.85&
+70 14 36.2&
$3.1 \pm 0.1$\\
&
b&
06 47 53.41&
+70 14 33.5&
$3.1 \pm 0.1$
\vspace{0.02in}\\
\hline
&
4a&
06 47 42.75&
+70 14 57.7&
$1.9 \pm 0.1$\\
&
b&
06 47 42.93&
+70 14 44.5&
$1.9 \pm 0.1$\\
&
c&
06 47 45.37&
+70 15 25.8&
$1.9 \pm 0.1$
\vspace{0.02in}\\
\hline
&
5a&
06 47 41.04&
+70 15 05.5&
$6.5 \pm 0.15$\\
&
b&
06 47 41.16&
+70 14 34.4&
$6.5 \pm 0.15$
\vspace{0.02in}\\
\hline
JD1&
6a&
06 47 55.74&
+70 14 35.7&
$11.0 \pm 0.3$\\
JD2&
b&
06 47 53.11&
+70 14 22.8&
$11.0 \pm 0.3$\\
JD3&
c&
06 47 55.45&
+70 15 38.0&
$11.0 \pm 0.3$
\vspace{0.02in}\\
\hline
&
7a&
06 47 50.91&
+70 15 19.9&
$2.2 \pm 0.15$\\
&
b&
06 47 47.73&
+70 14 23.2&
$2.2 \pm 0.15$\\
&
c&
06 47 48.69&
+70 14 59.8&
$2.2 \pm 0.15$
\vspace{0.02in}\\
\hline
&
8a&
06 47 48.61&
+70 15 15.8&
$2.3 \pm 0.1$\\
&
b&
06 47 47.34&
+70 15 12.5&
$2.3 \pm 0.1$
\vspace{0.02in}\\
\hline
&
9a&
06 47 43.79&
+70 15 00.4&
$5.9 \pm 0.15$\\
&
b&
06 47 44.98&
+70 14 23.2&
$5.9 \pm 0.15$\\
&
c&
06 47 49.06&
+70 15 37.7&
$5.9 \pm 0.15$
\\
\vspace{-0.1in}
\enddata
\tablenotetext{1}{Joint likelihoods for each system with approximate 68\% uncertainties.}
\end{deluxetable}